\providecommand{\columnwiselinenumbersfalse}{}
\providecommand{\runningpagewiselinenumbers}{}
\providecommand{\linenumbers}{}

\documentclass[desactivate]{aa}

\usepackage{graphicx,txfonts,lipsum,subcaption,lscape,placeins,xcolor}
                                
\newcommand{\htwo}{\ensuremath{{\rm H}_2}}
\newcommand{\cone}{C$\,$\resizebox{!}{5pt}{I}}
\def\gta{\ifmmode\stackrel{>}{_{\sim}}\else$\stackrel{>}{_{\sim}}$\fi}
\def\lta{\ifmmode\stackrel{<}{_{\sim}}\else$\stackrel{<}{_{\sim}}$\fi}

\begin{document}

   \title{R Coronae Borealis fadings: a dusty gas cloud eclipse model}

   \author{Mark A. Walker\inst{1}\corrauth{mark.walker@manlyastrophysics.org}   
        \and Artem V. Tuntsov\inst{1}\email{artem.tuntsov@manlyastrophysics.org}
        \and Arthur G. Suvorov\inst{1,2}\email{arthur.suvorov@tat.uni-tuebingen.de}
        }

   \institute{Manly Astrophysics, 15/41-42 East Esplanade, Manly, NSW 2095, Australia
   \and Theoretical Astrophysics, Eberhard Karls University, Auf der Morgenstelle 10, 72076 T\"ubingen, Germany}

   \date{Received 23 April 2026}

  \abstract
   {R Coronae Borealis stars (RCBs) are hydrogen-deficient supergiants that undergo deep fading events due to transient extinction by dust. First observed over two centuries ago, that behaviour remains poorly understood.
   Our aim is to investigate the possibility that the fading events are eclipses by orbiting, dusty gas clouds.
   We construct a simple physical model of the dust wind that is driven from the cloud surface by the stellar radiation field, and we calculate the extinction and scattered light from wind and cloud. We consider a broad range of dust compositions. For the model that best matches the data, we sketch out implications for the RCB phenomenon more broadly.
   Our calculations show that conventional dust materials like graphite and silicate produce eclipses whose ingress is too slow and whose morphology too symmetric to account for RCB fadings. But dust made of solid \htwo\ yields a good match: the eclipses are deep and strongly asymmetric, with rapid onset and slow recovery, and event timescales are about right if the clouds’ periastra are within a few tens of AU. Such close approaches imply tidal stripping/disruption of each cloud, with hydrogen-deficient debris accreting onto the star via a disk. Starlight incident on the disk creates a metal-ion plasma that emits a powerful brem\ss trahlung continuum, peaking in the mid-IR.
   We conclude that eclipses by orbiting, dusty clouds could be the cause of RCB fading events, but only if the dust is made of solid \htwo, in which case both the unusual, hydrogen-deficient nature of the stars and their mid-IR excess follow naturally. It is, however, challenging to account for the high eclipse rates that are observed; we propose a scenario in which the progenitor star is a wide binary inside a massive halo of clouds, suggesting a connection to the Galactic ``missing mass'' problem.}


   \maketitle

\section{Introduction}
RCBs are a class of variable stars similar to the prototype R Coronae Borealis. They are extremely rare, with the total number known in our own Galaxy only recently passing 100 stars \citep{2020A&A...635A..14T}. The observed characteristics of RCBs include \citep[e.g.][]{1996PASP..108..225C,2012JAVSO..40..539C}: (i) unpredictable fading events, lasting for months to years; (ii) high luminosities; (iii) an almost complete lack of hydrogen; (iv) a very high abundance of carbon; (v) a collection of optical and UV ``chromospheric'' emission lines; (vi) strong mid-infrared excess continuum emission; and, (vii) large values of the isotopic ratio ${}^{18}$O:${}^{16}$O. It was demonstrated long ago \citep{1973A&A....22....9T,1989ApJ...342..430I} that nuclear burning at the base of an extended, low-mass helium envelope, surrounding a degenerate carbon-oxygen core, can reproduce the positions of RCBs in the Hertzsprung-Russell diagram. And that configuration may readily account for the high carbon abundance if there is convective mixing of the nuclear burning products up into the photosphere.

Currently there are two main ideas for how a helium supergiant configuration may be reached: from a pair of white dwarfs (WDs) in a binary, one a high-mass CO-WD and the other a low-mass He-WD, which merge to form a single star \citep{1984ApJ...277..355W,2002MNRAS.333..121S}; or, as a final, helium shell-burning phase in the lives of giant stars \citep{1983ApJ...264..605I}. The observed isotopic ratio ${}^{18}$O:${}^{16}$O favours the double degenerate model \citep{2007ApJ...662.1220C}. However, although both final-flash and double-degenerate models provide evolutionary pathways to the helium supergiant configuration, neither of them explains the dust dimming events that are a defining characteristic of the RCB phenomenon. By contrast, in this paper we focus on modelling those dimming events and in doing so we uncover a possible new evolutionary pathway.

Because of the abundant \cone\ and C$_2$ seen in the stellar photospheres of RCBs \citep[e.g.][]{1975A&A....44..383S,2000A&A...353..287A} it has usually been assumed that the dust responsible for the fading events is itself some form of carbon \citep{1939ApJ....90..294O}. But the mechanism by which carbon particulates might sporadically condense and be ejected from the star remains obscure. Indeed there are some difficulties in trying to explain RCB fading events in that way, as follows. First, condensation of carbon particulates requires temperatures $\lta 1{,}500\,{\rm K}$, whereas the majority of RCBs have photospheric temperatures that are four or five times higher. That argues for nucleation at distances as large as $20\;R_*$ (stellar radii) from the star \citep[e.g.][]{1988MNRAS.233...65F}, whereas the typical spectrophotometric evolution that is observed during fading events seems to require dust formation much closer to the photosphere \citep{1993ASPC...45..115W}. In the latter case the necessary gas thermodynamic conditions may be met as a result of pulsation-induced shocks \citep{1996A&A...313..217W}. However, it remains difficult to {\it initiate\/} dust formation so close to the star because any grains exposed to the starlight would promptly evaporate \citep[][their footnote 3]{1996A&A...313..217W}. Secondly, the dust sprays in RCBs are inferred \citep{1972ApJ...178L.129F} and observed \citep{2012A&A...539A..56J,2011ApJ...743...44C} to be highly directional, and that is challenging to understand in the context of an isolated, spherically-symmetric body such as a star. Finally, it is unclear why carbon condensation should occur in discrete events, at apparently random times. There have been suggestions that the trigger for the fadings is linked to stellar pulsations \citep[e.g.][]{1999AJ....117.3007L,2007MNRAS.375..301C}, but when all the observational data are considered the evidence in support of that idea is not strong \citep{2025MNRAS.537.2635C}.

Here we describe a different model for RCB fading events, in which a directed, dusty outflow arises not from the star but from an orbiting gas cloud. This scenario avoids the difficulties just described: there are special times and directions that are associated with any given orbit, and the dust responsible for each eclipse is brought in ready-made within the gas cloud. It is then straightforward to generate a sequence of randomly-timed eclipses if there are many clouds on long-period orbits, each transiting at a different time. The possibility that orbiting, dusty gas clouds might be responsible for RCB fading events was actually suggested a long time ago \citep{1972PASP...84R.646W,1972ApJ...178L.129F}, but it was quickly rejected because the eclipses were anticipated to be unlike RCB light curves \citep{1973MNRAS.161..293F}. We will show that those eclipse properties were correctly anticipated for most, but not all dust compositions.

This paper is organised as follows. The next section describes the general picture we are using, and details some further background. Then \S3 evaluates how mass is lost from the cloud, and thus provides the launch conditions for the dust wind, whose structure is calculated in \S4. The resulting light curves are presented in \S5. Section 6 describes implications of the eclipse models --- specifically the tidal stripping/disruption of the cloud, and the resulting accretion of matter by the star. Up to that point all of the calculations are for dust made of solid \htwo, then in \S7 we assess the properties of the winds that arise when the dust has a more conventional composition. Some predictions of our eclipse model are given in \S8, followed by Discussion (\S9) and Conclusions (\S10). Relevant material that lies slightly outside the main development path is consigned to five appendices. Appendix A quantifies the optical properties of dust made of solid \htwo, along with two more conventional dust materials. Appendix B describes the production of atomic hydrogen as a result of photo-dissociation of solid \htwo\ that is exposed to far-UV from the RCB star. Appendix C presents light curves computed for large grains of solid \htwo\ (in contrast with the reference calculations in the main text, which employ small grains). Appendix D shows examples of extinction-only light curves for tidally deformed clouds. Finally, Appendix E sketches one possible scenario that can give rise to an intense shower of cloud disruptions.

\begin{figure*}[t]%
    \centering
    \hbox{\hskip 2cm{\includegraphics[width=9cm]{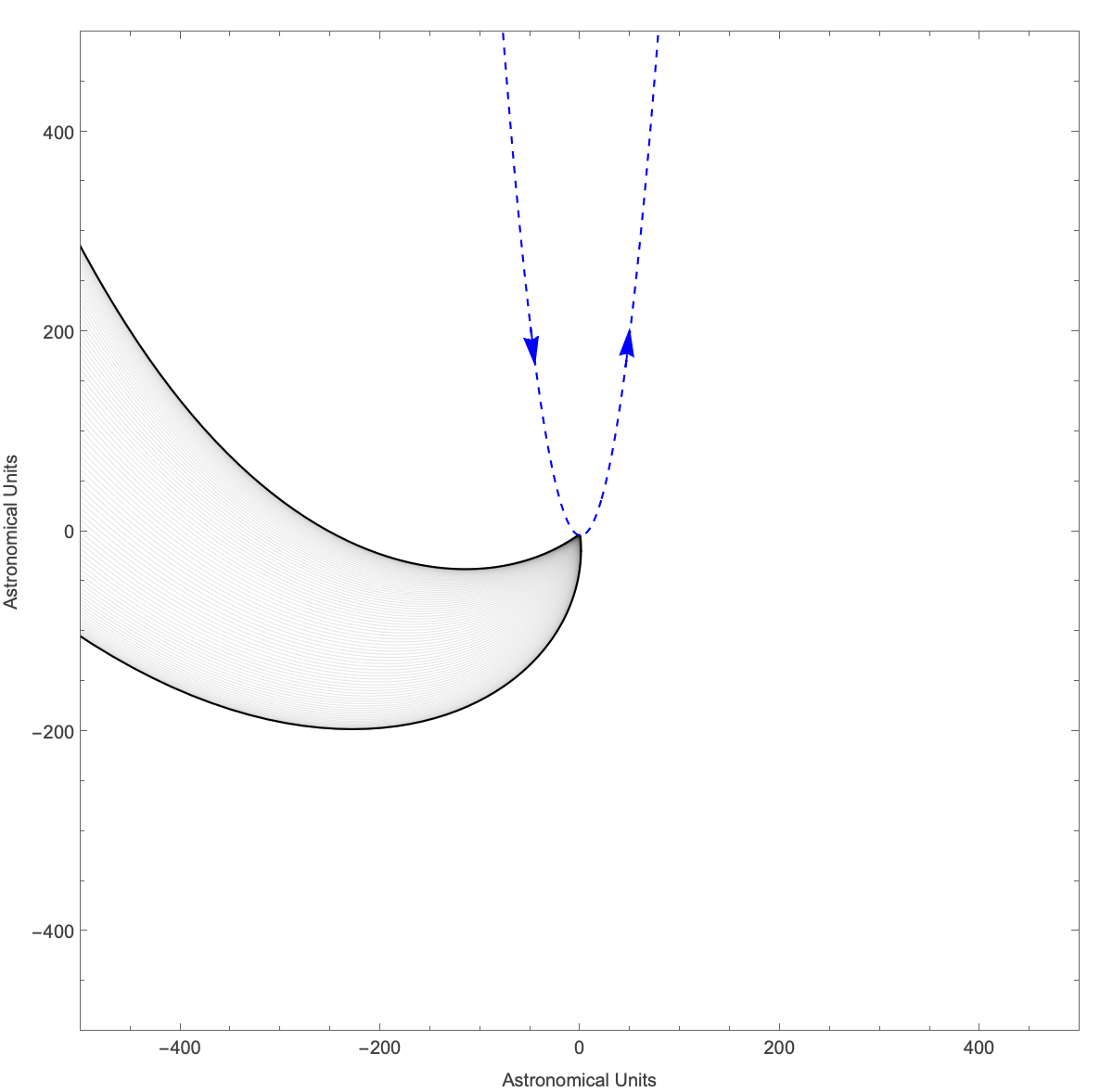}}\qquad{\includegraphics[width=5cm]{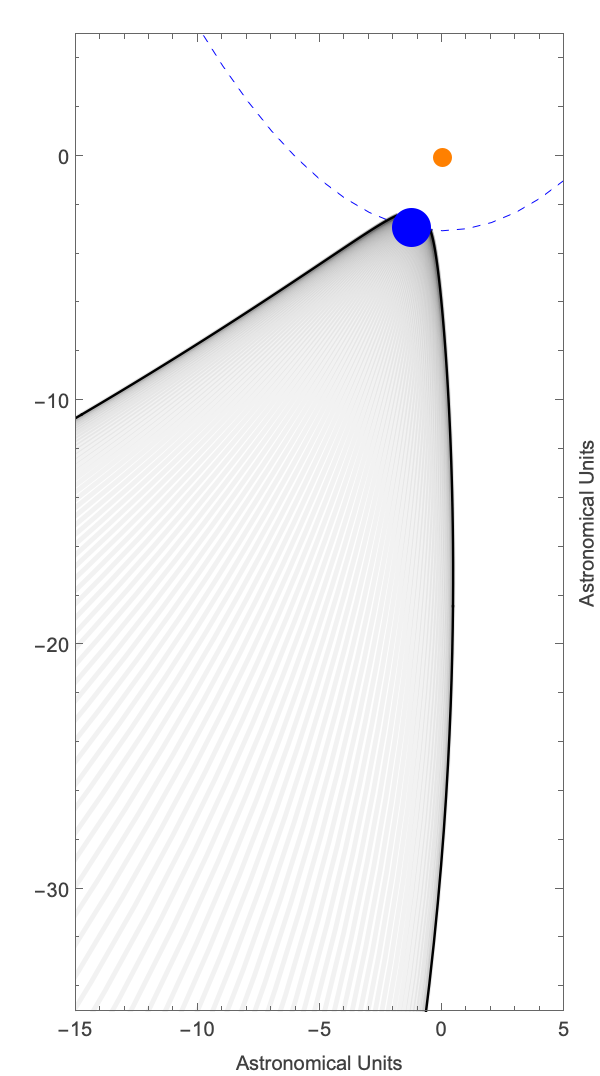}}}%
    \caption{Diagram of our model hydrogen dust wind structure, showing a slice in the orbital plane for an orbit with periastron at $3\;{\rm AU}$. This snapshot corresponds to the point of deepest eclipse for an observer at $(0,-\infty)$, with the occultation front just at the point of obscuring the entire star. The left-/right-hand panels show wide/narrow views of the wind, with distances marked in AU relative to the star at the origin. The cloud's orbit is shown as the dashed blue line; the cloud is shown as a blue disk, and the star as an orange disk. Solid, black lines show the hydrogen dust wind. For visual emphasis only, the interior of the wind is shaded.}%
    \label{fig:windstructure}%
\end{figure*}

\section{The overall picture}\label{sec:overallstructure}
In order to produce eclipses as brief as observed, the cloud radius cannot be large. For example: a cloud on a Keplerian orbit at a distance of $\sim 10\;{\rm AU}$ from the star would move only $\sim 2\;{\rm AU}$ in $1\;{\rm yr}$ --- which is the median eclipse duration for RCBs \citep{2025MNRAS.537.2635C}. Such tiny structures have not been directly observed in the neutral interstellar medium (ISM), but molecular gas at high densities and low temperatures is notoriously difficult to detect -- it is a type of baryonic dark matter -- and our Galaxy could harbour a lot of mass in this form \citep{1994A&A...285...79P,1994A&A...285...94P,1996ApJ...472...34G,1998ApJ...498L.125W}. Some guidance is available from theoretical models of molecular cloud structure which have demonstrated \citep{2019ApJ...881...69W} hydrostatic equilibria for tiny clouds (e.g. $\sim 1\;{\rm AU}$ in radius) with masses in the planetary range. Those model structures are all so cold and dense that they manifest phase equilibrium between gaseous \htwo\ and its condensate --- a possibility anticipated by \citet{1994A&A...285...94P}. Although the condensate can be liquid, the solid form is relevant to the major part of the mass-radius plane, and for brevity we use the terms ``solid'' and ``snow'' to refer to both types of \htwo\ condensate. We use the term ``snow cloud'' to refer specifically to a gas cloud in which \htwo\ condensates are present.

Particles of solid \htwo\ (``hydrogen dust'') are not part of the conventional picture of interstellar dust \citep[e.g.][]{2003ARA&A..41..241D}. But in recent years they have become interesting in that context as it was recognised that they should exhibit a rich collection of spectral features that are reminiscent of those seen from the ISM. None of these features arise from the pure solid; rather, they depend on the presence of ions.
Specifically, one expects mid-IR bands from the molecular ion H$_6^+$ and its isotopologues --- notably ${\rm (HD)_3^+}$, which is expected to dominate at low ionisation rates \citep{2011ApJ...736...91L}. Only five of the eleven vibrational modes of that molecule have been reliably characterised, and for those five the fundamental vibrational transitions coincide with mid-IR astronomical bands that are usually attributed to polycyclic aromatic hydrocarbons (PAHs) \citep{2011ApJ...736...91L}. It has been suggested that the observed astronomical bands might arise from ${\rm (HD)_3^+}$ rather than PAHs \citep{2011ApJ...736...91L}. And in the UV/optical/IR region, the rovibrational transitions of \htwo\ molecules situated in an electric field have been suggested to be the source of the Diffuse Interstellar Bands \citep{2022ApJ...932....4W}.
Furthermore, grain charging makes hydrogen dust far more durable than particles of the pure solid \citep{2013MNRAS.434.2814W}. Although hydrogen dust is unconventional it is used in our reference model here because it works well (see \S\ref{sec:modellightcurves}), whereas a cloud populated with silicates or carbonaceous grains does not (\S\ref{sec:conventionalgrains}). Specifically, our reference model employs particles of pure solid hydrogen, with zero ion content --- consistent with dust being driven from the cloud by a cool (non-ionising) stellar radiation field.

Detailed radiation hydrodynamic calculations are computationally intensive, so in order to sketch out results for a variety of possible circumstances we have instead employed a highly simplified model of the dust wind that is driven from the cloud by the star. The key simplifications we use are: (i) the flow is launched from the limb of the cloud (i.e. a one-dimensional initial condition); (ii) post-launch radiation forces are computed in the optically thin limit;  and, (iii) all the dust particles are characterised by a unique value for the opacity to momentum-transfer. Together these approximations yield a model where the dust flow is confined to a surface, rather than a volume, and the reduced dimensionality effects big computational cost savings. Those savings appear first in the construction of the dust flow itself, and secondly in the radiative transfer through that flow. It is helpful to have a mental picture of the dust wind at the outset, as shown in figure \ref{fig:windstructure} --- pre-empting the actual calculations.

We assume that each cloud is on a parabolic orbit. Near periastron, a parabola is a good approximation to all bound, high eccentricity orbits. High eccentricity orbits are a sensible choice for this investigation because no periodicities have been found in RCB fading events \citep{1996PASP..108..225C}, and that constraint is easy to satisfy if the orbital periods are long. We further assume that the observer is at the same longitude as the point of periastron; this assumption is arbitrary, but is used consistently throughout this paper so as to limit the dimensions of the model parameter space. We explore models within the remaining two-dimensional parameter space, defined by the periastron distance of the cloud and the latitude of the observer relative to the orbital plane.

We have adopted the following properties for the star \citep{1997MNRAS.284..489R}: mass $M_*=1\;{\rm M_\odot}$; radius $R_*=80\;{\rm R_\odot}$ ($\simeq 0.37\,{\rm AU}$); and, luminosity $L_*=10^4\;{\rm L_\odot}$. There is considerable uncertainty about the properties of the cloud and the dust. For our reference set of calculations we use hydrogen dust, with the properties described in Appendix A. And to limit the size of our parameter space we have utilised only a single cloud model --- the same cloud model as we previously employed \citep{suvorovwalker2025}, namely a polytrope of index $n=3/2$ with mass $M_c=3\times10^{-5}\,{\rm M_\odot}$ and radius $R_c=0.78\,{\rm AU}$. For simplicity we undertake our reference set of calculations under the assumption that each cloud is spherical; the influence of tidal deformation on the light curves is illustrated in Appendix \ref{sec:tidal}.

\section{Mass loss from the cloud surface}\label{sec:flowfromsurface}
Because of the high luminosity of RCBs, any type of dust at the surface of an orbiting gas cloud will experience strong radiation pressure forces and a dust-driven wind is expected to flow from the cloud.

\begin{figure}[t]
    \centering
    \includegraphics[width=8cm]{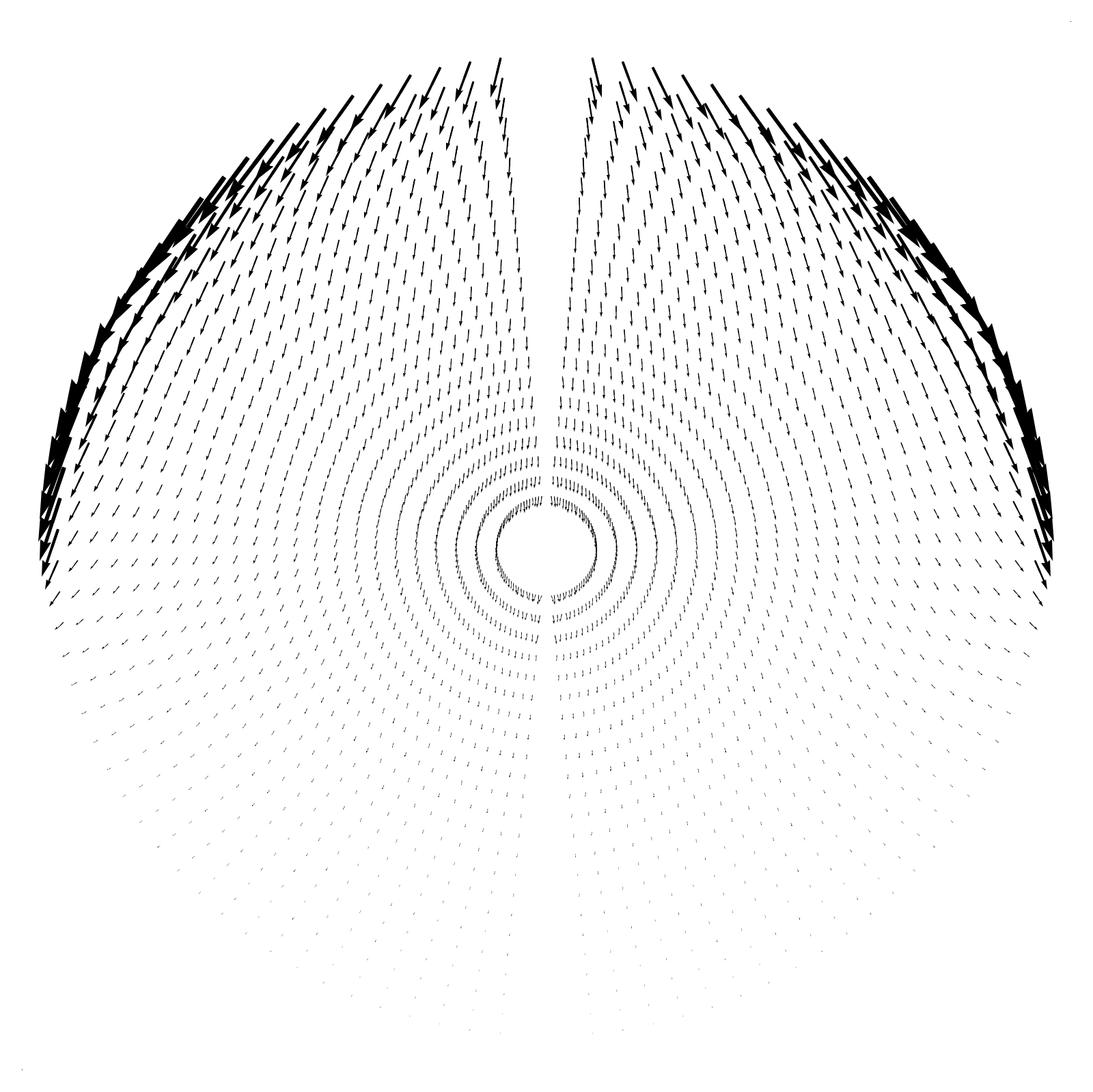}
    \caption{Meridional vector plot of radiation force-density in a spherical cloud with a uniform distribution of small (isotropically-scattering) \htwo\ grains, evaluated via Monte Carlo simulation. The optical depth between centre and surface is 10, in this example, and with $10^8$ input photons the total number of scatterings is $\sim 10^9$. The number of scatterings sampled becomes tiny at small radii and close to the symmetry axis, where the volume elements of the coordinate grid become vanishingly small, and for clarity those ``noisier'' samples have not been plotted. The star is assumed to be very distant, so that the incident photons are all propagating downward in this plot.}
    \label{fig:forcedensity}
\end{figure}

\subsection{Coupling between dust and gas}\label{sec:coupling}
In the optical, the opacity of hydrogen-helium gas is miniscule in comparison with that of the dust; so the gas experiences very little radiation force and dust streams through it. There is consequently a hydrodynamic drag force exerted by the gas on the dust, which couples the two components --- the gas decelerates the dust and the dust accelerates the gas. For the case under consideration here, where the dust and gas are initially in (or close to) phase equilibrium, the gas-grain coupling can also modify the grain size distribution because the drag force leads to heating of both dust and gas and that affects the condensation/sublimation balance. Finally, the gas also cools because it expands as it flows, and that promotes condensation. 

We have not studied the dynamics of the gas-grain coupling in depth. It is, however, clear from our cursory investigation that various different behaviours can arise --- depending on grain size and gas temperature, for example. At present we lack reliable estimates for the physical conditions at the surface of a snow cloud. However, the gas presumably has a very low density, so that the rate of collisions between grains and gas molecules is low, and we therefore proceed with the simple assumption that gas and grains are uncoupled. That should be a good approximation for our model at temperatures $T< 2\;{\rm K}$, if the gas pressure is comparable to that of saturated \htwo. Cloud surface temperatures below that of the cosmic microwave background are permitted if entropy increases with radius within a convective cloud, as is the case for the models of \citet{2019ApJ...881...69W}.

\begin{figure}[t]
    \centering
    \includegraphics[width=8cm]{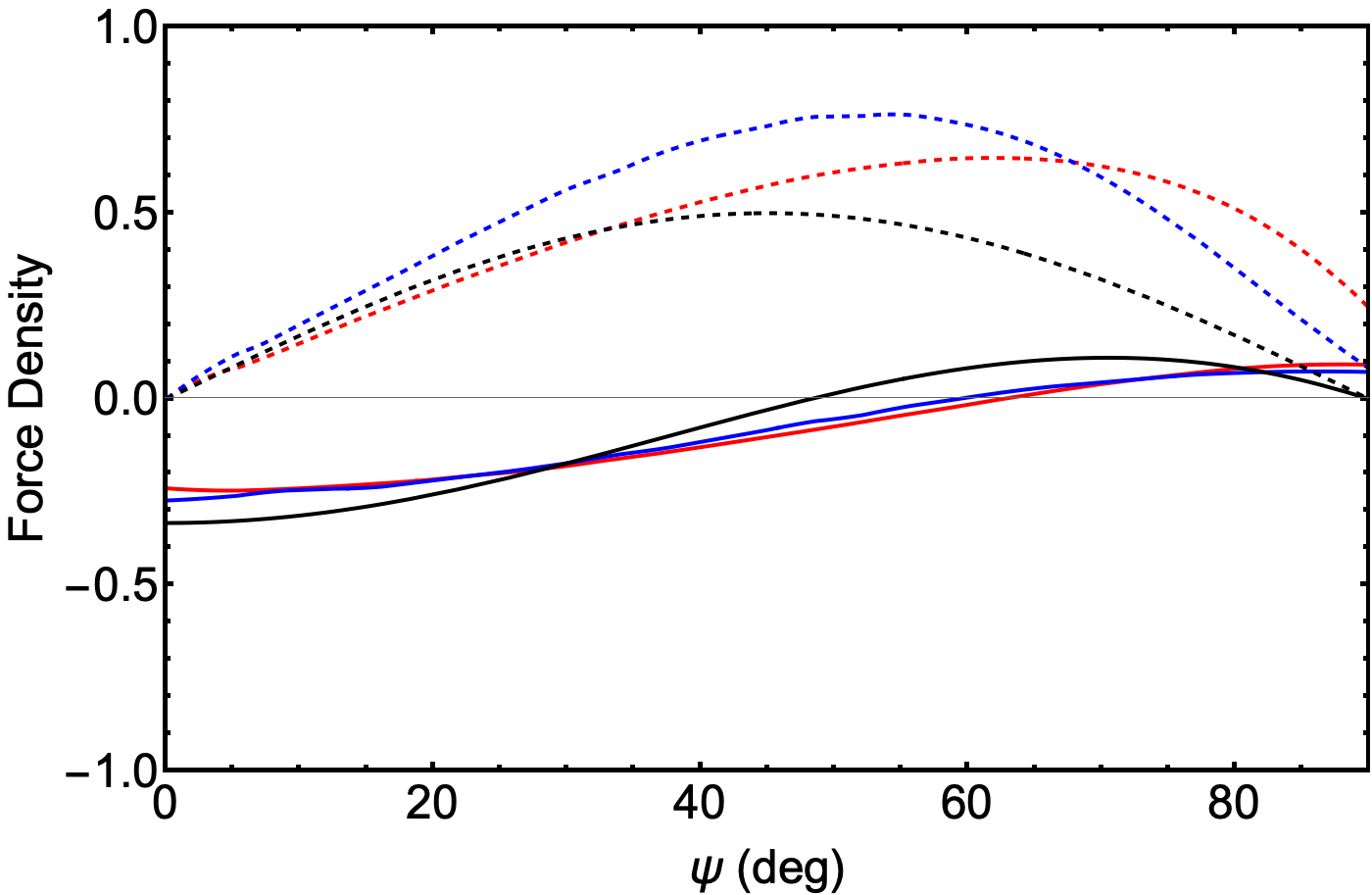}
    \caption{Radial (solid lines) and tangential (dashed lines) components of the radiation force density at the surface of the cloud. Black lines show the simple analytic approximation described in the text, while the red/blue lines show the results of Monte Carlo simulations for the case of small/large \htwo\ grains. The angle $\psi_\circ$ is defined by the point at which the radial force crosses zero: at $\psi_\circ\simeq 61^\circ$ (our adopted value) for the Monte Carlo results, whereas $\psi_\circ=\cos^{-1}(2/3)\simeq 48^\circ$ for the analytic curve.}
    \label{fig:surfaceforces}
\end{figure}

\subsection{Launching the dusty wind}\label{sec:windlaunch}
To gain some insight into the launching of a wind from a snow cloud we undertook Monte Carlo calculations of the radiation forces experienced by hydrogen dust in a spherical cloud illuminated by a distant star. The optical properties of the hydrogen dust particles are described in Appendix A; the most important point is that they scatter light, but do not absorb it. In the regime of large optical depth, the general character of the results did not change as the total optical depth and its radial profile were varied. Nor were there great differences between the case of isotropic scattering (small dust particles) and forward scattering (large dust particles). The main feature that was observed in all cases is that the radiation forces are high near the surface of the cloud on the side facing the star, with an orientation that varies systematically from the centre to the limb of the cloud (as seen from the star). One example Monte Carlo simulation is shown in figure~\ref{fig:forcedensity}, for isotropically scattering grains in a uniform cloud with a total optical depth (centre to surface) of $\tau=10$.

Noting that the force distribution is axisymmetric around the line joining the centre of the cloud to the centre of the star, we introduce the polar angle $\psi$, relative to that axis and measured at the centre of the cloud. Then for $\psi \ll 1$ the radiation force at the surface is approximately radial and directed inwards, pushing the dust towards the centre of the cloud. As $\psi$ increases the force vector rotates around, and for most of the illuminated surface the tangential component is larger than the radial component. As $\psi$ increases further the radial component of the force at the surface changes sign and is outward from there to the limb. We label the point at which the radial component changes sign as $\psi=\psi_\circ$. The value of $\psi_\circ$ depends on the grain characteristics, as described below. The radial and tangential components of the radiation forces at the cloud surface are plotted as a function of $\psi$ in figure \ref{fig:surfaceforces}. 

These features can be understood in a simple, approximate model in which the incident number-flux of photons from the star, at the surface of the cloud, is locally balanced by an outward number-flux of scattered photons that are isotropically distributed over the outgoing hemisphere. For this model the radiation force becomes tangent to the surface at $\psi=\psi_\circ=\cos^{-1}(2/3)$ ($\psi_\circ\simeq 48^\circ$), where the normal component of the incident momentum flux from the star balances that of the emergent, roughly isotropic, scattered radiation field. A radiation-driven dusty wind is expected to arise between $\psi=\psi_\circ$ and the limb. Our Monte Carlo calculations, for both small and large grains, indicate a somewhat larger $\psi_\circ\simeq 61^\circ$, and that is the value we adopt. At $\psi=\psi_\circ$ the tangential force computed via Monte Carlo simulation is approximately 60\% larger than our approximate analytic model. In the development below we employ the analytic form, so as to avoid unnecessary clutter in the equations, but for numerical work we use the larger figure indicated by the Monte Carlo results.

The radiation forces acting on the dust particles are large and will have a substantial influence on the structure of the outer layers of the cloud --- e.g. the cloud will no longer be spherical. We have not attempted to model that restructuring. Instead we simply note that the surface dust will be sheared-/blown-off in the region $\psi\gta \psi_\circ$. As described in Appendix \ref{sec:opticalproperties}, individual dust grains can be characterised by ``efficiencies'', $Q$, that are their cross-sections for interaction with radiation, in units of the geometric cross-section, $\pi a^2$. The momentum transfer (radiation pressure force) efficiency is $Q_{pr}$, and over the entire acceleration zone we approximate the net force experienced by a single dust particle with the constant value 
\begin{equation}\label{eq:radiationforce}
Q_{pr}\pi a^2\;\frac{L_*}{4\pi c D^2}\,\sin\psi_\circ\,\cos\psi_\circ
\end{equation}
(where $L_*$ is the stellar luminosity, and $D$ the distance of the cloud from the star), directed at an angle $\pi/2-\psi_\circ$ relative to the cloud-star axis. Integrating over the acceleration zone -- whose length is $\sim R$, the cloud radius -- gives a launch speed $V$:
\begin{equation}\label{eq:flowvelocity}
V^2\simeq U^2\,\times\frac{R}{D}\,\sin\psi_\circ\,\cos\psi_\circ,
\end{equation}
where we have introduced the characteristic speed $U$, such that $U^2=\kappa_{pr} L_* /2\pi c D$. With our adopted radiation pressure opacity (grain cross-section per unit mass) $\kappa_{pr}=10^4\;{\rm cm^2\,g^{-1}}$ (from Appendix A) and $\psi_\circ=61^\circ$  these speeds evaluate to: $U\simeq 3{,}760\;\sqrt{L_4 / D_{AU}}\;{\rm km\,s^{-1}}$, and $V\simeq 2{,}450\; \sqrt{L_4\, R_{AU}}/D_{AU}\;{\rm km\,s^{-1}}$, where $L_* = 10^4\, L_4\;{\rm  L_\odot}$, $D = D_{AU}\;{\rm  AU}$, and $R = R_{AU}\;{\rm  AU}$. Thus high speed dusty flows are expected from snow clouds near RCB stars, with $L_4\sim 1$.

Equations \ref{eq:radiationforce} and \ref{eq:flowvelocity} are appropriate to the cloud surface; meaning $\tau_{pr}\lta 1$, where $\tau_{pr}$ is the optical depth (in the direction of the star) to momentum transfer. (Equivalently, for a pure scattering opacity, this can be thought of as the optical depth to isotropisation of the incoming radiation field.) However, the radiation forces are still large for optical depths $\tau_{pr}\sim{\rm few}$, and sub-surface dust is also driven from the cloud --- just at a lower speed than the surface particles. As noted earlier, our simplified description of the wind assigns the same flow speed to all particles --- i.e. the optically thin value (equation \ref{eq:flowvelocity}). We therefore allow for the sub-surface flow by imposing an optical-depth cutoff that is larger than unity; specifically, we take $\tau_{pr}{\rm (launch)}=5$. Beyond the fact that this numerical value is $\sim\,{\rm few}$, it is arbitrarily chosen; it corresponds to radiation forces $\sim 0.01\times$ and launch speeds $\sim 0.1\times$ their surface values. The assumed value of $\tau_{pr}{\rm (launch)}$ affects both the extinction and the scattered light levels in our computations. On a point-wise basis -- i.e. each point on the dust sheet -- the optical depth to extinction scales in proportion to $\tau_{pr}{\rm (launch)}$. There is no such simple scaling for the level of scattered light, except in the optically thin limit (far from the cloud), where the surface brightness is expected to be proportional to $\tau_{pr}{\rm (launch)}$. Obtaining accurate values for those quantities will require a much more sophisticated model than the one we present here.

The corresponding optical depth perpendicular to the flow is $\tau_{pr,\perp}{\rm (launch)}=\tau_{pr}{\rm (launch)}\,\cos\psi_\circ$, and the dust (\htwo) mass loss rate from the cloud is 
\begin{equation}\label{eq:masslossrate}
\begin{split}
-\dot{M}\; & \simeq\; \frac{2\pi R}{\kappa_{pr}}\,V\,\tau_{pr,\perp}{\rm (launch)} \\
           & \simeq\;\; 8.7\times 10^{-8}\;\frac{R_{AU}^{3/2}\; L_4^{1/2}}{D_{AU}}\;\;\;\;{\rm M_\odot\;yr^{-1}}.
\end{split}
\end{equation}
With the cloud mass and radius that we have adopted in this paper, and $L_4\sim 1$, the mass loss rate in equation (\ref{eq:masslossrate}) implies depletion of the \htwo\ on a timescale of $375\, D_{AU}\;{\rm yr}$.

\section{Dust wind from a snow cloud in orbit}\label{sec:orbit}
After the wind has been launched the radiation field emanating from the cloud rapidly becomes unimportant --- because the cloud is very optically thick, by hypothesis, so relatively little radiation emerges from the side that faces away from the star, and because the cloud subtends a rapidly decreasing solid angle as the dust moves away from the surface. To a good approximation, then, once the wind has been launched from the cloud the radiation force on the dust is simply that due to the star. Consequently there is a large force component normal to the dust sheet, and the dust will promptly separate from any gaseous component of the wind. Gas fluid elements then move under gravity alone, and the trajectories are just the usual hyperbolic Keplerian orbits --- approximately just straight lines, with flow velocity equal to the launch velocity. We set aside the gaseous component for now.

To describe the dynamics of the dust we proceed as follows. With a purely radial force we have two conserved quantities: the angular momentum and the total energy. The latter is the sum of the kinetic energy and the potential energy, $\Phi\,$:
\begin{equation}\label{eq:potential}
\Phi = \frac{1}{2}m U^2\, \frac{D}{r} \left(1-\frac{L_d}{L_*}\right),
\end{equation}
where $L_d$ is a critical luminosity, analogous to the Eddington luminosity, at which radiation pressure just balances the force of gravity on each dust particle, of mass $m$, i.e.
\begin{equation}\label{eq:criticalluminosity}
L_d\equiv \frac{4\pi G\,M_*\,c}{\kappa_{pr}}.
\end{equation}
Numerically we have $L_d\simeq 1.2\;(M_*/{\rm M_\odot})\;{\rm L_\odot}$, for our adopted $\kappa_{pr}$, so gravity is dominant over radiation pressure for stars on the lower main-sequence. But for RCB stars, with mass $M_*\sim 1\;{\rm M_\odot}$ and luminosity $L_*\sim 10^4\;{\rm L_\odot}$, the radiation force exceeds the gravitational force even for macroscopic lumps of solid \htwo, up to about a centimetre in size, and for the micron-sized particles in the wind gravity is negligible in comparison with the radiation force.

It is convenient to introduce the rescaled radius $\tilde{r}\equiv r/D$, angle $\tilde{\theta}\equiv \theta/\tilde{V}_\theta $, time $\tilde{t}\equiv U\,t/D$, and the radial and tangential components of the launch velocity, $\tilde{V}_r\equiv V_r/U$ and $\tilde{V}_\theta\equiv V_\theta/U$, respectively. We can then write the two evolution equations as
\begin{equation}\label{eq:velocities}
\begin{split}
\frac{{\rm d}\,\tilde{\theta}}{{\rm d}\,\tilde{t}}             & = \frac{1}{\tilde{r}^2}, \\
\left( \frac{{\rm d}\,\tilde{r}}{{\rm d}\,\tilde{t}} \right)^2 & = \left(1 - \frac{1}{\tilde{r}}\right) +\tilde{V}_\theta^2\left(1-\frac{1}{\tilde{r}^2}\right) + \tilde{V}_r^2.
\end{split}
\end{equation}
Examining these equations we see that the flow has an approximately constant velocity at small radii, $(\tilde{r}-1)\ll 1$, then accelerates radially in the region $\tilde{r}\sim 1$, and at large radii, $\tilde{r}\gg 1$, it approaches a coasting solution with ${{\rm d}\,\tilde{r}}/{{\rm d}\,\tilde{t}}\rightarrow \sqrt{1+\tilde{V}_r^2+\tilde{V}_\theta^2}$ and $\tilde{\theta}\rightarrow \tilde{\theta}_\infty$ ($\simeq 2$). 

We note that the asymptotic angular radius of the wind is $\sim 2\tilde{V}_\theta \sim \sqrt{R/D}$, which is large compared to the angular size of the cloud if $D \gg R$. Consequently, for a randomly placed observer, the wind is far more likely than the cloud itself to cause an eclipse. This circumstance comes about because $\theta$ increases rapidly at early times when the dust velocity vector makes a large angle to the radial direction.

It is possible to obtain analytic results for $\tilde{\theta}(\tilde{r})$ and $\tilde{t}(\tilde{r})$. But to determine the wind structure, and the resulting light curves for a distant observer, we need $r$ and $\theta$ as functions of the flow-time. Without the inverse function $\tilde{r}(\tilde{t})$ one can either use analytic approximations, followed by a rapid iteration to the solution, or else a wholly numerical approach. We use both methods in this paper: the former to construct extinction curves in the presence of tidal deformation, and the latter for all other light curves.

For a stationary, spherical cloud the geometric structure of the dust wind would be a radially oriented, hollow tube with a circular cross-section that increases with radius. In practice, though, each cloud is orbiting the star, so the axis of the tube doesn't keep a fixed orientation in space but becomes wound up as the cloud moves in its orbit (figure 1).

\subsection{Optical depth of the dust wind}\label{sec:windopticaldepth}
In the present work we make no attempt to describe the dust density profile normal to the surface that forms our model dust wind; instead we work with quantities integrated through the surface, and we approximate the dust distribution as an infinitely thin sheet. As we wish to model the obscuration of the star a key quantity of interest is the optical depth to extinction. As noted earlier, the normal optical depth to momentum transfer for the wind at launch is $\tau_{\perp, pr}({\rm launch})\simeq \tau_{pr}({\rm launch})\,\cos\psi_\circ$. And the normal optical depth to extinction is just a factor $Q_{ext}/Q_{pr}$ larger. The ratio of extinction to momentum-transfer cross-sections, $Q_{ext}/Q_{pr}$, is computed in Appendix A.

To proceed further we make the assumption that the dust particles do not evolve, either in number or size, as they move away from the cloud. In that case the normal optical depth scales directly with the surface density of the dust sheet, and that declines according to the stretching of each element of the flow since launch.

To describe the areal stretching of the sheet it is convenient to introduce the position vector, $\vec{\xi}(t,\varphi)$, of a general point in the surface, where $\varphi$ denotes the azimuth of the launch point around the limb of the cloud. We note that the orientation of the surface is important in determining the optical depth as seen along a particular direction, and therefore we define the areal stretch factor as a vector product:
\begin{equation}\label{eq:stretchfactor}
\vec{S}(t,\varphi) = \frac{1}{V R}\,\left(\frac{\partial\,\vec{\xi}}{\partial\,t} \right)_{\!\varphi} \times \,\left(\frac{\partial\,\vec{\xi}}{\partial\,\varphi} \right)_{\!t}.
\end{equation}
The optical depth through any portion of the dust sheet is then simply
\begin{equation}\label{eq:extinction}
\tau(\hat{\boldsymbol{k}},t,\varphi) =  \frac{\tau_\perp({\rm launch})}{|\hat{\boldsymbol{k}} \cdot \vec{S}(t,\varphi)|}\; ,
\end{equation}
as seen in the direction specified by the unit vector $\hat{\boldsymbol{k}}$. Numerical values of $\tau_{ext}$ are required at many points across the face of the star, as seen by the observer, and we therefore constructed an interpolation function for that purpose, taking as input the flow pattern evaluated for the orbiting cloud.

\subsection{Scattered light}\label{sec:scatteredlight}
In our reference model the star is obscured by hydrogen dust particles; they scatter light, but do not absorb it. Consequently there can be no change in luminosity as the star is eclipsed. Rather, the effect of each eclipse is to change the angular distribution of the emergent light, so that some observers (at infinity) measure a lower flux, while others measure a higher flux --- depending on their location relative to the orbit of the eclipsing cloud. It is therefore essential to estimate the amount of light that is scattered towards any hypothetical observer, in addition to calculating the extinction of the light that is received directly from the star.

We can make a crude estimate of the scattered flux if we assume that (i) the scattered light is isotropically emitted from the material that eclipses the star, and (ii) the scattering material is only just extended enough to completely cover the star, as seen by the observer. With those approximations the observed, scattered flux, in units of the unobscured stellar flux, is $\sim R_*^2 / 2D^2$; thus a dimming event in which the measured flux falls to $\lta 10^{-3}$ of the unobscured stellar flux requires the eclipsing material to be at least $8\;{\rm AU}$ from the star. More generally the deep eclipses that RCBs are observed to undergo place strong constraints on the amount of scattered light, and thus on the model presented here. We therefore undertake a proper calculation of the scattered flux for our dusty wind model, as described below.

The flux, $F$, that is received can be calculated in the usual way, given the intensity, $I$, in the direction of the observer at every point in the dust wind:
\begin{equation}\label{eq:scatteredflux}
F_{obs} \,=  \int\!{\rm d}\Omega_{obs}\,I_{obs}\; ,
\end{equation}
where ${\rm d}\Omega_{obs}$ is an element of solid angle as seen by the observer. (Note that the entire dust-wind structure subtends a solid angle $\Delta\Omega_{obs} \ll 1$, and thus the usual cosine factor in the integral can be set to unity with negligible error.) To evaluate the intensity of scattered light then requires two distinct steps. Each element of area on the surface of the dust sheet can be locally approximated as a plane-parallel slab, so the first step is to undertake a Monte Carlo simulation of the intensity of scattered light for a beam of photons incident at a specified angle upon a slab of specified optical depth. That simulation tells us the angular distribution of the scattered light for one particular combination of incident angle and optical depth. And by repeating the process for many different optical depths and angles we arrive at an approximate, but comprehensive description (an interpolation function) of the scattered intensity for monodirectional incident light.

Our Monte Carlo simulations employed $10^6$ input photons for each of 40 values of the normal optical depth, and each of 40 values of the angle, $\theta_\circ$, between the slab normal vector and the input photon wavevector. For the sake of efficiency we forced each input photon to be scattered within the slab; but the subsequent scatterings and escape were randomly selected. We recorded the number of emerging photons in each of $10^4$ elements of solid-angle, $\Delta \Omega$, on the sphere, described by the usual spherical polar coordinates $(\theta,\phi)$. Then, if $p(\theta,\phi)$ is the fraction of all input photons that emerge in that direction, the quantity
\begin{equation}\label{eq:scatteredintensity}
\begin{split}
g & \equiv \frac{I_{MC}(\theta,\phi)}{F_{in}} \\
  & = \left\{ \frac{p(\theta,\phi)}{\Delta\Omega}\;\frac{1}{|\cos\theta\,|}\right\}|\cos\theta_\circ|\;\big(1-\exp(-\tau_\perp/|\cos\theta_\circ|)\big) ,
\end{split}
\end{equation}
is our Monte Carlo estimate for the scattered intensity, $I_{MC}(\theta,\phi)$, per unit incident flux, measured normal to the incident beam, $F_{in}$. We used the part of this expression in curly braces to define an interpolation function, with the remaining portion of the expression being evaluated exactly for whatever combinations of optical depth and incident angle are needed in practice. The slab is symmetric about its mid-plane, so we restricted the incident angle to the range $0\le \theta_\circ \le \pi/2$. It is also axisymmetric, so the azimuth of the incident photons can be fixed to zero. One further symmetry -- reflection symmetry about the plane containing the slab normal and the incident wavevector -- allows us to ignore the sign of the output azimuth. 

The second step is to break down the flux into a hierarchy: $F^{(0)}+F^{(1)}+F^{(2)}+F^{(3)}+...$, where $F^{(0)}$ is the unscattered light (i.e. received direct from the star), $F^{(1)}$ is light that has been scattered by one element of the dust wind, and $F^{(2)}$ is light that has been scattered by two different elements of the dust wind, etc. In our calculations we have included only the first three terms in this hierarchy, because under the conditions of interest to us the third term is typically found to be small compared to the second. In other words: direct illumination of the dust wind by the star contributes much more to the observed, scattered light than does indirect illumination, where light propagates directly from the star to one element of the dust wind, is scattered, then propagates to another element of the dust wind, is scattered again, and then proceeds directly to the observer. It is in fact difficult to go any further up the hierarchy of such a calculation, because each additional step involves an extra integral over all elements in the dust wind, with a correspondingly large increase in the compute time. 

Obviously, for the first term in the hierarchy -- the unscattered light -- we need only integrate over the stellar photosphere, accounting for any extinction along the line-of-sight to each point thereon. The optical depth of the dust wind is determined in the way described in \S\ref{sec:windopticaldepth}, whereas the cloud itself is assumed to be completely opaque.

Evaluation of the second term requires us to integrate over the full extent of the wind. In practice the wind structure is determined numerically, as a grid of points, $\vec{r}_j$, with corresponding (normal) optical depth $\tau_{\perp,j}$, and element of area $\Delta\vec{A}_j=\Delta A_j\;\hat{\boldsymbol{n}}_j$, so that the integral in equation \ref{eq:scatteredflux} can be approximated by a discrete sum:
\begin{equation}\label{eq:scatteredfluxdiscrete}
\frac{F^{(1)}}{F_*}\simeq\sum_j \;g(\tau_{\perp,j},\hat{\boldsymbol{n}}_j,\hat{\boldsymbol{r}}_j,\hat{\boldsymbol{r}}_{obs})\; \frac{\Delta A_j}{r_j^2}\;|\hat{\boldsymbol{r}}_j\cdot \hat{\boldsymbol{r}}_{obs}|\;\exp(-\tau).
\end{equation}
In this expression: all hatted quantities are unit vectors; $\vec{r}_{obs}$ is the location of the observer; and, $F_*$, is the flux from the star that is measured by the observer in the absence of the dust wind (the ``baseline'' flux). The factor $\exp(-\tau)$ accounts for extinction either before or after it is scattered at the $j$-th element of the wind. In the form that equation \ref{eq:scatteredfluxdiscrete} is written, the arguments of the function $g$ appear to span a seven-dimensional space; but four dimensions suffice because of the symmetries mentioned earlier --- we used the variables $\tau_{\perp}$, $\cos\theta_\circ$, $\cos\theta$, and $\cos\phi$. Rapid evaluation of the singly-scattered flux, $F^{(1)}$, via equation \ref{eq:scatteredfluxdiscrete}, is facilitated by the implicit assumption that the illuminating source (i.e. the star) is point-like.

Calculation of $F^{(2)}$ proceeds in an analogous way, except that the second scattering, by the $k$-th element of the wind, accounts for the combined illumination of that element by light scattered from all other elements, $j$, in the wind, which in turn have been illuminated only by the star. Computation of the doubly-scattered flux, $F^{(2)}$, thus requires a double-sum, $\sum_k \sum_j$, which is slow to evaluate as there are many ($\sim 4\times 10^4$) points in the mesh that describes the dust wind geometry. To speed up the calculation of the doubly-scattered flux we employ two forms of sparse sampling. First, we evaluate the contribution from only 4\% of the points in the dust wind mesh -- randomly sampled at each epoch -- and then multiply by 25. And secondly, we only evaluate the doubly-scattered flux for one in every 20 of the epochs in each light curve; linear interpolation is used to infer appropriate values in-between. Typically we find that $F^{(2)} \ll F^{(1)}$, and therefore we neglect $F^{(3)}$ and higher orders of scattering.

\section{Model eclipse light curves}\label{sec:modellightcurves}
\begin{figure}[h]%
    \centering
    \includegraphics[width=0.47\textwidth]{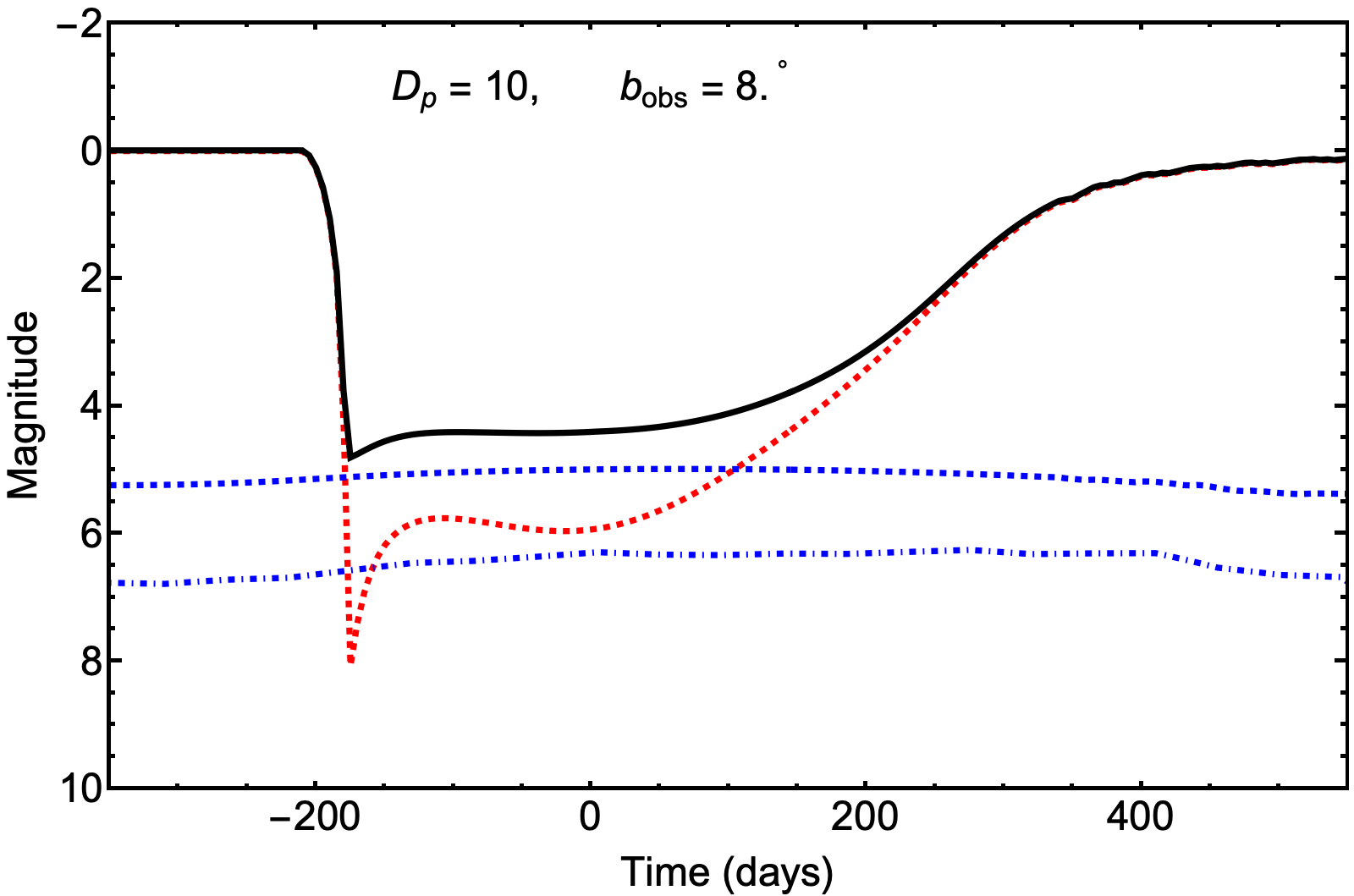}%
    \caption{Light curve for a cloud with periastron at $D_p=10\;{\rm AU}$ (at epoch $0$). The observer is at the same longitude as the periastron, and at latitude $b_{obs}= 8.0^\circ$. The curves are: total flux (black, solid line), made up of:  unscattered light from the star (red, dashed line); light that is scattered once in the dusty wind (blue, dashed line); and, twice-scattered light (blue, dot-dashed line).}%
    \label{fig:medianeclipse}%
\end{figure}

Figure \ref{fig:medianeclipse} shows an example light curve for our model (here, and subsequently, $\lambda=600\,{\rm nm}$ magnitudes are shown with 5-day sampling). This example has properties that are similar to the average values that have been reported for the dimming events of RCBs \citep{2025MNRAS.537.2635C}, which are: median event duration 0.9 year; median event depth 4.6 magnitudes (in V-band); and, for the prototype R CrB itself, a median decline time of about 30 days. 

The contributions to the theoretical eclipse light curve that are included in this calculation are: (i) light coming directly from the star, subject to extinction where it passes through dust; (ii) light that is scattered by the dust wind, and which then propagates to the observer; (iii) light that is scattered by the dust wind, then scattered again at a second, different location in the wind, which then propagates to the observer; and, (iv) light scattered from the illuminated face of the cloud. The last of those contributions is very small in the example shown here, and consequently it is off-scale; it becomes more prominent for orbits with smaller periastra, as will be seen below. In addition to demonstrating a match to typical characteristics of RCB fading events (depth, onset time and duration), the example shown in figure \ref{fig:medianeclipse} illustrates two points made earlier. First, the twice-scattered flux is small in comparison with the singly-scattered flux, so truncating the sequence there -- i.e. ignoring $n$-times scattered flux, with $n\ge3$ -- is an acceptable approximation. Secondly, if it were not for the light scattered by the dusty wind the model eclipse would be several magnitudes deeper. 

\begin{figure*}[t]%
    \centering
    \hbox{\qquad\includegraphics[width=8cm]{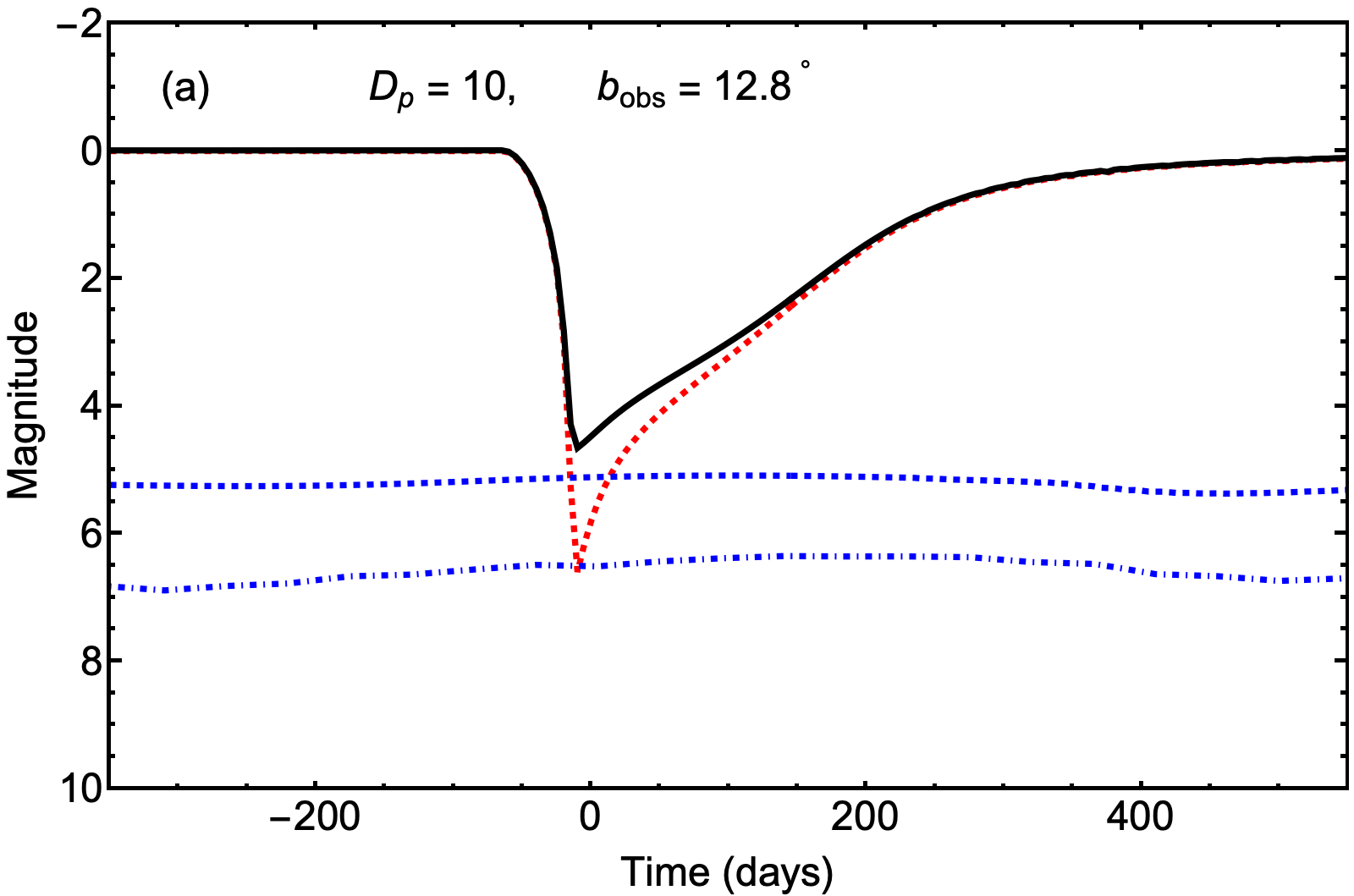}\qquad\includegraphics[width=8cm]{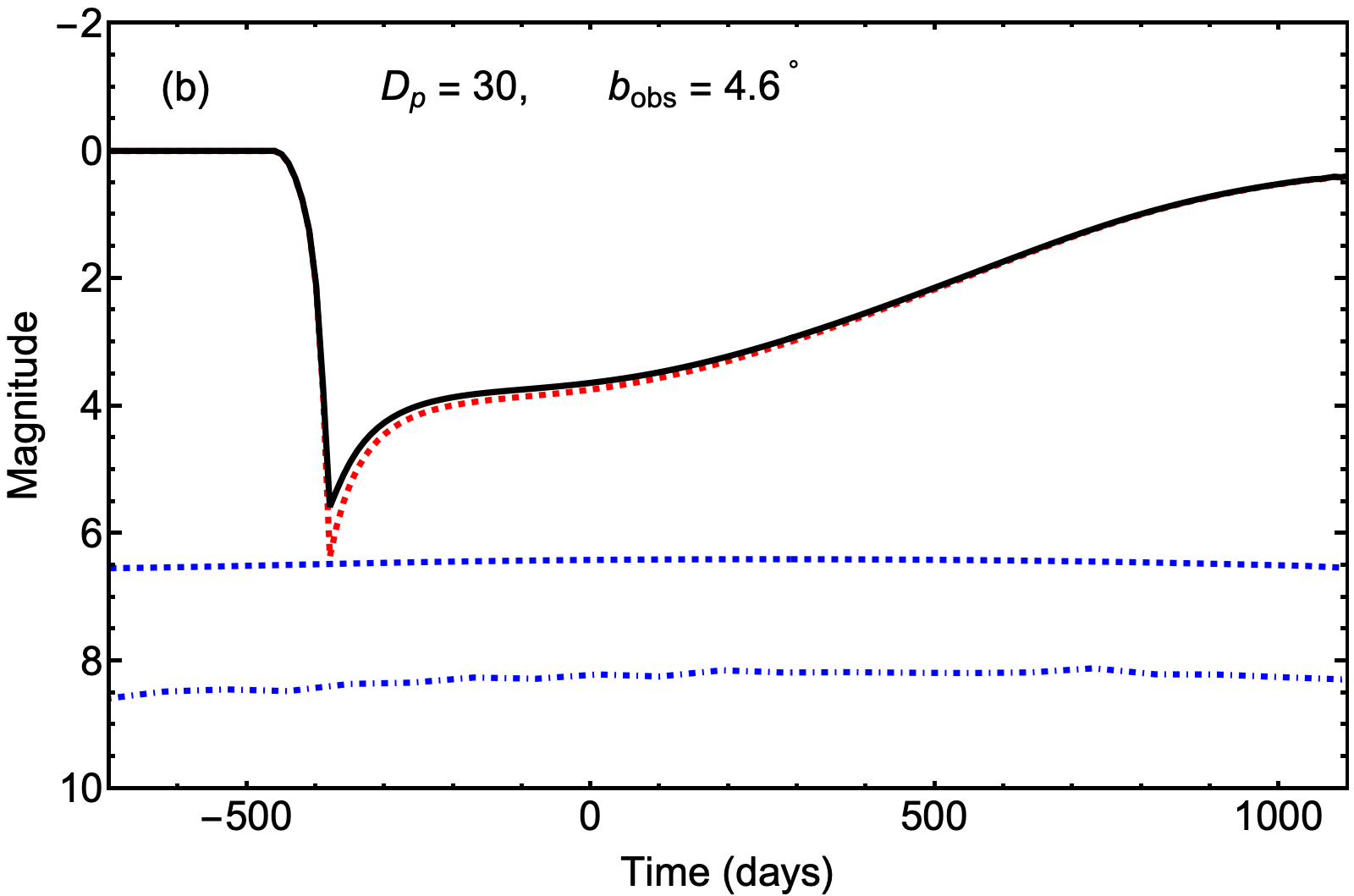}}%
    \medskip
    \hbox{\qquad\includegraphics[width=8cm]{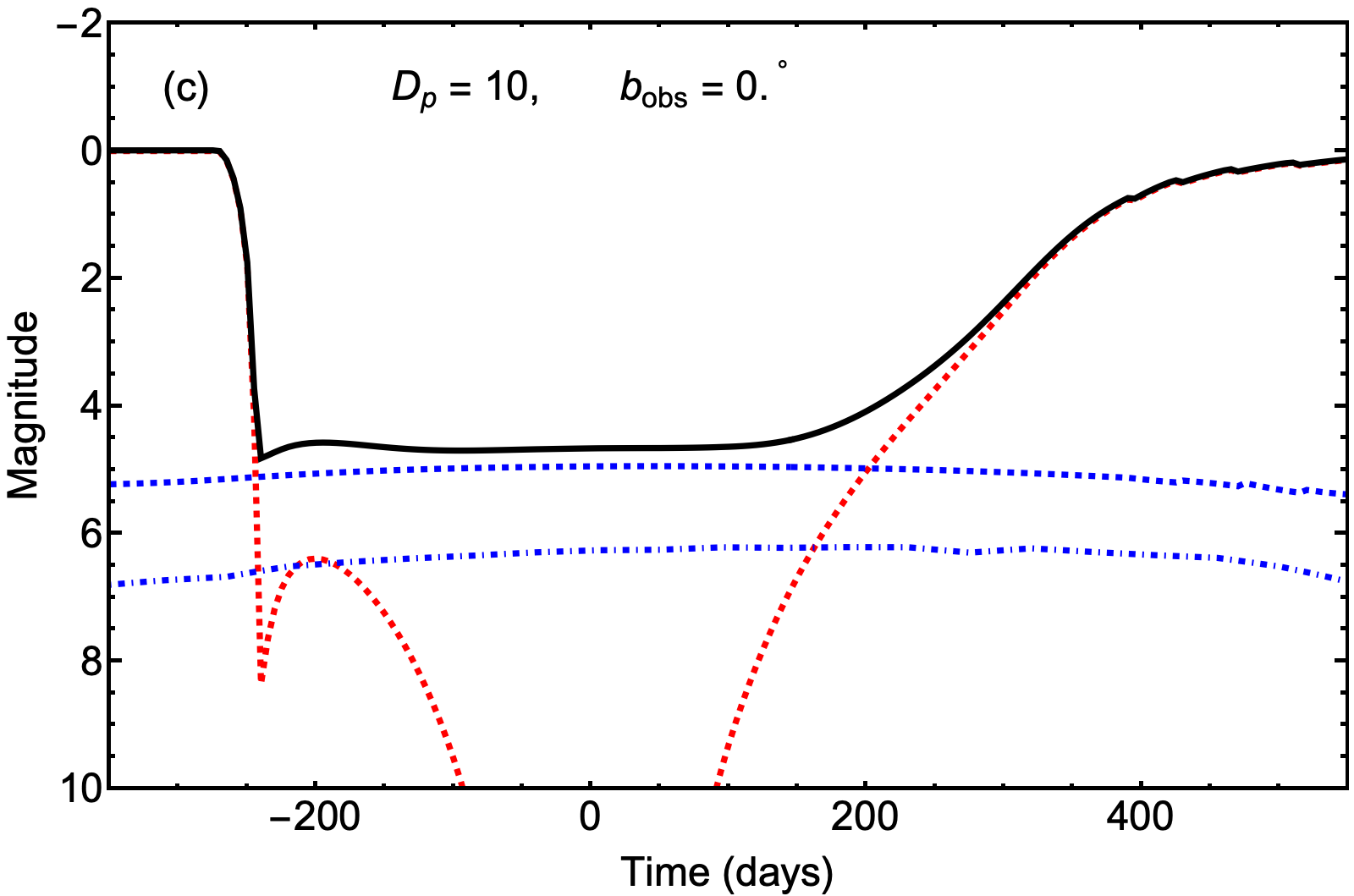}\qquad\includegraphics[width=8cm]{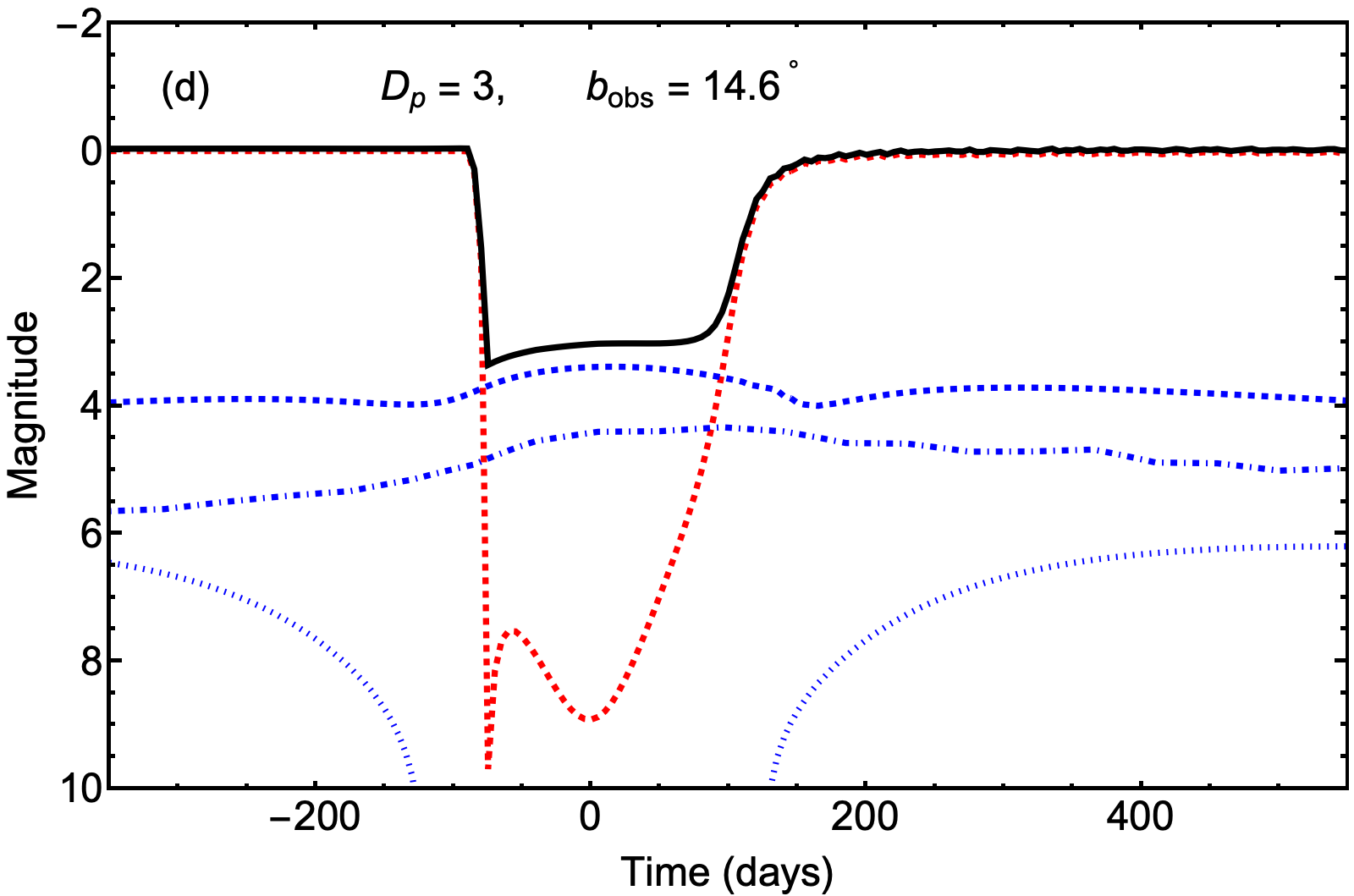}}%
    \caption{Model light curves for different parameter combinations. Left column: cloud periastron at $D_p={\rm 10\;AU}$; observer at zero latitude (c, lower panel) and 80\%\ of the maximum latitude for which eclipses can be observed (a, upper panel). Time is relative to the epoch of periastron. Right column: cloud periastron at $D_p={\rm 3\;AU}$ (d, lower panel), and $D_p={\rm 30\;AU}$ (b, upper panel) for an observer at the median (absolute) latitude for eclipses. Lines are coded in the same way as in Figure \ref{fig:medianeclipse}, but with the addition of a blue, dotted line that shows the light scattered from the illuminated face of the cloud; that contribution is computed in all cases, but is faint and off-scale except for the $D_p={\rm 3\;AU}$ case. Note that the top right panel covers twice the temporal extent of the others.}
    \label{fig:distancelatitudevariation}%
\end{figure*}

Varying the input parameters of the model -- specifically, the periastron distance and the latitude of the observer -- yields light curves that differ in duration, depth and onset time, as set out in table \ref{tab:lightcurvemetrics} and illustrated in Figure \ref{fig:distancelatitudevariation}. These plots show that when the eclipse is observed at low-/mid-latitudes the unscattered light typically goes through two distinct minima, arising from extinction in well-separated regions in the dusty flow. The first, sharp minimum is due to the occultation front, i.e. the leading edge of the dust wind which crosses the star ahead of the cloud's conjunction. From the observer's perspective the occultation front is a fold in the dust sheet, so although the sheet has a small perpendicular optical depth at that point it is seen almost edge-on and causes a great deal of extinction. The sharpness of that first minimum is partly a result of approximating the dust wind as a curved sheet rather than a fully three-dimensional flow. The second, broader mininum in the unscattered light is caused by the flow close to the cloud (and to some extent by the cloud itself), where the dust sheet has substantial perpendicular optical depth.

It is no surprise that Figure \ref{fig:distancelatitudevariation} demonstrates that both eclipse duration and decline time increase with periastron distance, as the cloud's orbital speed is lower at larger distances from the star. However, it is worth mentioning that the speed at which the occultation front crosses the star is not actually the cloud's Keplerian speed. Rather, at low latitudes, it can be as high as the geometric mean of the Keplerian speed and the non-radial component of the wind launch speed, $V_\theta$ --- which is a lot larger. Thus the onset time of the eclipse is much less than would be seen if the cloud itself, rather than the dusty wind, were entirely responsible for the occultation. That difference comes about because the wind structure rotates around the star at the same, Keplerian angular rate as the cloud, but the occultation front is several times further from the star, as can be seen in figure \ref{fig:windstructure}, and its speed is higher in direct proportion. The onset time also increases with the observer's latitude, as the occultation front becomes increasingly skew relative to its direction of motion.

\begin{table}[t!]
\centering
\begin{tabular}{|c|c||c|c|c|c|c|}
\hline
$D_p$ & $b_{obs}$ & Depth & Onset & Duration & Centroid \\
 & & & Time & & Shift \\
$({\rm AU})$ & $({}^\circ)$ & (mag)&(days) & (years) & (AU)\\
\hline
\hline
3 & 14.6 & 3.4 & 13 & 0.6 & 47\\
\hline
10 & 0 & 4.8 & 26 & 1.7 & 109\\ 
\hline
10 & 8.0 & 4.8 & 29 & 1.4 & 105\\ 
\hline
10 & 12.8 & 4.7 & 44 & 0.8 & 85 \\
\hline
30 & 4.6 & 5.6 & 66 & 3.4 & 110\\
\hline
\end{tabular}
\caption{Light curve metrics for the events shown in figures \ref{fig:medianeclipse} and \ref{fig:distancelatitudevariation}. Depth, onset time, and duration are measured as in \citet{2025MNRAS.537.2635C}. For the first, third, and fifth models listed the value of $b_{obs}$ is approximately half the maximum value of $|b_{obs}|$ for which eclipses can be seen. The final column is the maximum shift, over the course of the event, in the apparent centroid of light (see \S\ref{sec:astrometry}).}
\label{tab:lightcurvemetrics}
\end{table}

The timing of ingress varies strongly with the observer's latitude. It is well ahead of conjunction at low- and mid-latitudes, where the occultation front precedes the cloud. But for observers at high latitudes any obscuration is due to dust that is launched far out of the cloud's orbital plane, and thus ingress tends to be close to conjunction -- as in panel (a) of figure \ref{fig:distancelatitudevariation} -- or even later. Finally, as expected on the basis of simple estimates (see \S\ref{sec:scatteredlight}), the contribution of scattered light decreases as the periastron distance increases, and thus deeper eclipses are possible for larger periastra. Eclipses by distant clouds are, however, less probable, because the latitudinal extent of the wind is smaller: for an orbit with periastron at $3\;{\rm AU}$ the maximum latitude for an eclipse is approximately $30^\circ$, whereas it is only about $9^\circ$ if $D_p=30\;{\rm AU}$.

\subsection{Sensitivity to other parameters}
Up to this point we have kept free parameters to a minimum, in so far as possible  --- attempting a disciplined approach to the modelling, with unvarnished results and no fine tuning. However, it may be of interest to readers to get a feel for how sensitive our results are to some of the variables that we have not explored. To that end, in this section we provide a summary of how the eclipse properties change when the context is modified. Our reference point is the calculation shown in figure \ref{fig:medianeclipse}, whose light-curve metrics are close to those of the median observed for RCB fading events \citep{2025MNRAS.537.2635C}.

Some dimensions are difficult to explore in a self-consistent way. For example: the ratio of efficiencies for extinction and momentum-transfer, $Q_{ext}/Q_{pr}$, is tied up with the degree of forward-scattering for the dust grains, and thus in principle changing its value requires a whole new set of Monte Carlo simulations for the scattered light distribution (\S\ref{sec:scatteredlight}). But that has not been attempted; rather, the parameter value has been changed in isolation. Some parameter dependencies are trivial --- e.g. the cloud mass has no influence at all on the dust dynamics, and all choices of cloud mass yield the same light-curves if tidal deformations (Appendix \ref{sec:tidal}) are neglected. And, of course, there is some parameter degeneracy. For example a given fractional change in either the stellar luminosity, $L_*$,  or the dust opacity, $\kappa_{pr}$, yields exactly the same outcome. Ditto the stellar radius, $R_*$, and $T_*^2$, the square of the photospheric temperature (assuming that $L_*\propto R_*^2 T_*^4$ is fixed). Finally, changing the efficiency ratio, $Q_{ext}/Q_{pr}$, by a given fraction has the same effect on the light-curves as a change in the optical depth at launch, $\tau_{pr}({\rm launch})$.

Excluding the trivial or degenerate cases just mentioned, there are six parameters to explore. They are: stellar luminosity, $L_*$; stellar mass, $M_*$; stellar radius, $R_*$; cloud radius, $R$; the longitude of the observer, $l_{obs}$, relative to periastron; and, the line-of-sight (to the star) optical depth of the wind at launch, $\tau_{pr}({\rm launch})$. Table \ref{tab:metricsensitivity} shows the percentage change in eclipse metrics that are found when varying each of those parameters by $\pm 20$\% ($20^\circ$ in the case of $l_{obs}$), while keeping all other parameters fixed.

By far the most common outcome -- seen in over 2/3 of the entries in the table -- is little-or-no change ($\lta 5$\%) in the result. Nor are there any instances of high sensitivity to the background parameters, with the strongest dependencies being $\sim$linear. In the case of the eclipse depth, none of the six parameters had much effect at all. Eclipse onset-time, on the other hand, shows some clear dependencies; unsurprisingly, it is $\sim$linearly dependent on the stellar radius. Na\"ively, the speed of the occultation front is expected to vary in proportion to $(M_*L_*R)^{0.25}$, which accounts for much of the observed parameter dependence of the onset time. Similarly, the $\sim$inverse-square-root dependence of eclipse duration on $M_*$ is presumably just a reflection of the Keplerian speed of the snow cloud.

The dependencies shown in table \ref{tab:metricsensitivity} suffice to characterise our model in the immediate vicinity of the chosen set of background parameters. For the stellar properties that is sufficient, as they are likely to be close to the true values. Of the other three quantities, a wider range of values for the longitude of the observer should certainly be explored in future work; we anticipate that $l_{obs}$ is likely to play an important r\^ole when tidal effects are included (Appendix \ref{sec:tidal}) --- because tides cause the size and shape of the cloud to evolve systematically through periastron. It would also be useful to explore the influence of different cloud radii, beyond the $\pm 20\%$ range considered here, because that parameter is very poorly constrained. There appears to be little prospect of interesting models with clouds that are {\it much\/} larger in radius, because the resulting eclipses would be too long-lasting. But clouds that are a lot smaller may be acceptable, as the briefer eclipses that result can be lengthened by choosing larger periastron distances.

\begin{table}[t!]
\centering
\begin{tabular}{|c||c|c|c|c|c|}
\hline
Parameter & Depth & Onset & Duration & Centroid \\
changed & & Time & & Shift \\
  & \% & \%  & \%  & \% \\
\hline
\hline
Reference case & 4.8 mag& 29 day& 1.4 yr& 105 AU\\ 
\hline
\hline
$L_*=10^4\;{\rm L_\odot}+\Delta$ & $0$ & $-6$ & $0$ & $+1$ \\
\hline
$L_*=10^4\;{\rm L_\odot}-\Delta$ & $0$ & $+4$ & $0$ & $0$ \\
\hline
$M_*=1\;{\rm M_\odot}+\Delta$ & $0$ & $-2$ & $-9$ & $-1$ \\
\hline
$M_*=1\;{\rm M_\odot}-\Delta$ & $0$ & $+11$ & $+12$ & $0$ \\
\hline
$R_*=80\;{\rm R_\odot}+\Delta$ & $-1$ & $+25$ & $0$ & $-5$ \\
\hline
$R_*=80\;{\rm R_\odot}-\Delta$ & $+1$ & $-24$ & $0$ & $+4$ \\
\hline
$R=0.78\;{\rm AU}+\Delta$ & $-4$ & $-14$ & $+17$ & $+12$ \\
\hline
$R=0.78\;{\rm AU}-\Delta$ & $+4$ & $+12$ & $-19$ & $-19$ \\
\hline
$l_{obs}=0^\circ+\Delta$ & $0$ & $-3$ & $+4$ & $0$ \\
\hline
$l_{obs}=0^\circ-\Delta$ & $+2$ & $+10$ & $+3$ & $-3$ \\
\hline
$\tau_{pr}=5+\Delta$ & $+1$ & $0$ & $+3$ & $+23$ \\
\hline
$\tau_{pr}=5-\Delta$ & $-3$ & $0$ & $-3$ & $-32$ \\
\hline
\end{tabular}
\caption{Sensitivity of eclipse metrics to assumed parameters. Parameter values are given in the leftmost column in the form: reference value $\pm$ change. For $l_{obs}$ we use $\Delta=20^\circ$, and for all other parameters $\Delta=20\%$. The optical depth, $\tau_{pr}$, is the value at launch. The four rightmost columns give the percentage changes (rounded to the nearest whole number) in eclipse metrics, relative to the values for the reference case.}
\label{tab:metricsensitivity}
\end{table}

\section{Cloud disruption and accretion}\label{sec:disruptionaccretion}
When the snow cloud gets close to the star it experiences tidal distortion, and that affects the eclipse light curves. For simplicity, we have neglected that effect and assumed spherical clouds for all of the calculations in the main body of this paper. Examples of extinction-only eclipse light curves that include tidal distortion are given in Appendix \ref{sec:tidal}. But in addition to the influence on eclipse light curves there is a qualitatively new feature of the model that tides introduce, as follows.

On orbits such as those yielding the light curves shown in figures \ref{fig:medianeclipse} and \ref{fig:distancelatitudevariation}, the tidal gravitational forces of the star are actually large enough to disrupt our model snow cloud \citep{suvorovwalker2025}. So, after periastron, the cloud no longer exists as a gravitationally bound entity --- it becomes a stream of tidal debris, some of which will be accreted onto the star via an accretion disk. In this section we sketch the implications of cloud disruption and accretion for the evolution of the star, for the origin of the mid-IR flux and its temporal variations.

\subsection{A novel evolutionary path}\label{sec:evolution}
For high eccentricity (approximately parabolic) cloud orbits, half of any tidal debris is initially bound and under gravity alone that material would accrete onto the star, whereas the other half is unbound and escapes to infinity \citep[e.g.][]{1988Natur.333..523R}. In our case, though, expansion of the debris stream leads eventually to complete condensation of the \htwo, and the intense radiation field of the star modifies the potential in which those particles move. The combined radiation+gravitational potential is repulsive for particle radii $\lta 1\,{\rm cm}$ (see Appendix \ref{sec:opticalproperties}). The largest lumps of condensate in the pre-disruption cloud have been estimated to be up to about $10\,{\rm cm}$ in size \citep{suvorovwalker2025}, and those particles become unbound if the initial cloud orbit has eccentricity $\gta 0.9$. Our model for the fading events of RCBs thus implies that the central stars should be accreting helium, but not hydrogen. That is encouraging, because hydrogen deficiency is one of the key characteristics of RCBs.

The model we have presented does not require the complete disruption of every cloud that causes an eclipse --- occultations may arise from orbits with larger periastra in which the clouds undergo only tidal stripping of some of their outer layers. Indeed, for sufficiently distant encounters there could be eclipses without even any stripping. On the other hand, observers see no eclipses from clouds on low inclination orbits, regardless of whether they disrupt. So there can be both eclipses without accompanying disruptions, and disruptions without accompanying eclipses. Absent a clear picture of how those two competing effects influence the overall event statistics, we use the eclipse rate as an estimate of the cloud disruption rate. For our model cloud -- with a mass of $3\times 10^{-5}\;{\rm M}_\odot$, one quarter of which is assumed to be helium -- the observed rate of fading events ${\cal R} \sim 0.3\;{\rm yr^{-1}}$ for RCBs \citep{2024MNRAS.527.9274S,2025MNRAS.537.2635C} then implies an accretion rate $\dot{M}_*\sim 10^{-6}\;{\rm M}_\odot\, {\rm yr^{-1}}$ in helium. That is high enough to permit stable nuclear burning of helium on the surface of a degenerate core \citep{1983ApJ...264..605I}, yielding a luminosity (some from accretion power, but more from nuclear fusion) $\sim 10^4\;{\rm L}_\odot$ for as long as cloud disruptions continue at the observed rate.

The foregoing considerations reveal that the broader picture of the model under study here is potentially a very simple one: an intense shower of snow-cloud disruptions by a CO-WD may be able to account for both the unusual stellar type and the dramatic eclipses of RCBs.

\subsection{Emission from the accretion disk}\label{sec:diskaccretion}
Assuming that the population of disrupting clouds has non-zero mean (vector) angular momentum about the star, the accretion we have just described will proceed via a disk. As explained in point (iii), below, the gas temperature in the inner disk is expected to be $\sim 1{,}000\;{\rm K}$, so the disk will be geometrically thin. In order to quantify other properties of the disk we adopt the ``alpha-disk'' model as summarised by \cite{1981ARA&A..19..137P}, and we assume an accretion rate $\sim 10^{-6}\;{\rm M}_\odot\, {\rm yr^{-1}}$ (as above). In that case the inner disk has the following characteristics: radial inflow speed $|v_R|\lta 0.16\;{\rm km\,s^{-1}}$; surface density $\gta 10^2\;{\rm g\,cm^{-2}}$ (predominantly helium, but presumably also containing metals); and, vertical scale-height $\sim 3\times 10^{11}\;{\rm cm}$. Thus the helium density in the mid-plane of the disk is $n({\rm He})\gta 5\times 10^{13}\;{\rm cm^{-3}}$.

Helium is essentially transparent in the optical/near-UV and will remain neutral. But that is not true of all the metals, and through them the disk absorbs the incident starlight. A metal-ion plasma results. The relevant radiation processes were modelled by \citet{1996A&A...313..217W} for a parcel of gas adjacent to an RCB photosphere --- they did not have an accretion disk in mind, but the radiation interactions are the same modulo the unknown metal abundances in the disk. Based on that model we expect the following three points to apply.\hfill\break
(i) Irrespective of the gas temperature, atomic carbon will be essentially all in the form C$^+$ \citep[\S3.2.1 in][]{1996A&A...313..217W}.\hfill\break
(ii) For high gas densities -- as we anticipate in the accretion disk -- free-free radiation from the metal-ion plasma is expected to be important \citep[\S3.3 in][]{1996A&A...313..217W}.\hfill\break
(iii) When free-free is the dominant emission process, the radiative-equilibrium temperature for gas near the photosphere of an RCB star should be $\sim 1{,}000\;{\rm K}$ \citep[\S3.4 in][]{1996A&A...313..217W}.

Metal abundances in the accretion disk are not predicted by our model, but if metals are present then carbon is presumably one of the most abundant. Point (i), above, then leads us to expect that the free-electron density in the disk mid-plane should be $n_e\sim n({\rm C^+})$. As a numerical example: if the carbon abundance in the disk is carbon:helium~$\simeq0.004$ by number, as in the Sun, then $n_e\sim 2\times 10^{11}\,{\rm cm^{-3}}$. In turn that corresponds to a free-free optical depth, normal to the plane of the disk, that exceeds unity for wavelengths $\lambda \gta 6\;{\rm \mu m}$.

It is a straightforward exercise to calculate how much photospheric light impinges on a geometrically thin accretion disk; the result is 25\% of all the emitted light. If the accretion rate is high, as we are assuming, then the incident photospheric light will almost all be absorbed and subsequently re-emitted as mid-IR radiation.

RCBs do indeed exhibit mid-IR flux in excess of the photospheric emission, typically at power levels $\sim30$\%\ of the stellar light \citep[e.g.][]{2011ApJ...739...37G,2018AJ....156..148M} and often quite well approximated by a $T\sim 1{,}000\;{\rm K}$ Planck function\footnote{RCBs are not unique in that respect: binary post-AGB stars commonly have strong mid-IR excess emission with a black-body slope at long wavelengths \citep{2006A&A...448..641D}. And in one case the excess has been shown to conform accurately to a $T \simeq 1{,}150\;{\rm K}$ black body \citep{2003A&A...397..595D}. Their infrared excess is usually interpreted as emission from a dusty disk, but metal-ion brem\ss trahlung is potentially interesting as an alternative interpretation.} \citep{2001ApJ...555..925L,2011ApJ...739...37G}. Thus it appears that brem\ss trahlung from a metal-ion plasma in an accretion disk may be able to account for the general character of the mid-IR continuum excess observed from RCBs.

\subsubsection{Non-photospheric spectral features}
The brem\ss trahlung opacity is strongly wavelength dependent, being much smaller in the optical and UV where spectral structure arising from free-bound and bound-bound emission ought to be observable when the stellar photosphere is eclipsed. In particular, optical and UV line emission from metals in the disk may be seen prominently against a weak continuum. Keplerian rotation introduces shifts of up to $\pm 50\;{\rm km\,s^{-1}}$ on any emission lines, and we therefore suggest that the observed ``E1'' and ``E2'' groups of RCB ``chromospheric'' emission lines \citep{1963ApJ...138..320P,1972MNRAS.158..305A,1990MNRAS.244..149C} arise from the accretion disk in our model.

The accretion disk should also contain various species of small molecules --- anything present in the pre-disruption molecular cloud that is robust enough to survive at $\sim 1{,}000\;{\rm K}$, plus whatever forms in situ. The rotational levels of these molecules will be thermally populated to high quantum numbers, but the pure rotational transitions all lie at long wavelengths, where the continuum optical depth is large, and would therefore be difficult to detect. Excited vibrational states should also be significantly populated at temperatures $\sim 1{,}000\;{\rm K}$, and correspondingly we expect radiative transitions in the form of various molecular rovibrational bands in the mid-IR. In the flux emerging from the disk (i.e. what would actually be observed), the mid-IR spectral structure should be only small in amplitude because the continuum optical depth is of order unity. Absent a detailed model of the irradiated accretion disk structure, which is beyond the scope of this work, it is not possible to say whether any given mid-IR spectral feature should be seen in emission or absorption.

The available mid-IR data reveal that most RCBs do display some weak spectral structure. In the literature those features have been interpreted in terms of various carbonaceous dust materials \citep{2001ApJ...555..925L,2011ApJ...739...37G,2013ApJ...773..107G} -- e.g. amorphous carbon grains, and PAHs -- consistent with the idea that the dust particles responsible for the fading events are carbon grains ejected from the star. In the context of the model under discussion here, though, we currently lack the detailed quantitative description that would be required to predict the mid-IR spectral structure arising from the accretion disk. We therefore suggest no specific interpretations of any mid-IR spectral features in terms of disk emission.

Notwithstanding the fact that our model predicts powerful mid-IR continuum from the accretion disk, we note here the possibility that the observed mid-IR spectral features might actually arise elsewhere. Three points support that idea. First, the entrance apertures of the mid-IR instruments encompass a region around each target star that is much larger than the accretion disk. Thus the spectrographs may have captured emission from solid \htwo\ -- dust particles or larger lumps in the tidal debris -- that is remote from the star. Secondly, although pure \htwo\ dust is expected to absorb almost nothing in the optical band,  once grains are outside the cloud they are exposed to far-UV from the star, and some of those photons cause ionisation in the solid. As mentioned in \S\ref{sec:overallstructure}, that leads to the formation of ${\rm(HD)_3^+}$, and that molecular ion is known \citep{2011ApJ...736...91L} to have several mid-IR transitions that coincide with the almost ubiquitous mid-IR astronomical bands that are usually attributed to PAHs. The similarity between those bands and the mid-IR features of some RCBs prompts us to suggest that the latter features actually arise from ${\rm(HD)_3^+}$. Finally, although only a small number of RCB stars have been identified as having "PAH-like" mid-IR features \citep{2013ApJ...773..107G}, those examples are in fact the ones where the features are strongest (highest equivalent widths), and thus clearest (e.g. least affected by continuum fitting). Moreover, although that group of RCB stars is small in number, it includes two of the four hottest RCBs known \citep[DY~Cen and HV2671,][]{2002AJ....123.3387D} --- stars that can be expected to create higher concentrations of molecular ions in any circumstellar solid \htwo.

\subsection{Time-dependent mass flow through the disk}\label{sec:timedependence}
If mass supply to the disk is interrupted -- as would happen if cloud disruptions cease for any reason -- then material in the inner disk will accrete onto the star on a timescale $t_{acc}\equiv R_*/|v_R|$. In the previous section we arrived at the estimate $|v_R| \lta 0.16 \;{\rm km\, s^{-1}}$, implying $t_{acc}\gta 11\;{\rm yr}$. For active RCB stars the number of disruptions within one accretion timescale is ${\cal R}\, t_{acc} \gta 3$, indicating that the mass supply is effectively continuous --- so the accretion disk is present at all times, and likewise the reprocessed radiation it produces (\S\ref{sec:diskaccretion}). However the flux in reprocessed radiation will vary somewhat as the disk mass (etc.) fluctuates on timescale $t_{acc}$. By contrast, low-activity RCB stars with ${\cal R}\, t_{acc} \ll 1$ will usually not have an accretion disk: it will form only after a disruption occurs, and then it will gradually disappear again on the timescale $t_{acc}$. In this low-activity regime the reprocessed radiation will appear as bright ``flares'' lasting for a time $\sim t_{acc}$. A detailed model of the irradiated disk would be required in order to attempt a quantitative comparison with the available data \citep[e.g.][]{1997MNRAS.285..317F,2022A&A...667A..83T}.

\section{Conventional grain materials}\label{sec:conventionalgrains}
The model described above works well in many respects. But hydrogen is an unconventional choice of dust material, and readers may wonder whether orbiting gas clouds permeated by more conventional types of dust might also be able to account for the fading events? In this section we show that they cannot.

\begin{figure}[t]
    \centering
    \includegraphics[width=8cm]{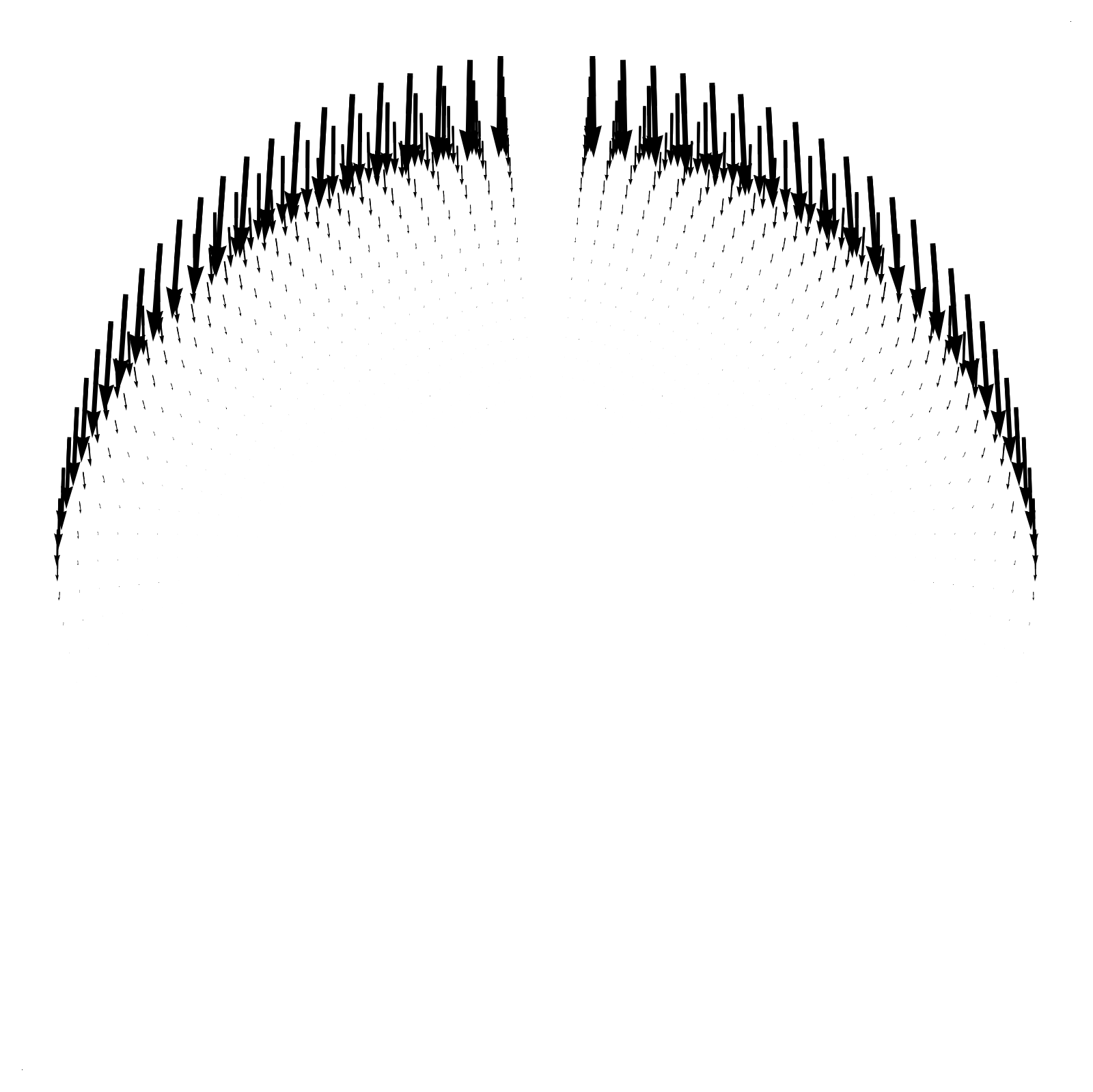}
    \caption{As figure \ref{fig:forcedensity}, but for large grains of conventional composition (silicate/graphite), with scattering albedo $Q_{sca}/Q_{ext}= 0.6$.}
    \label{fig:forcedensitysilicate}
\end{figure}

\begin{figure}[t]
    \includegraphics[width=8cm]{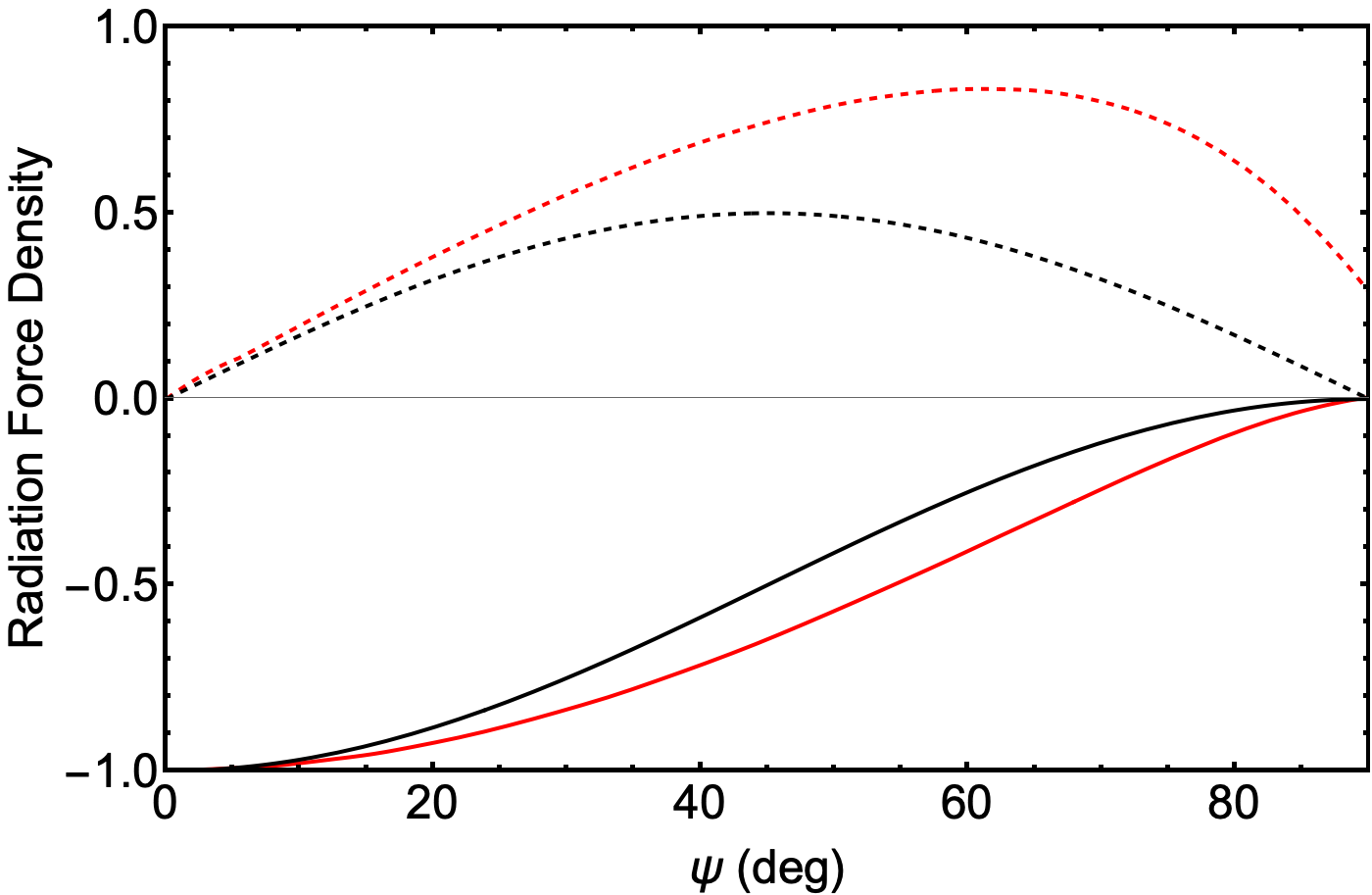}
    \centering
    \caption{Radial (solid lines) and tangential (dashed lines) components of the radiation force-density at the cloud surface. Red lines show the Monte Carlo results for large grains of conventional composition with a scattering albedo of $0.6$. Black lines show the analytic approximations appropriate to pure absorption.}
    \label{fig:surfaceforcedensitysilicate}
\end{figure}

\begin{figure*}[t]%
    \centering
    \hbox{\hskip 2cm{\includegraphics[width=9cm]{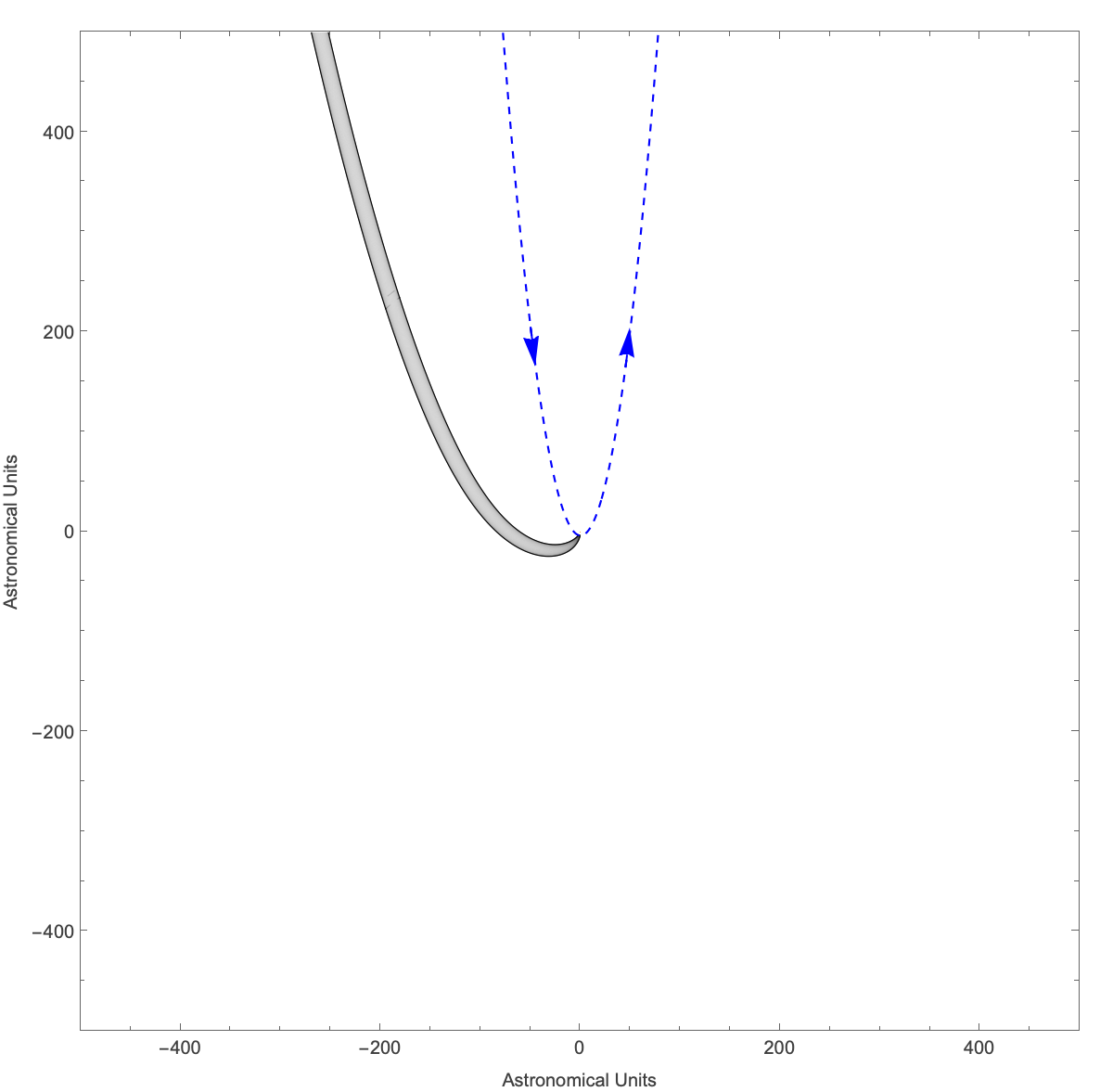}}\qquad{\includegraphics[width=5cm]{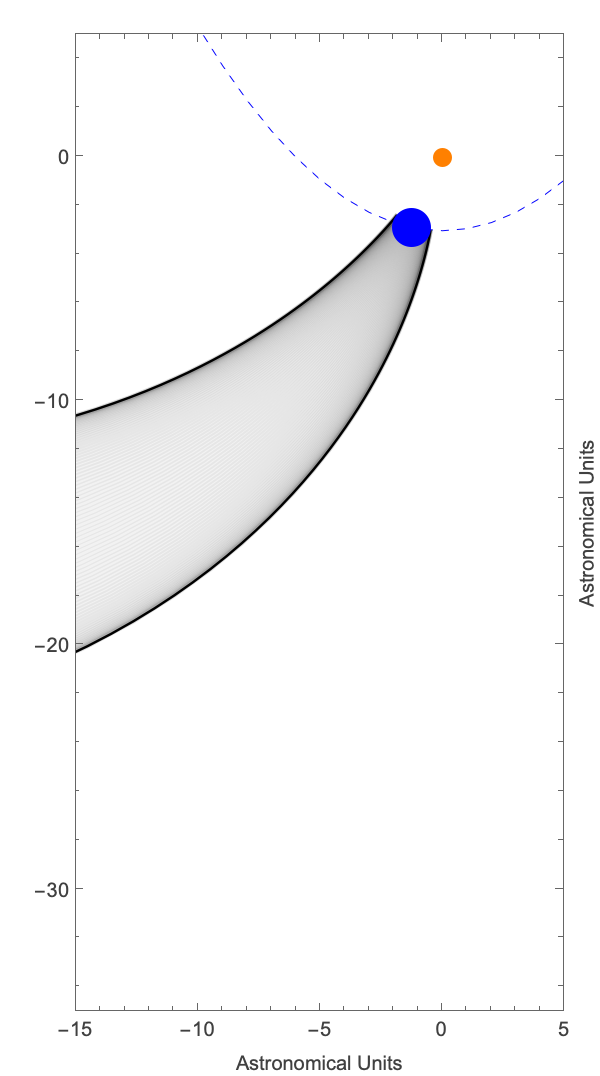}}}%
    \caption{As figure \ref{fig:windstructure}, but for large ($a\sim 10\,{\rm \mu m}$) grains of silicate or graphite (albedo$\,= 0.6$, hence $\psi_\circ\simeq 89^\circ\!.1$). The opacity is assumed to be $\kappa_{pr}=300\,{\rm cm^2\,g^{-1}}$. Note the small opening angle of the flow in comparison with that shown in figure \ref{fig:windstructure}.}%
    \label{fig:silicatewindstructure}%
\end{figure*}

The optical properties of silicate and graphite dust grains are shown in Appendix A (figure \ref{fig:opticalproperties}). Although several aspects differ markedly from those of solid \htwo\  (e.g. the peak opacity for graphite is about five times larger than that for hydrogen), the most important point here is that both of the conventional types of dust exhibit significant absorption, with a scattering albedo that is typically far from unity. Indeed the smallest particles of both graphite and silicate are absorption dominated. That makes a big difference to the wind launch conditions because it means that there is no scattered radiation field emerging from the cloud interior. Instead the emerging radiation will be only thermal photons, at much longer wavelength, which have very low probability of interacting with the dust. Thus, when absorption dominates, the radiation forces on the grains at the cloud surface are just those due to the radiation coming direct from the star, and the radial and tangential force components are $-\cos^2\psi$ and $\cos\psi\,\sin\psi$, respectively. The radial component only vanishes at the limb of the cloud, so $\psi_\circ=90^\circ$.

\begin{figure}[h]%
    \centering
    \includegraphics[width=0.47\textwidth]{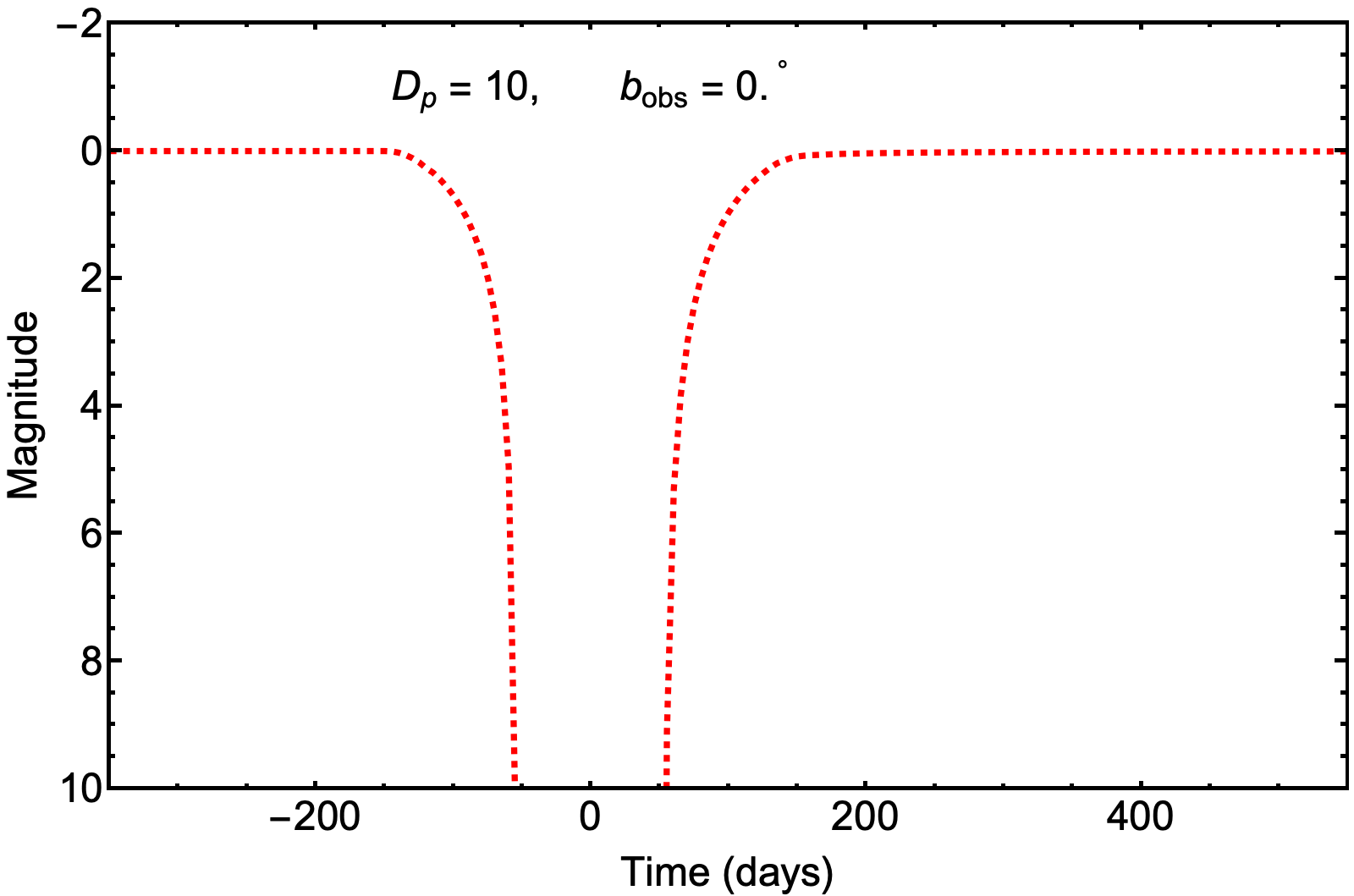}%
    \caption{The unscattered light curve for a cloud containing large silicate/graphite grains with albedo$\,=0.6$ (so $\psi_\circ=89^\circ\!.1$), $Q_{ext}/Q_{pr}=2$ and $\kappa_{pr}=300\,{\rm cm^2\,g^{-1}}$ (appropriate to $a\sim 10\,{\rm \mu m}$ grains). Scattered light is not computed. Note the symmetric form and the slow ingress (approximately 90 days). This figure can be compared with the unscattered light component shown for hydrogen dust in the lower-left panel (c) of figure \ref{fig:distancelatitudevariation}.}%
    \label{fig:silicateeclipse89}%
\end{figure}

\begin{figure}[h]%
    \centering
    \includegraphics[width=0.47\textwidth]{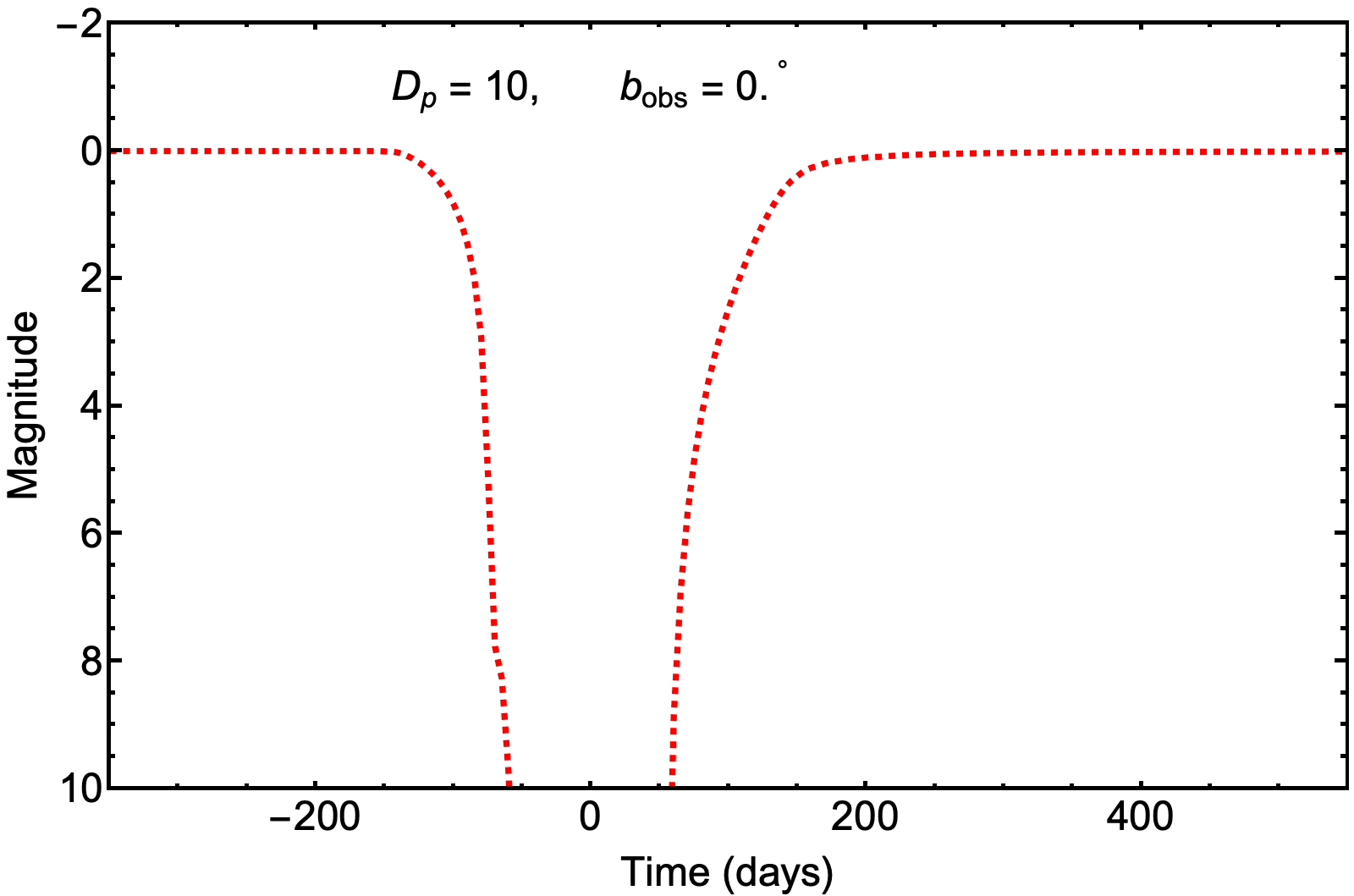}%
    \caption{As figure \ref{fig:silicateeclipse89}, but for silicate grains with $a\sim 0.2\,{\rm \mu m}$, so albedo$\,=0.8$ ($\psi_\circ=86^\circ\!.8$),  $Q_{ext}/Q_{pr}=2$ and $\kappa_{pr}=10^4\;{\rm cm^2\,g^{-1}}$.}%
    \label{fig:silicateeclipse87}%
\end{figure}

Such a large value of $\psi_\circ$ is problematic for two reasons. First, with $\psi_\circ= 90^\circ$ the perpendicular optical depth of the wind at launch is vanishingly small, and the wind can only cause significant extinction where the dust sheet is seen edge-on. But that condition is only met at the limb, where the cloud itself (presumed opaque) contributes to the extinction. Thus the eclipse light curves are in effect just those due to obscuration by the cloud alone, and the profiles are very nearly time-symmetric. That is in stark contrast to the observed fading events, which exhibit rapid onset and slow recovery. Secondly, with $\psi_\circ=90^\circ$ the occultation front moves with exactly the same speed as the cloud itself --- i.e. just the Keplerian speed of the orbit. That is in conflict with the data. The difficulty is that the time required for the star's flux to decline, at the onset of a fading event, can be as little as twenty days \citep{2025MNRAS.537.2635C} (for R CrB itself), and in that time the cloud is required to move a distance of at least $2R_*$. That corresponds to a speed of $64\;{\rm km\,s^{-1}}$, whereas the orbital speed for a parabolic orbit at $10\;{\rm AU}$ periastron (for example) is only about one fifth of that value. These two problems will be familiar to some readers, as they have been raised previously as objections to an orbiting gas cloud interpretation of RCB fading events \citep{1973MNRAS.161..293F}.

Very small grains of conventional composition are ruled out by the arguments above. At the other end of the size spectrum one finds that large grains of either silicate or graphite have an albedo $\sim 0.6$, so that photons scatter almost twice as often as they are absorbed. This case has no simple analytic solution so we have repeated the Monte Carlo photon transport through the cloud, as described in \S 3 but with an albedo of $0.6$. (For simplicity, our Monte Carlo simulations for conventional grain materials assume isotropic scattering.) The results of that experiment are shown in figures \ref{fig:forcedensitysilicate} and \ref{fig:surfaceforcedensitysilicate}; they demonstrate that the surface radiation forces are inward until very close to the limb of the cloud ($\psi_\circ\simeq 89^\circ\!.1 $). To compute the wind structure and the light curves we specify $\kappa_{pr}=300\,{\rm cm^2\,g^{-1}}$ and $Q_{ext}/Q_{pr}=2$, appropriate to $a\sim 10\,{\rm \mu m}$ grains for either of the conventional materials. The resulting wind is shown in figure \ref{fig:silicatewindstructure}, illustrating the much smaller opening angle of the flow as compared with figure \ref{fig:windstructure}. An example light curve\footnote{We have not computed the scattered light contribution to the light curves for silicate/graphite dust winds, because each different value of the albedo requires a separate set of slab Monte Carlo simulations, which are computationally intensive, as described in \S\ref{sec:scatteredlight}.} is shown in figure \ref{fig:silicateeclipse89} --- explicitly demonstrating the almost symmetric form and the slow ingress expected for the eclipses. In short: the problems with small grains of conventional composition are shared by large ones.

Figure \ref{fig:opticalproperties} shows that the albedo for graphitic dust is largest for the largest particles, and it follows that graphite grains of all sizes are ruled out.

For silicates, on the other hand, the absorption is not as strong as in graphite, and there is a peak in the albedo at $a\simeq0.2\,{\rm \mu m}$. Allowing for a spread in sizes of a factor of a few around that peak we can contemplate a population of silicate grains with $Q_{sca}/Q_{ext}\simeq 0.8$. Repeating the Monte Carlo experiment, but with an albedo of $0.8$, shows that the radial component of the radiation force at the cloud surface changes sign at $\psi_\circ\simeq 86^\circ\!.8$. Thus even the best of the conventional grains are problematic, yielding a perpendicular optical depth of the wind at launch that is much less than unity, and a wind launch velocity that is almost radial with respect to the star ($V_\theta \ll V_r$). An example light curve for $a\sim 0.2\,{\rm \mu m}$ silicate grains (albedo$\,= 0.8$, $\kappa_{pr}=10^4\,{\rm cm^2\,g^{-1}}$, $Q_{ext}/Q_{pr}=2$) is shown in figure \ref{fig:silicateeclipse87}. Unlike figure \ref{fig:silicateeclipse89} this example is noticeably asymmetric, but not nearly as asymmetric as most of the observed RCB eclipses; and, like figure \ref{fig:silicateeclipse89}, ingress is slow.

The calculations in this section have been undertaken for silicate and graphite. However, it is clear that the problems that arise with those materials are a direct result of their non-negligible optical absorption. Those problems would, therefore, be shared by any other material with that characteristic.

\begin{figure*}[h!]%
    \centering
    \hbox{\qquad\quad{\includegraphics[width=5.5cm]{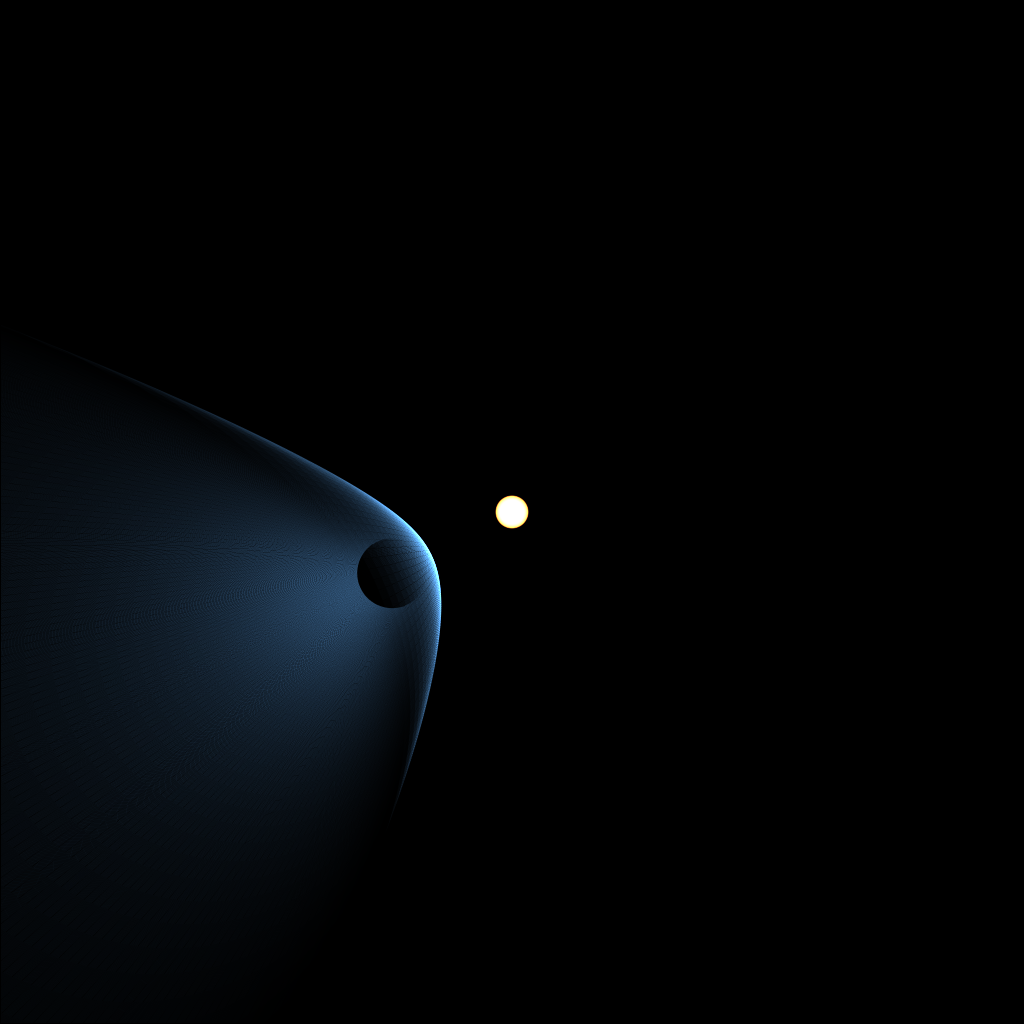}}{\includegraphics[width=5.5cm]{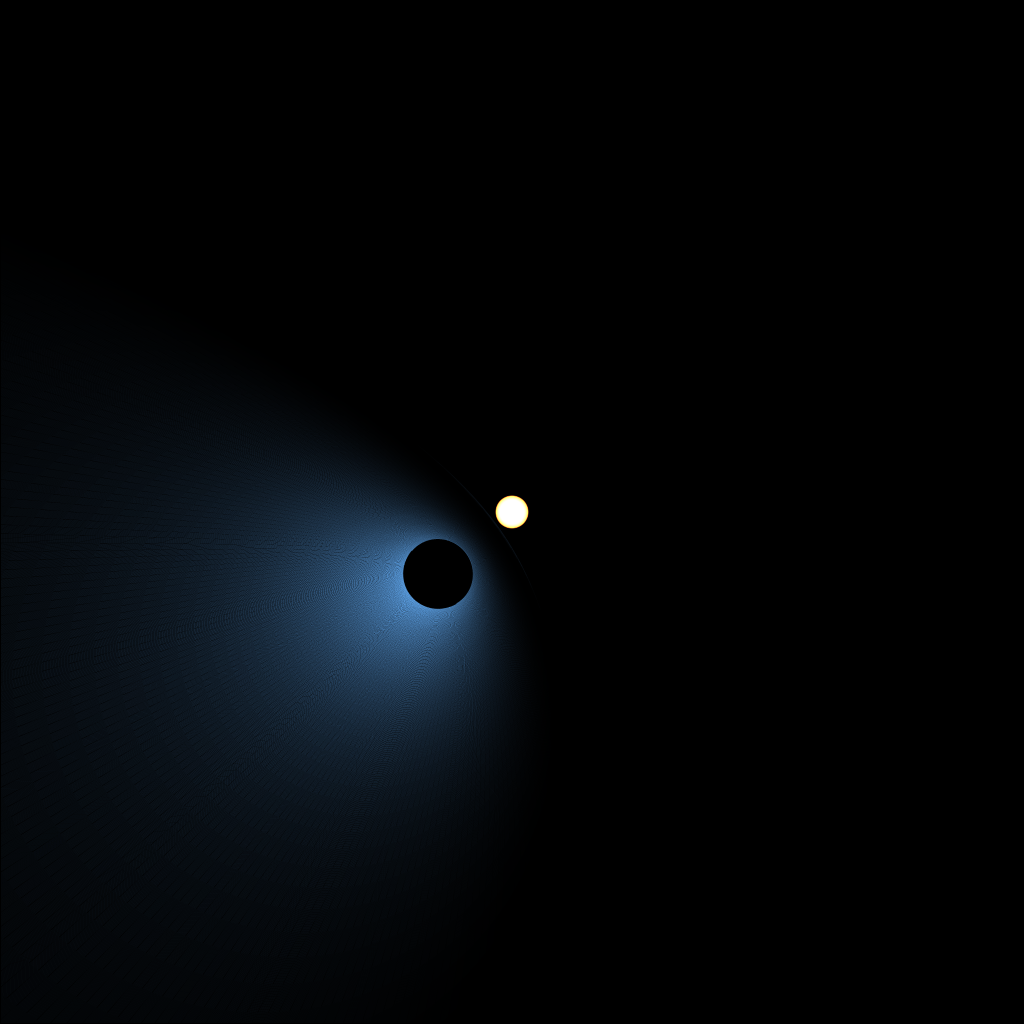}}{\includegraphics[width=5.5cm]{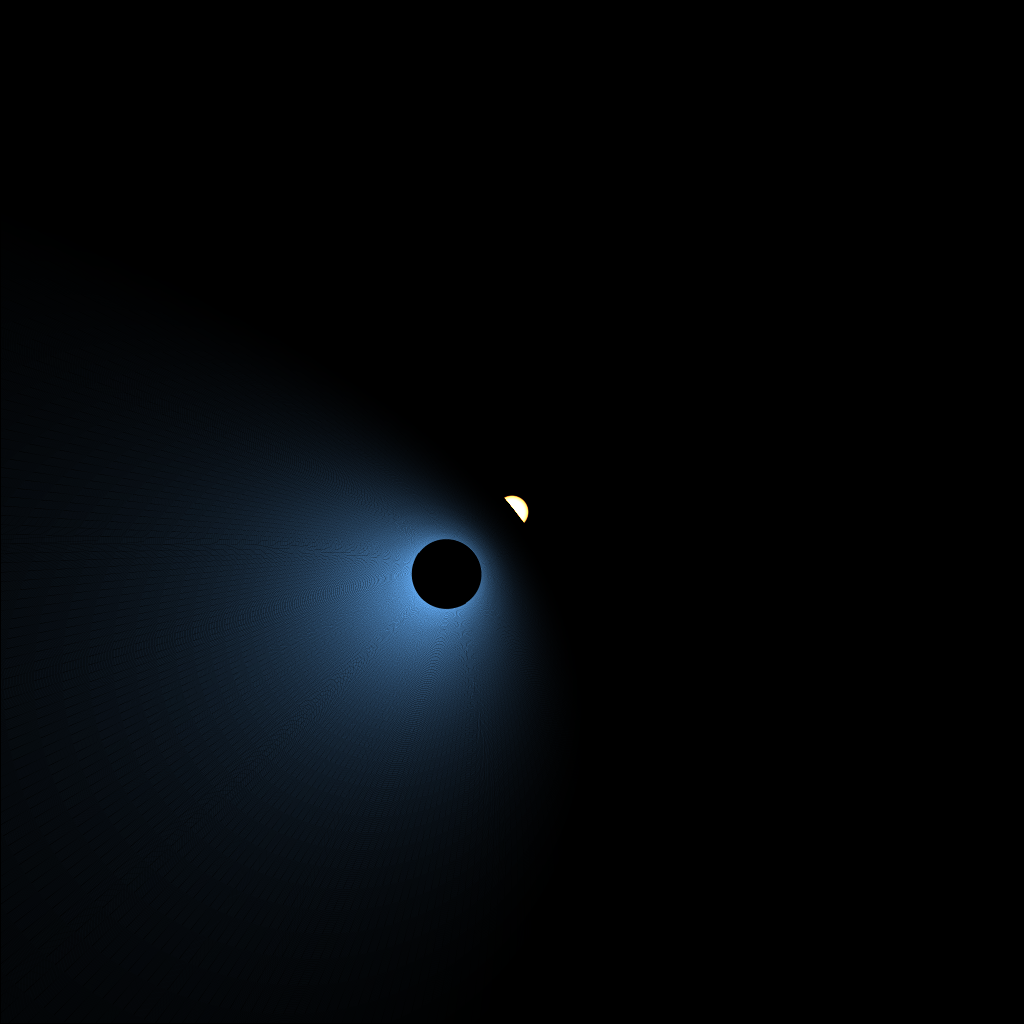}}}%
    \vskip -0.04cm
    \hbox{\qquad\quad{\includegraphics[width=5.5cm]{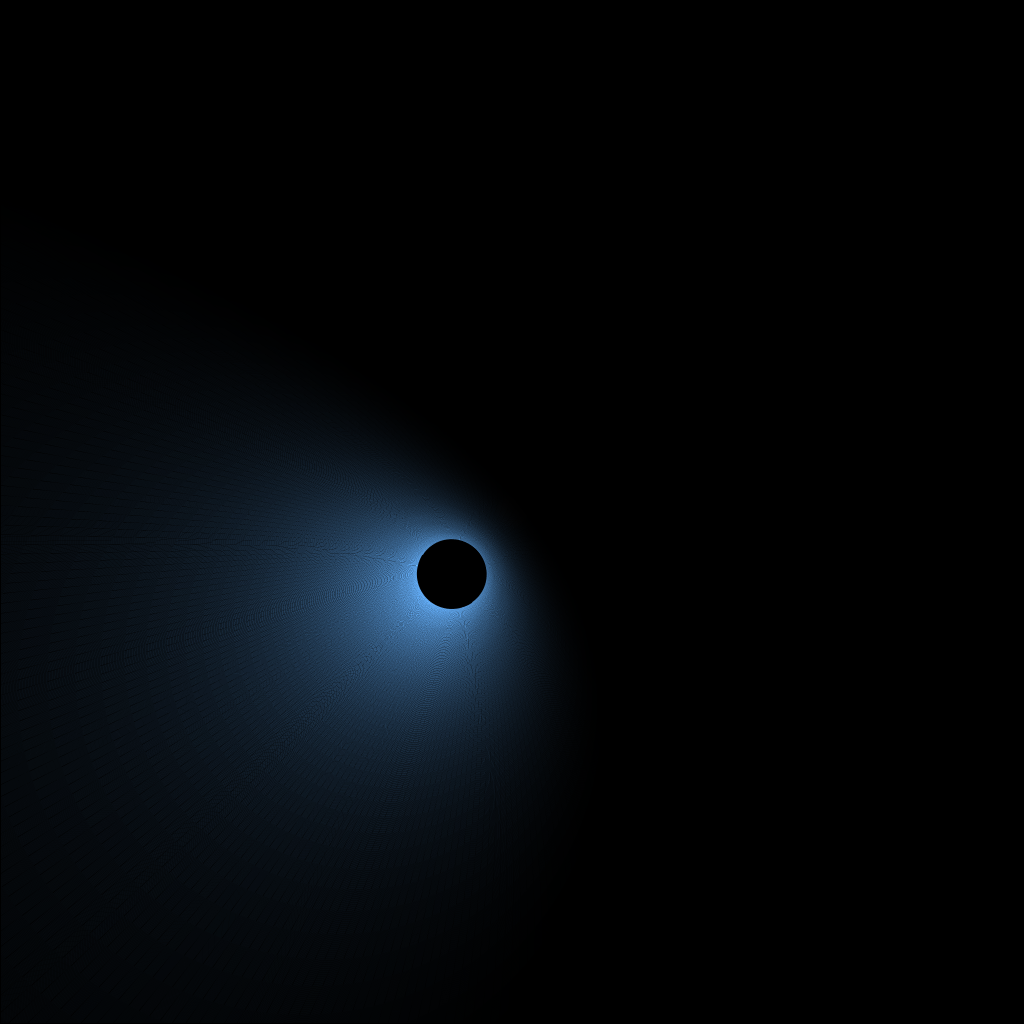}}{\includegraphics[width=5.5cm]{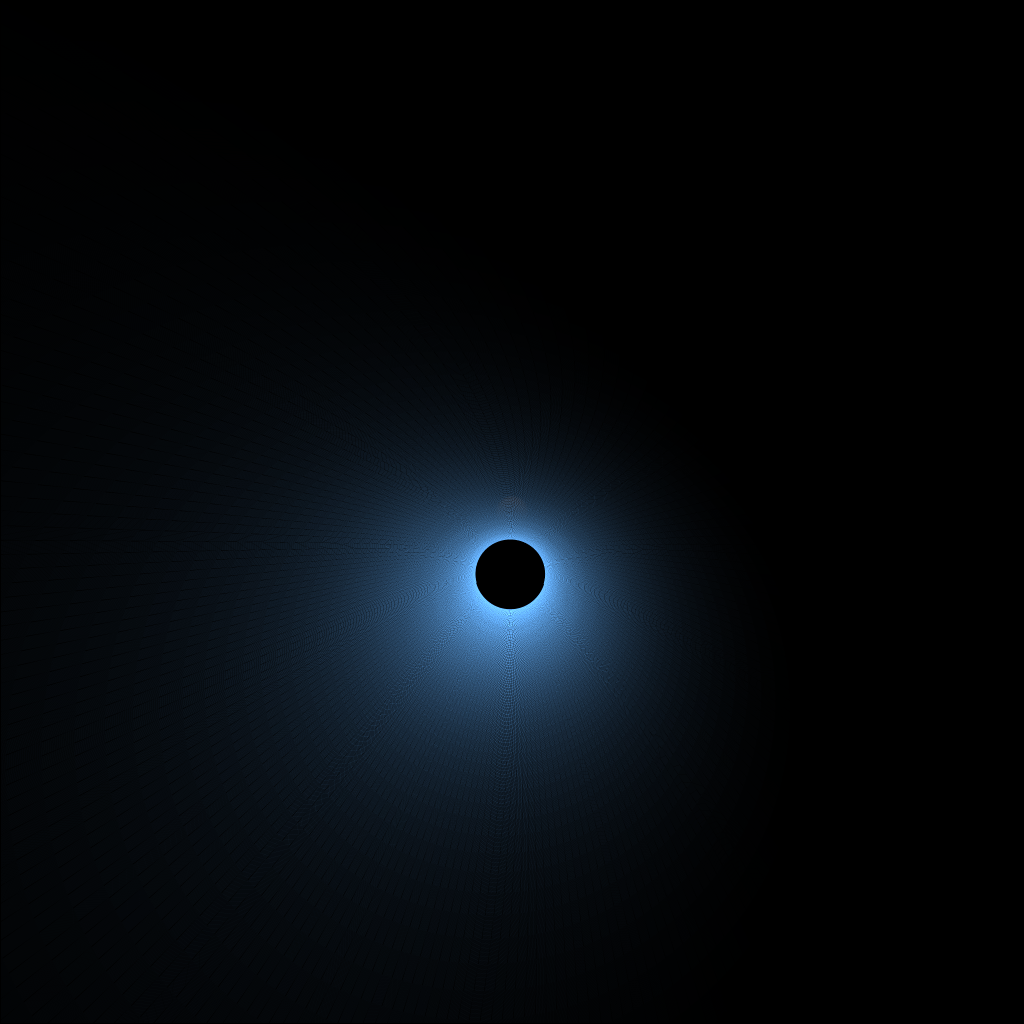}}{\includegraphics[width=5.5cm]{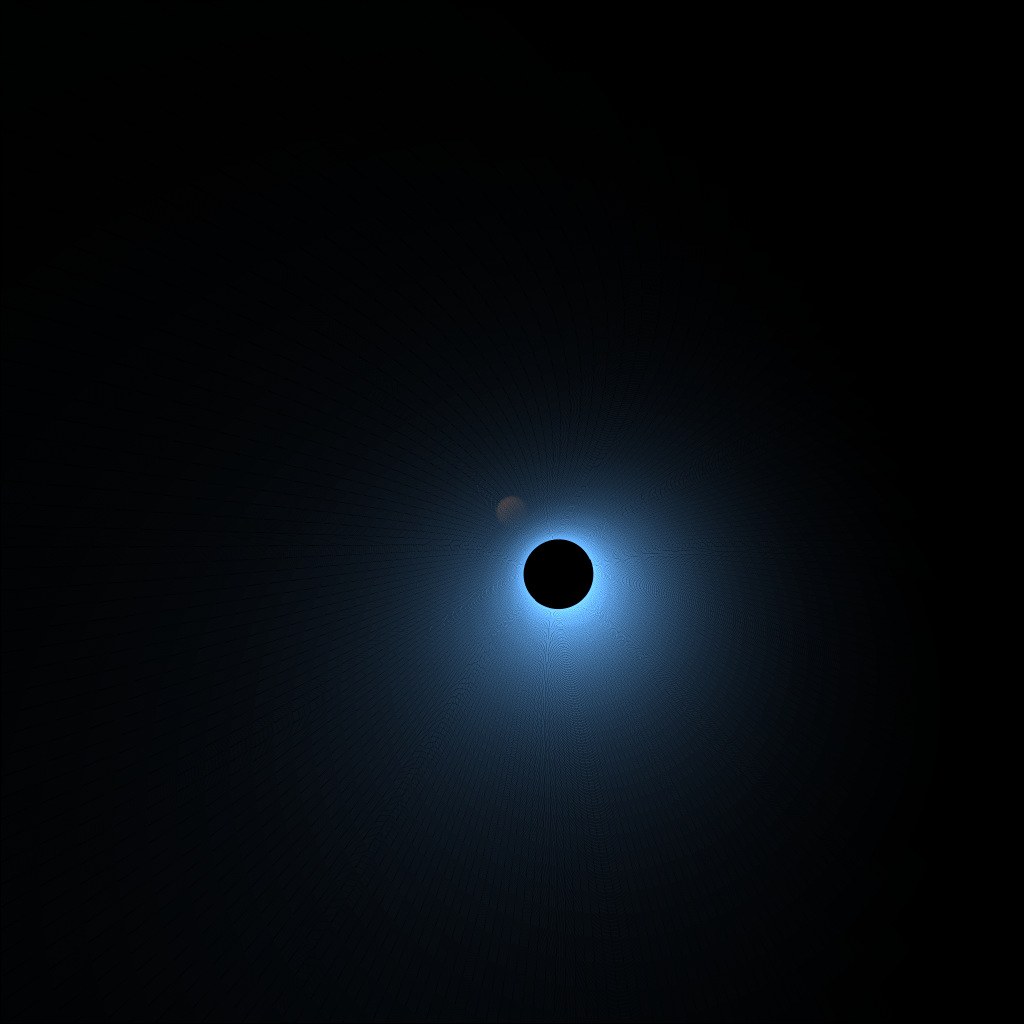}}}%
    \vskip -0.04cm
    \hbox{\qquad\quad{\includegraphics[width=5.5cm]{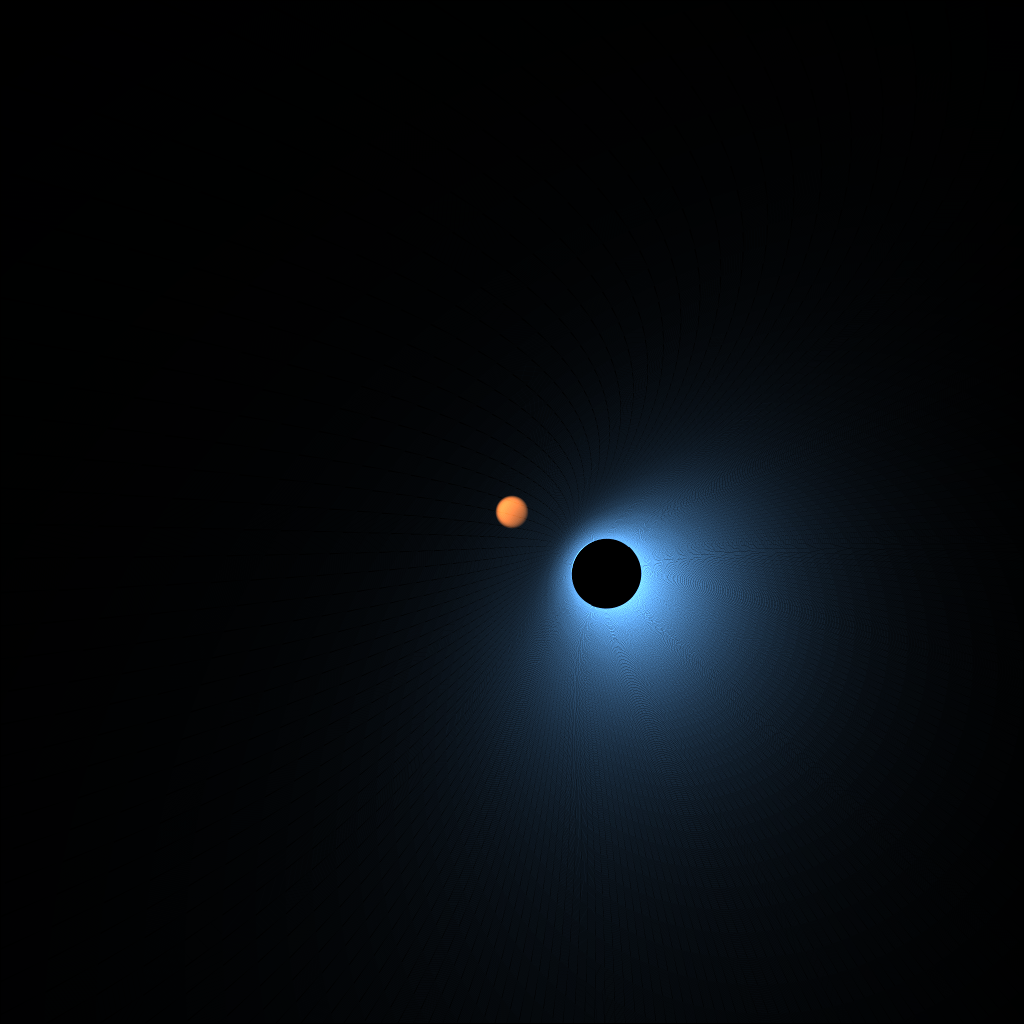}}{\includegraphics[width=5.5cm]{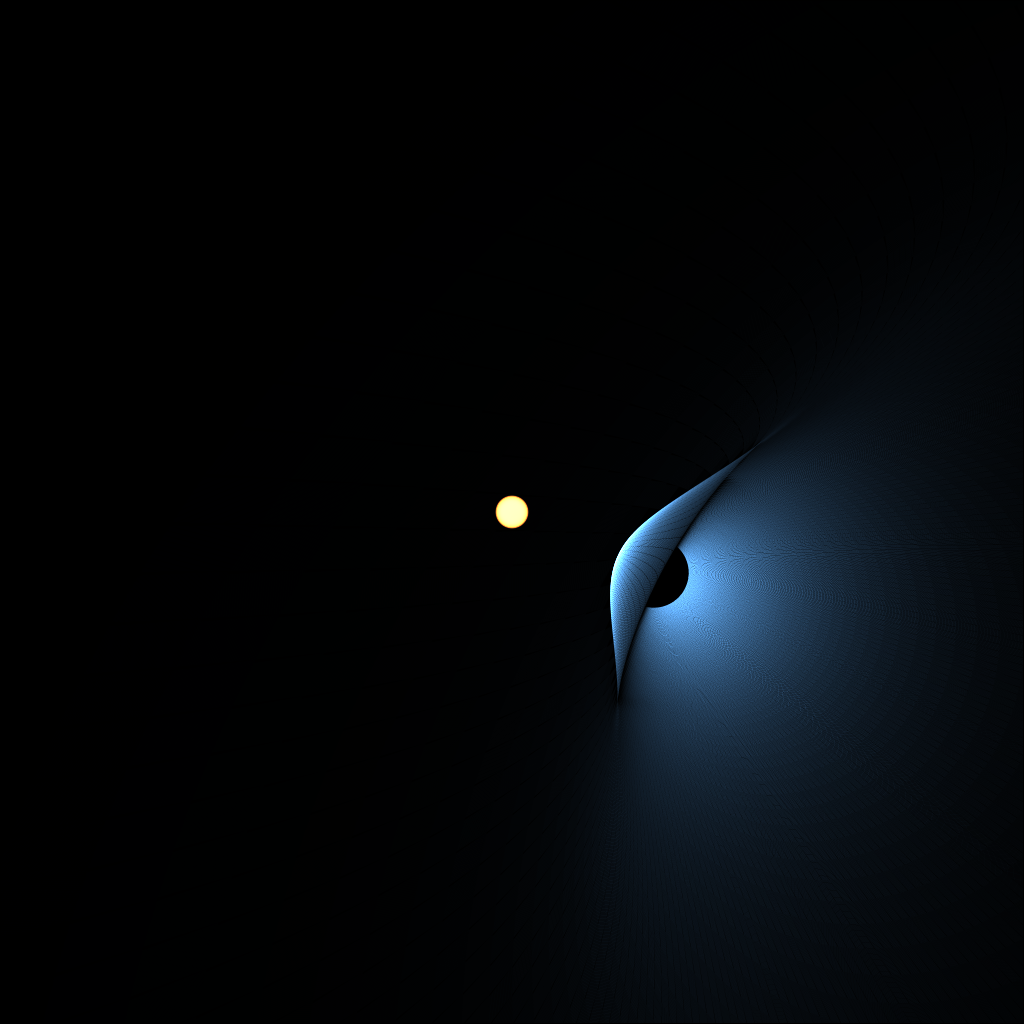}}{\includegraphics[width=5.5cm]{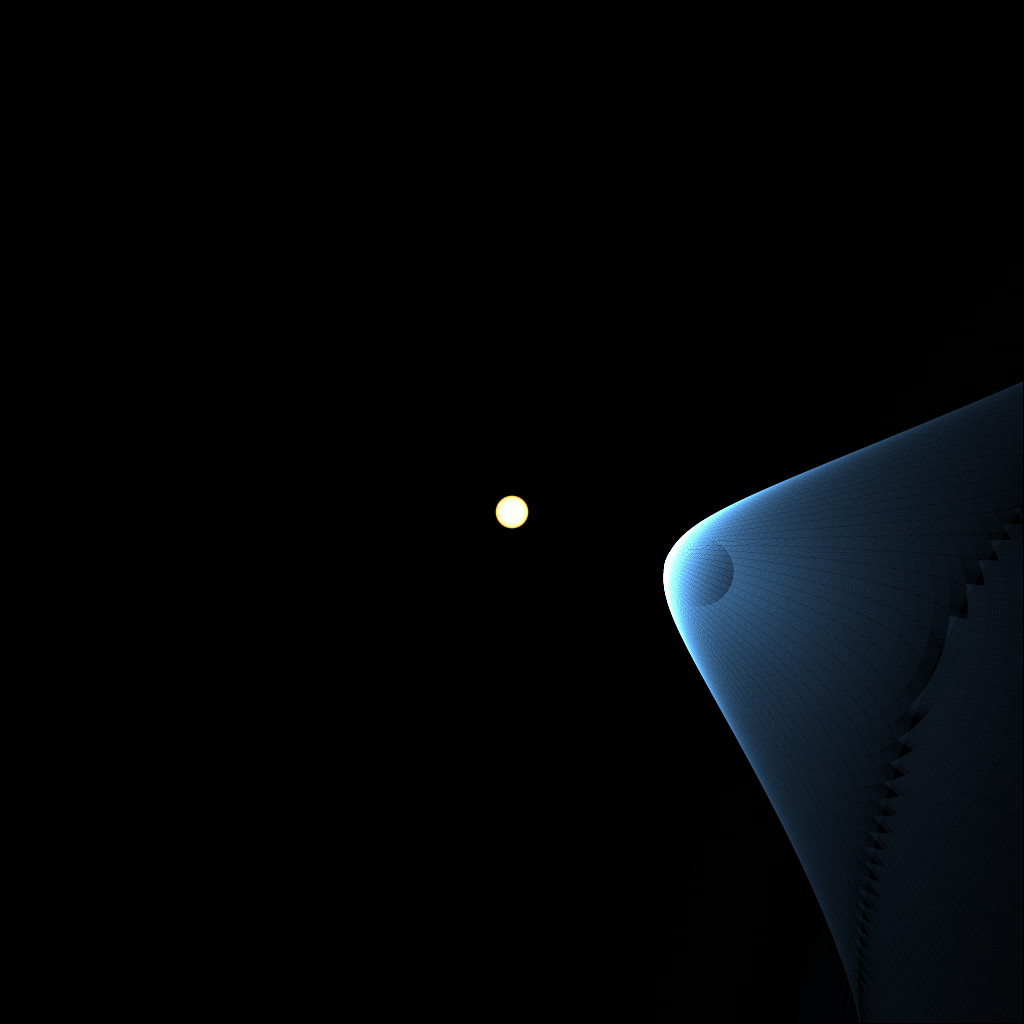}}}%
    \caption{How the eclipse of figure \ref{fig:medianeclipse} would appear to the observer if it were imaged at high resolution. These snapshots correspond to the following epochs (top-left to bottom-right): (i) the first epoch in the light curve ($t=-350\,$d); (ii) contact of the occultation front with the limb of the cloud ($t=-210\,$d); (iii) half-way through the initial decline ($t=-190\,$d); (iv) at flux minimum ($t=-175\,$d); (v) at conjunction ($t=0$; periastron); and, (vi)-(ix) recovery to maximum light ($t=140,\;280,\;420,\;550\,$d). The image scale can be gauged from the $1.56\;{\rm AU}$ diameter of the cloud.}%
    \label{fig:windimages}%
\end{figure*}

A final point of difference for conventional grain materials, relative to \htwo\ snow clouds, arises from the considerations set out in \S\ref{sec:evolution}: if the clouds are not condensing \htwo\ then we no longer expect the star to accrete a hydrogen-deficient envelope from tidal debris. At a minimum this means that if hydrogen dust is replaced by conventional grains then the simple evolutionary path set out in \S\ref{sec:evolution} is not available, which in turn makes the eclipse model less appealing. Worse, any tidal stripping or disruption of the orbiting clouds must lead to pollution of the stellar photosphere with hydrogen, and that is likely to be in conflict with the extreme levels of hydrogen deficiency observed in RCBs.

\section{Predictions of the model}
In order to test our interpretation of RCB fadings as eclipses by dust clouds, it is useful to sketch some predictions, as follows.

\begin{figure*}[t]%
    \centering
    {\includegraphics[width=17cm]{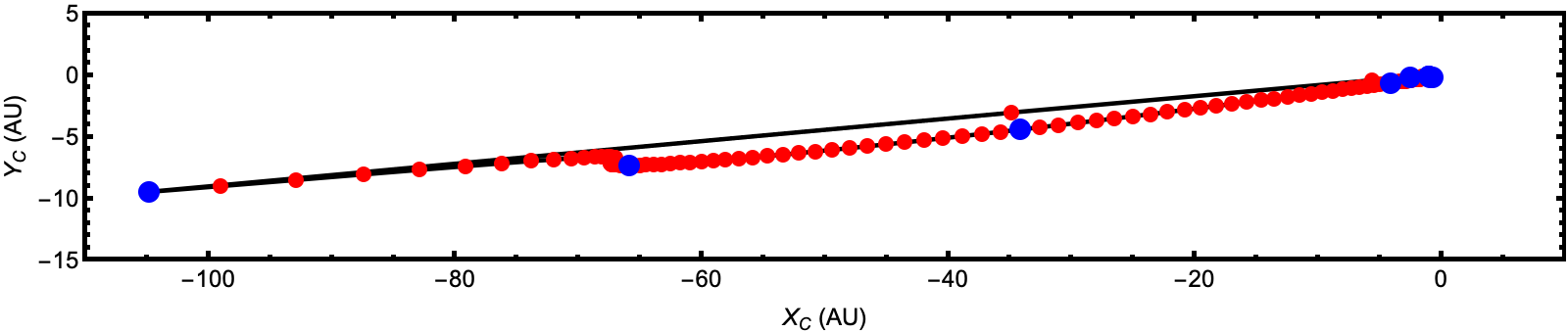}}%
    \caption{Observed centroids of light $(X_C(t), Y_C(t))$ for the model eclipse shown in figure \ref{fig:medianeclipse}. The sky coordinates are chosen such that the vector $(0,1)$ is parallel to the orbital angular momentum, as seen from the observer's location, and the point $(0,0)$ corresponds to the centre of the star. As with the photometry, the sampling interval is 5 days. Most of the 181 samples are plotted in red; the nine plotted in blue correspond to the epochs shown in figure \ref{fig:windimages}, with the three leftmost corresponding to the middle row of that figure.}%
    \label{fig:centroidshift}%
\end{figure*}

\subsection{Imaging and astrometry}\label{sec:astrometry}
The dust wind is a large structure in comparison with the star (see figure \ref{fig:windstructure}), and it may be possible to image the scattered light. Figure \ref{fig:windimages} shows the appearance of the system at nine epochs during the event shown in figure \ref{fig:medianeclipse}. We note that at present we cannot reliably calculate the small amount of light that diffuses through the cloud and emerges from the side that faces away from the star; we have therefore set the intensity to zero everywhere on that hemisphere. Although the dust wind is large compared to the star, very high angular resolution (of order a few milli-arcseconds) would nevertheless be required to resolve the structures shown in figure \ref{fig:windimages}.

Even if it is not possible to resolve the wind structure an astrometric shift of the integrated light may be measurable. Figure \ref{fig:centroidshift} shows the evolution of the centroid of light for the event in figure \ref{fig:medianeclipse}. This calculation includes only unscattered and singly-scattered light, because the twice-scattered flux is estimated via sparse sampling (\S \ref{sec:scatteredlight}). The largest centroid shift that arises in this event, and the events shown in figure \ref{fig:distancelatitudevariation}, is given in table \ref{tab:lightcurvemetrics}.

The behaviour in figure \ref{fig:centroidshift} is easily understood. The centre of unscattered light is always close to (within $R_*$ of) the origin, whereas the centre of scattered light is displaced $\sim 10^2\;{\rm AU}$ to the left and moves slowly to the right --- reflecting the overall sweep of the wind as the cloud orbits the star (figure \ref{fig:windstructure}). And the combined centroid is just the flux-weighted sum of those two locations. By comparison the conventional picture of RCB fading events, in which carbon dust is ejected from the star towards the observer \citep[e.g.][]{1996PASP..108..225C}, implies only tiny astrometric shifts --- comparable to the size of the stellar photosphere.

Two caveats must be given in relation to our computed centroid shifts. First, the position of the centroid of the scattered light is sensitive to the optical properties of the dust: we find that large \htwo\ grains, which are strongly forward scattering, exhibit peak centroid shifts that are about half those produced by small grains (see Table \ref{tab:largegrainmetrics}). Secondly, because the centroid is a position-weighted sum of flux contributions, which converges slowly with increasing distance from the star, the calculated centroid may be sensitive to the treatment of the dust dynamics.

The current observational situation is as follows. In the {\it Gaia\/} DR3 astrometric reduction RCBs that have undergone fading events are poorly described by the resulting astrometric solutions \citep{2024A&A...684A.131T}. One possible contribution to those poor fits is a genuine astrometric wander, such as discussed above. However, the DR3 results assume a constant colour for each star, whereas RCBs change colour during declines, and consequently the DR3 astrometric solutions cannot be relied on for RCBs \citep{2024A&A...684A.131T}. Useful constraints on the true level of astrometric wander of RCBs must therefore wait until {\it Gaia\/} astrometric solutions with appropriate, epoch-dependent colours are determined from the upcoming DR4 data release.

\subsection{Polarisation}
Linear polarisation is expected to arise in our model, as a result of scattering from an asymmetric structure. We have not calculated Stokes parameters in this, first presentation of an entirely new model, but it would be useful to do so in future work to compare with existing \citep[e.g.][]{1992AJ....103.1652W,1997ApJ...476..870C,2007AJ....134.1877K} and new data. We anticipate that the linear polarisation is likely to be high because our model is scattering dominated. And, because the degree of polarisation is expected to increase with increasing scattering angle, we also anticipate that during an eclipse the centroid of linearly polarised light will be further offset from the star than that of the unpolarised light.

\subsection{Molecular hydrogen winds}
Perhaps the most confronting aspect of our model is that RCBs, despite being extremely hydrogen deficient stars, are supposed to exist in an environment where hydrogen is plentiful. Hydrogen in the vicinity of R~CrB has been searched for via the 21cm line of atomic H, with only an upper limit obtained \citep{2015AJ....150...14M}: $M_{\rm H} \lta 0.3\;{\rm M_\odot}$. Our model predicts that an active RCB phase lasting $10^4\,t_4\;{\rm yr}$ would populate the ISM, local to the star, with a total of $\sim 0.06\, t_4\;{\rm M}_\odot$ in the form of particles of solid \htwo. A fraction of that material will undergo photo-dissociation to atomic hydrogen, primarily as a result of far-UV radiation from the RCB star itself while the debris is close to it. Appendix \ref{sec:photodiss} attempts to estimate that fraction, but the calculation is hampered by lack of information on the RCB photospheric far-UV intensity. Adopting a black-body spectrum provides an upper limit to the column of dissociated \htwo, and for $T_*=6{,}750\;{\rm K}$ that limit corresponds to a depth $\sim 250\;{\rm \mu m}$ in a macroscopic lump ($\gta 1\;{\rm cm}$ in size) of the solid. Evidently only a tiny fraction of any such lump will undergo photo-dissociation. For smaller particles radiation pressure is important and they are exposed to intense far-UV for a much shorter time, so a much smaller dissociated column results. We estimate that, for a $T_*=6{,}750\;{\rm K}$ black-body spectrum, only particles on the small-side of the opacity peak will be completely dissociated; those on the large side experience progressively smaller fractional material loss with increasing size, in proportion to $1/\sqrt{a}$. Thus if the \htwo\ particles in the debris stream are predominantly well above a micron in size the expected level of atomic hydrogen in the vicinity of R~CrB should be far below the observational upper limit.

We are not aware of any direct constraints on molecular hydrogen in the immediate vicinity of RCB stars. In fact cold \htwo\ is very difficult to constrain, primarily because the optical/IR transitions are quadrupolar and thus extremely weak. Unfortunately, strong spectral features are not expected in the \htwo\ dust extinction curve \citep{2015MNRAS.450.1032K}. However, the dust wind ought to carry with it some gas. Although our model neglects the dust-gas coupling in the wind (see \S\ref{sec:coupling}), and therefore makes no predictions for the accompanying gas flow, the \htwo\ column-density in dust is large ($\gta 10^{19}\;{\rm cm^{-2}}$ for unit optical depth at $\lambda=600\;{\rm nm}$), so even a tiny gaseous fraction could amount to a substantial column. Furthermore, a fiducial gas column $\sim 10^{12}\;{\rm cm^{-2}}$ in (metastable) He atoms has been inferred from observations of RCB winds \citep{2011ApJ...743...44C}, and in our model the gaseous \htwo\ column could be comparable or even greater. It therefore seems possible that some of the stronger far-UV lines of \htwo\ might be detectable in RCB winds.

\section{Discussion}\label{sec:discussion}
An aspect of cloud disruption that did not receive attention in \S6 is the likely appearance of the unbound \htwo\ debris stream. In the pre-disruption cloud, only tiny (e.g. micron-sized) particles of solid \htwo\ can remain in suspension within the fluid near the surface, whereas much larger lumps can be anticipated in the core where the gas densities are much higher; \citet{suvorovwalker2025} estimated sizes up to $10\;{\rm cm}$. Presumably, then, the size distribution of \htwo\ solids in the debris stream will also be broad, with a correspondingly wide spread in radiation pressure force and terminal speed, but all directed radially away from the star. Thus the \htwo\  debris is expected to take the form of a radially elongated stream of particles whose total mass is $\sim 2\times 10^{-5}\;{\rm M_\odot}$. The macroscopic lumps within the stream would be difficult to observe directly, but if particles in the size range $1$-$100\;{\rm \mu m}$ amount to a significant fraction of the total mass then they should be readily detectable in scattered light. These large dust grains are strongly forward-scattering (Appendix  \ref{sec:opticalproperties}) and thus what would be seen is strongly biased towards the $\sim 1$\% of debris streams lying inside a $\sim 10^\circ$ cone towards the observer. We identify these \htwo\ debris streams with the radially elongated structures reported in scattered light for the RCB stars R~CrB and UW~Cen \citep{2011ApJ...743...44C,2012A&A...539A..56J}.

Not all of the known characteristics of RCBs are addressed by the calculations herein. Even amongst the short (seven point) list of RCB properties given in the introduction there is one item -- (vii) the oxygen isotope ratio -- that we can say nothing about. Accretion and the nuclear fusion of the accreting material are not within the scope of our modelling. However, to the best of our knowledge there is no reason to expect that the observed isotope ratio cannot be reproduced, and the absence of a prediction is not in itself objectionable.

The outstanding issue with our eclipse model is the lack of an astrophysical context. Given a large number of \htwo\ snow clouds on orbits that bring them close to the star there is the potential to explain much of the observed RCB phenomenology, but what would give rise to such a circumstance? It cannot be a steady-state situation, as the clouds are stripped/disrupted near periastron. What seems to be needed is (i) a large population of circumstellar clouds, and (ii) a mechanism that brings about a rapid change in their orbital elements, creating an intense shower of eclipses and disruptions. Proposition (i) seems surprising, initially, but it is less so when we recall the cometary globules seen in planetary nebulae --- most clearly in the Helix Nebula \citep[e.g.][]{1996AJ....111.1630O}. In the Helix, individual clump masses have been estimated at $\sim 10^{-5}\,{\rm M_\odot}$ \citep{1992MNRAS.255..177M}, and \htwo\ imaging has revealed $\sim 40{,}000$ molecular clumps \citep{2009ApJ...700.1067M}. It is conceivable that these clumps are ``primordial'' in some sense --- i.e. already present when the progenitor star was on the main-sequence. The mechanism (ii) might perhaps be similar to that responsible for showers of comets \citep{1981AJ.....86.1730H}? Alternatively it could involve the dissolution of a binary or triple star system, when the primary sheds its envelope to become a CO-WD, with the subsequent orbit of the WD taking it through the surrounding ``halo'' of clouds. In Appendix \ref{sec:eventshower} we present a simplified $N$-body calculation that illustrates one specific example of the latter scenario. The simulation yields a high rate of cloud disruptions, but it does so from a starting point where the halo of clouds has a total mass that is almost half that of the progenitor binary star. If that, or a similar scenario is to explain the observed RCB population the simplest underlying hypothesis is that most/all intermediate mass stars have similar haloes of clouds while they are on the main sequence, with RCBs arising as a brief phase in the lives of a small subset of the binaries in the stellar population. In that case the aggregate snow cloud population is relevant to the dynamics of the stellar population as a whole (i.e. the Galaxy), and the Galactic missing mass/dark matter problem, as originally suggested by \citet{1994A&A...285...79P}. A careful study of our proposed disruption-shower scenario, and other possible scenarios, would be valuable but is beyond the scope of this work.

The success of the present model in explaining RCB attributes encourages us to consider whether it might also offer insight into other peculiar stars? That is a question with a broader scope than can be tackled here; we confine our discussion to two groups of stars that appear closely related to the RCBs. First, there is a population of luminous stars that are spectroscopically similar to RCBs, but which do not exhibit fading events: these are the dustless hydrogen-deficient carbon stars \citep[e.g.][]{2022A&A...667A..83T} (dLHdCs). Unlike RCBs, dLHdCs do not have a strong infrared excess, and in our model that means they are not accreting tidal debris. With no eclipses and no accretion there is nothing to suggest that dLHdCs currently have any orbiting hydrogen snow clouds. So it is natural to identify dLHdCs with those stars that have been through an RCB phase in the past, and which have destroyed most/all of their orbiting snow clouds, but which have not yet burned all of the helium envelope that was accreted. A crude estimate of the lifetime in this phase can be arrived at from the steady-state models of \cite{1989ApJ...342..430I}: if we assume an envelope mass of $\sim 10^{-2}\;{\rm M_\odot}$ and a nuclear burning rate of $\sim 10^{-6}\;{\rm M}_\odot\, {\rm yr^{-1}}$ the implied lifetime of dLHdCs is $\sim 10^4\,{\rm yr}$. That lifetime is much greater than the disk accretion timescale, $t_{acc}\gta 11\;{\rm yr}$ (\S\ref{sec:timedependence}), indicating that the reprocessed radiations (\S\ref{sec:diskaccretion}) are expected to disappear fairly promptly once disruptions cease.

Secondly, there is another group of stars that, like RCBs, show both hydrogen deficiency and, in some cases, the presence of dust: the Wolf-Rayet (WR) stars \citep{2007ARA&A..45..177C}. Indeed some WRs are known from interferometric imaging to have dust winds arising from a companion at a separation of a few AU, with the dust distribution following a spiral morphology \citep{1999Natur.398..487T} --- quite similar to our RCB model, but with circular/elliptical companion orbits instead of parabolae. In contrast to RCBs, though, WRs have high temperature photospheres\footnote{Hot RCBs do exist, with estimated photospheric temperatures up to $\sim 20{,}000\,{\rm K}$ \citep{2002AJ....123.3387D}, but they are very few in number \citep{2020A&A...635A..14T}, and somewhat diverse in their properties \citep{2002AJ....123.3387D}.} \citep[$30{,}000$-$150{,}000\,{\rm K}$,][]{2007ARA&A..45..177C} and are luminous in the far-UV. In that situation we expect that the illuminated face of a cloud would be ionised, and if a wind is driven from the limb of the cloud it would be an ionised wind, not a dust wind. However, existing snow cloud models \citep{2019ApJ...881...69W} indicate high column-densities, so that the bulk of the neutral gas is self-shielding in the far-UV. We suggest, therefore, that when a snow cloud is adjacent to a hot star a hydrogen dust wind may be launched from the shielded, rear hemisphere of the cloud where the radiation field is benign. Our simplified dust wind model cannot be used to make quantitative predictions for this circumstance --- a new description is required. However, figure \ref{fig:forcedensity} demonstrates that the radiation forces on the dark side of the cloud are tiny in comparison with those on the illuminated side. And furthermore the cloud is transparent to only a small fraction of the luminosity of a WR star. In combination these two effects lead us to expect dust wind launch speeds that are tiny in comparison with those for a cool star of the same luminosity.

\section{Conclusions}
When a dusty gas cloud is close to a luminous star, the character of the resulting dust wind is sensitive to the optical properties of the grains. For conventional compositions the dust flows almost radially away from the star, even at launch, and therefore the wind modifies the eclipse light curves only slightly. The same would be true for any hypothesised dust composition in which the grains exhibit significant optical absorption; for that entire class of materials, eclipses by orbiting, dusty gas clouds are unlike RCB fading events. Pure hydrogen dust, however, has negligible optical absorption. In that case the scattered light emerging from the cloud exerts a strong lateral radiation force, and the wind is launched with a large opening angle. Eclipses by hydrogen snow clouds show a good match to the properties of RCB fading events. 

That is encouraging in itself. But what is really striking about that interpretation is that it also appears to account for other key characteristics of RCBs, even though they have no obvious connection to the fading events. First is the hydrogen deficiency of the star, which in our model arises as a result of the accretion of hydrogen-deficient tidal debris. Second is the strong mid-IR excess, which in our model is expected because of free-free emission from a metal-ion plasma in the accretion disk. Both aspects emerge naturally from our model of RCB fading events. Observationally neither feature is unique to RCBs, and our novel interpretations may be of interest in relation to other types of stars exhibiting mid-IR excess or hydrogen deficiency.

There are, therefore, strong motivations to investigate these ideas further. For RCBs the most pressing aspect to study is the astrophysical context: how might a large number of snow clouds suddenly appear on disrupting orbits? We sketched one scenario that can produce that result, but at the cost of a massive halo of clouds around the progenitor of the RCB star. If that scenario, or something like it, is indeed the evolutionary path that gives rise to RCB stars, then it seems likely that \htwo\ snow clouds are a dynamically important component of our Galaxy.

\begin{acknowledgements}
MAW thanks Stan Owocki for discussions relating to the launching of the wind, and Tim Bedding for feedback.
\end{acknowledgements}

\bibliographystyle{aa}
\bibliography{RCBPhenomenon}{}

\onecolumn

\begin{appendix}

\section{Optical properties of dust}\label{sec:opticalproperties}

\begin{figure*}[t!]%
    \centering
    \hbox{\qquad\includegraphics[width=5.1cm]{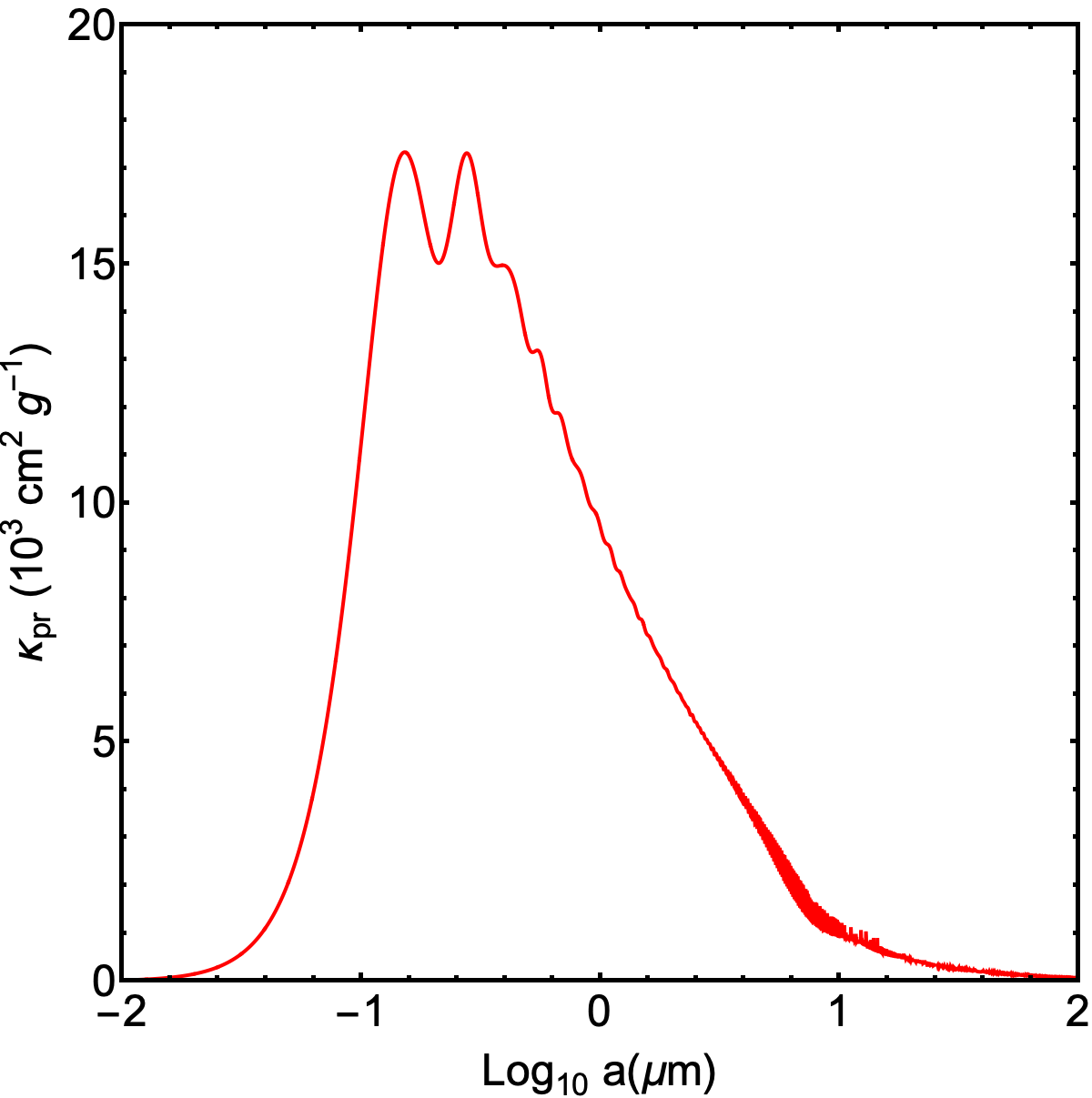}
    \quad\includegraphics[width=5.1cm]{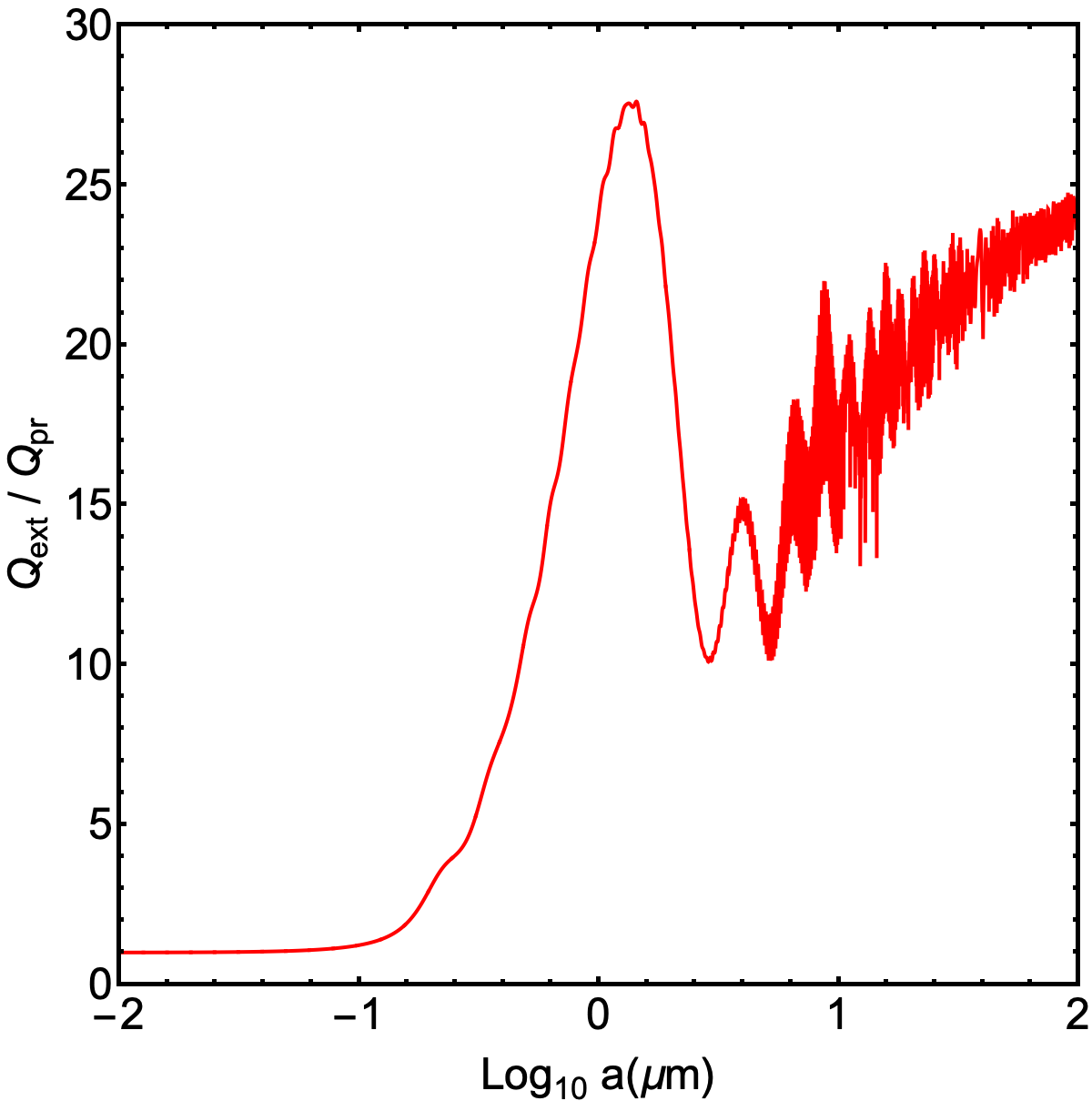}
    \quad\includegraphics[width=5.1cm]{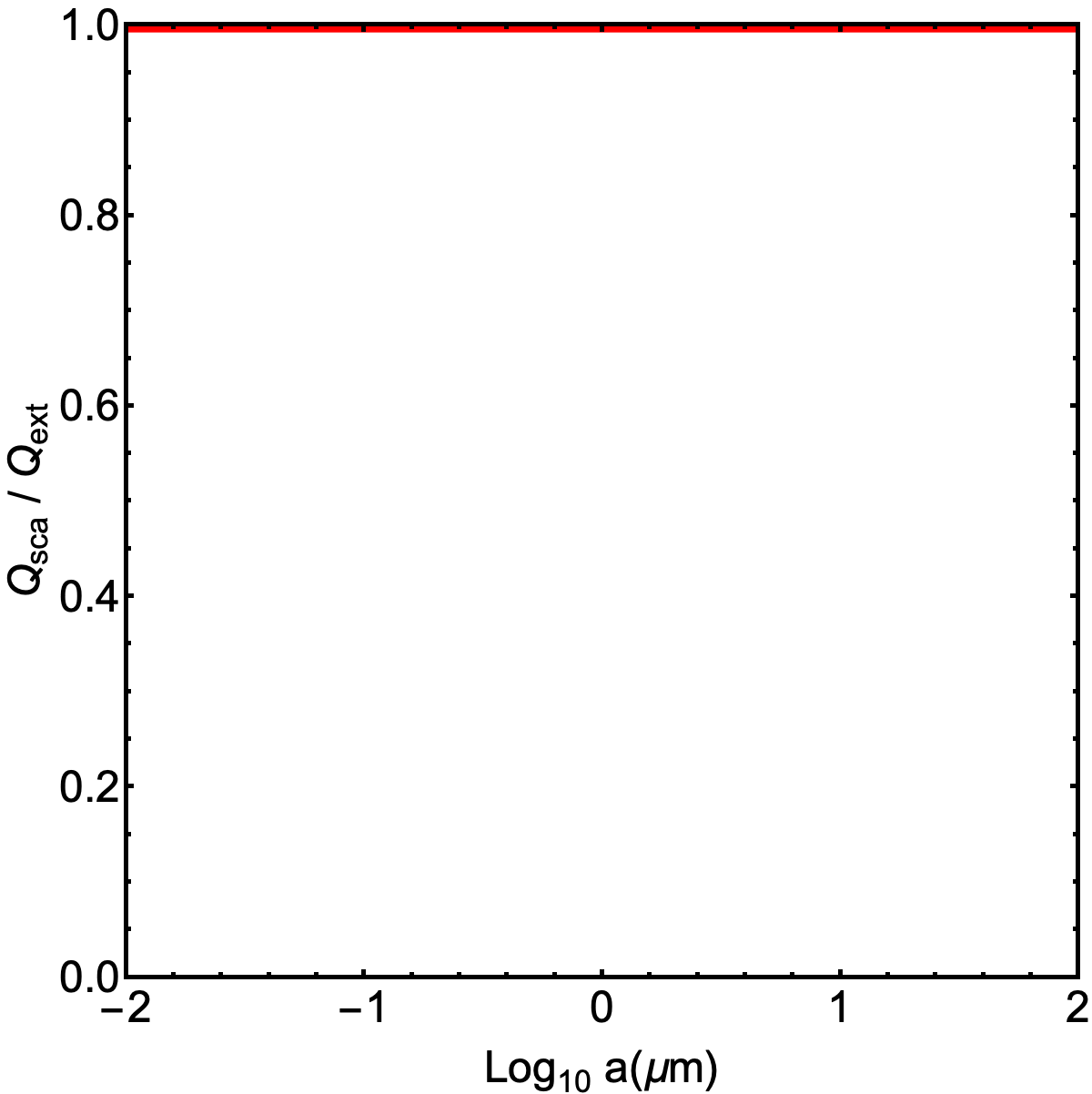}}%
    \medskip   
    \hbox{\qquad\includegraphics[width=5.1cm]{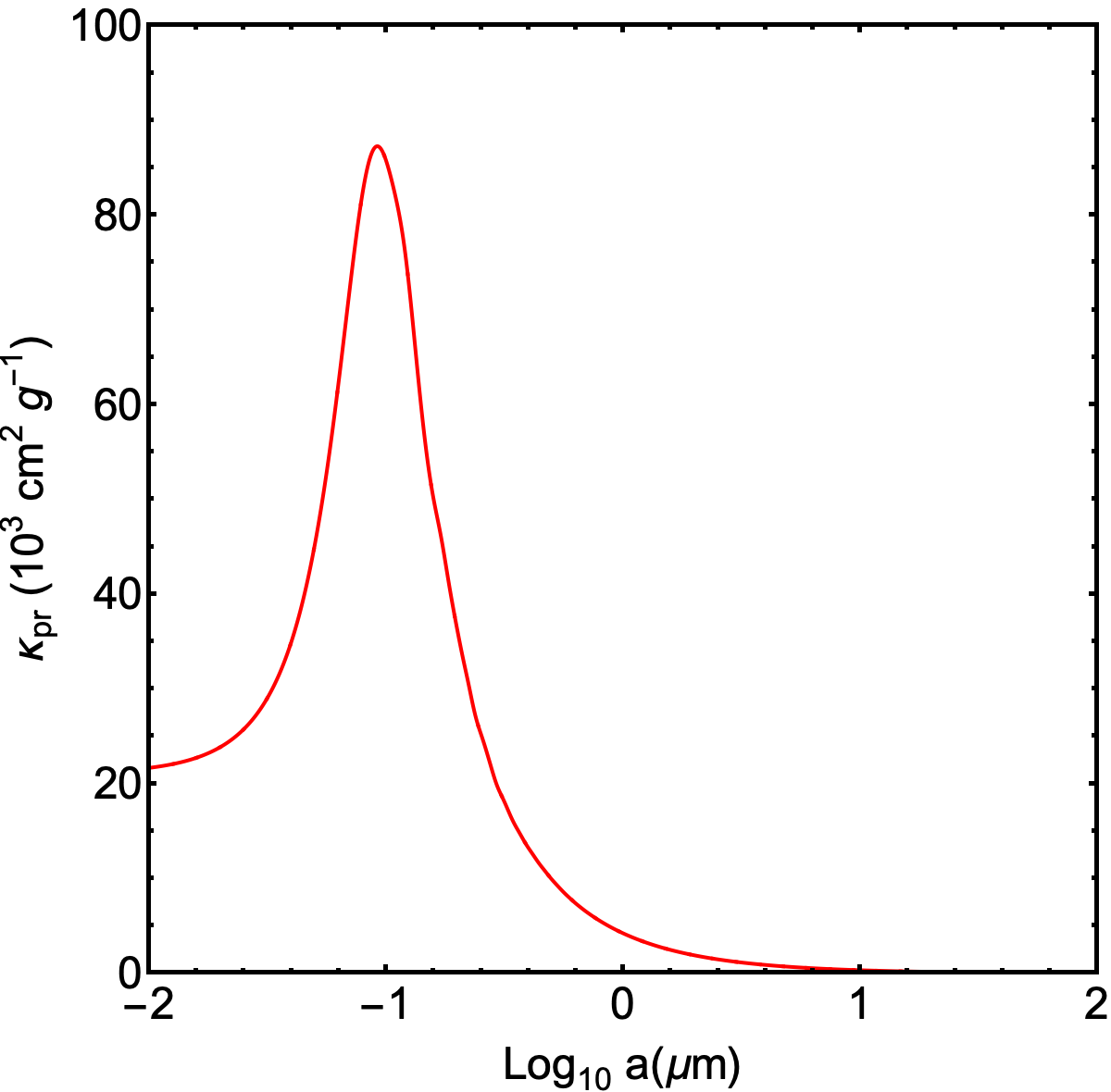}
    \quad\includegraphics[width=5.1cm]{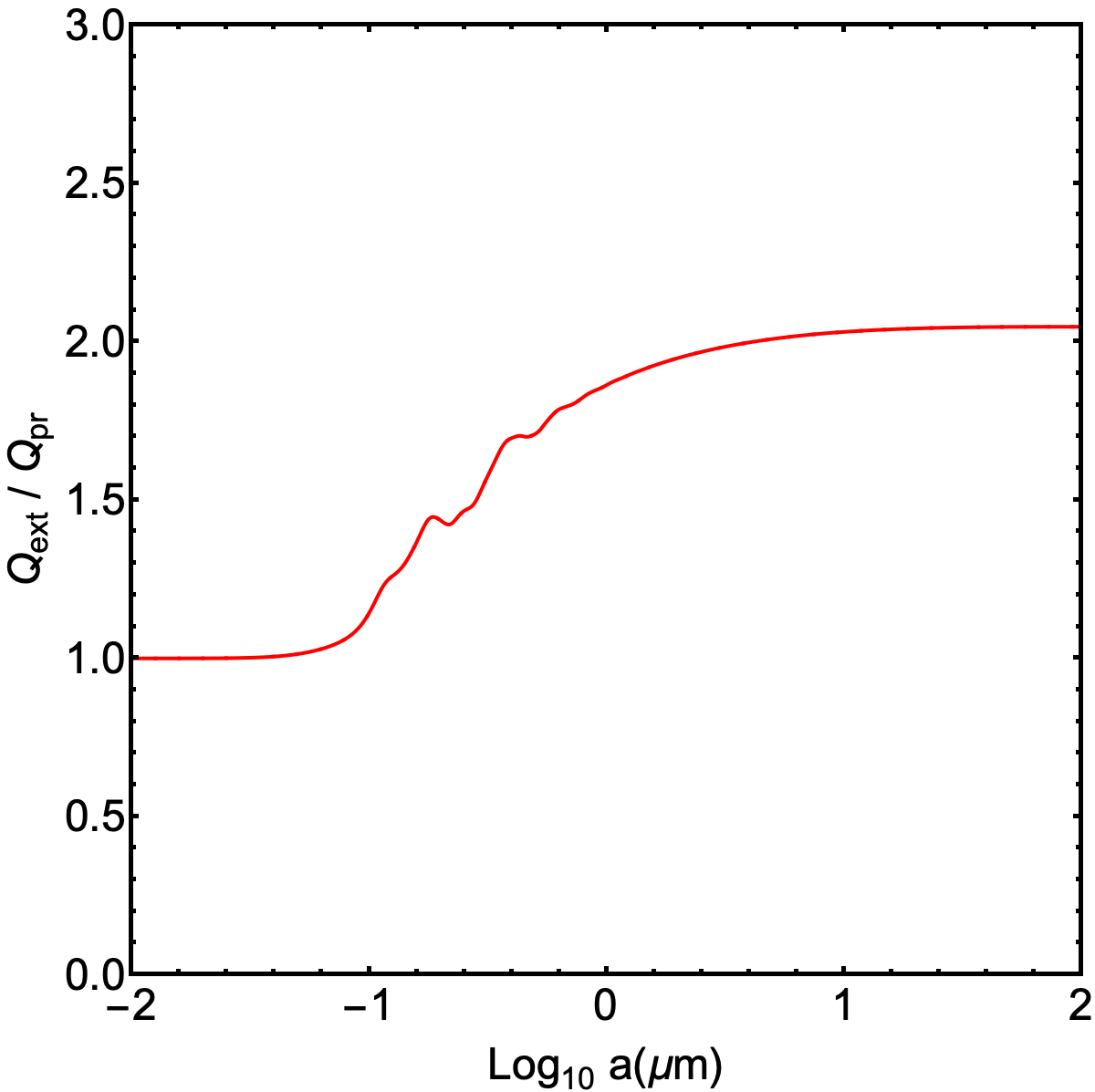}
    \quad\includegraphics[width=5.1cm]{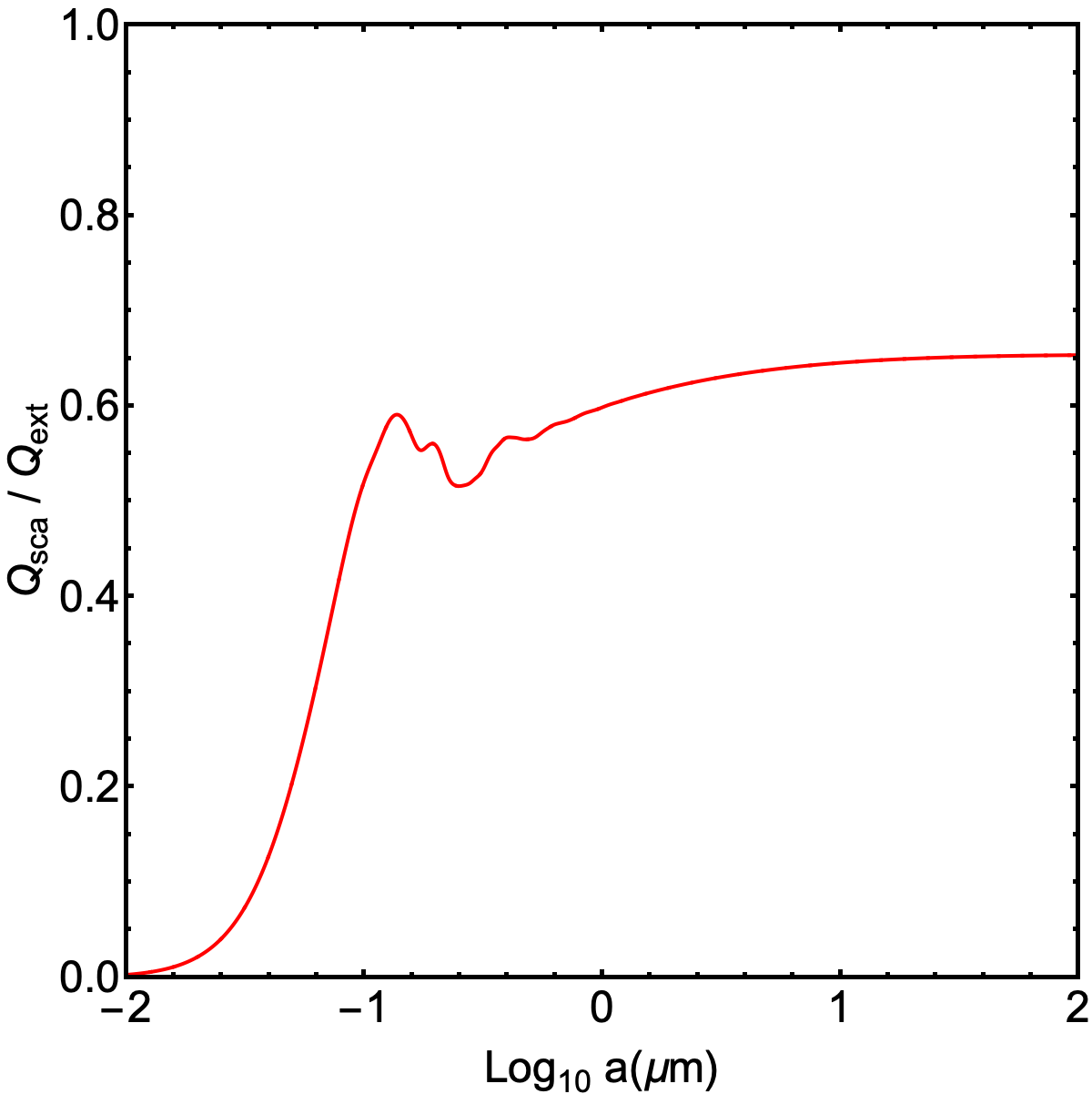}}%
    \medskip   
    \hbox{\qquad\includegraphics[width=5.1cm]{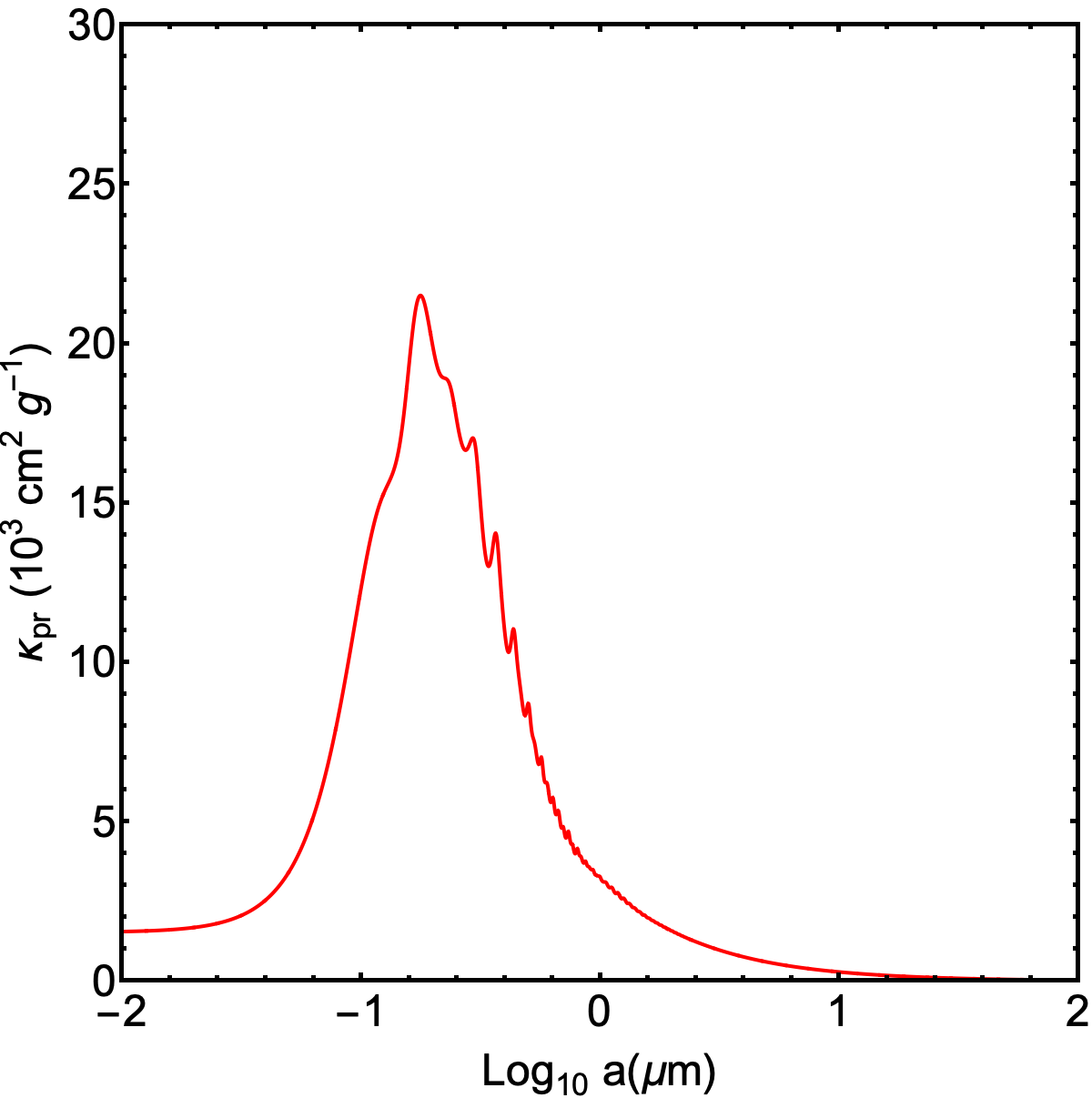}
    \quad\includegraphics[width=5.1cm]{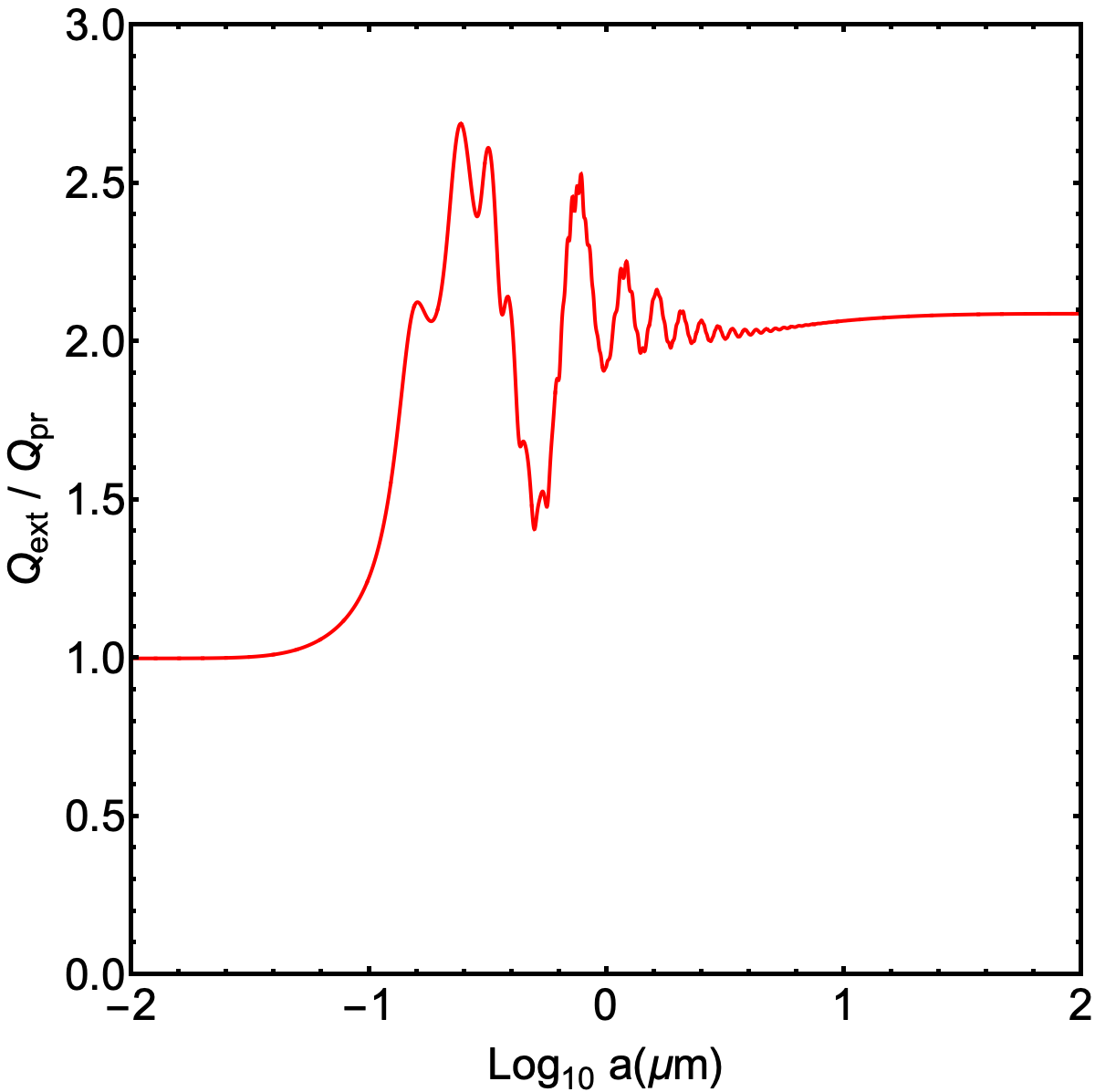}
    \quad\includegraphics[width=5.1cm]{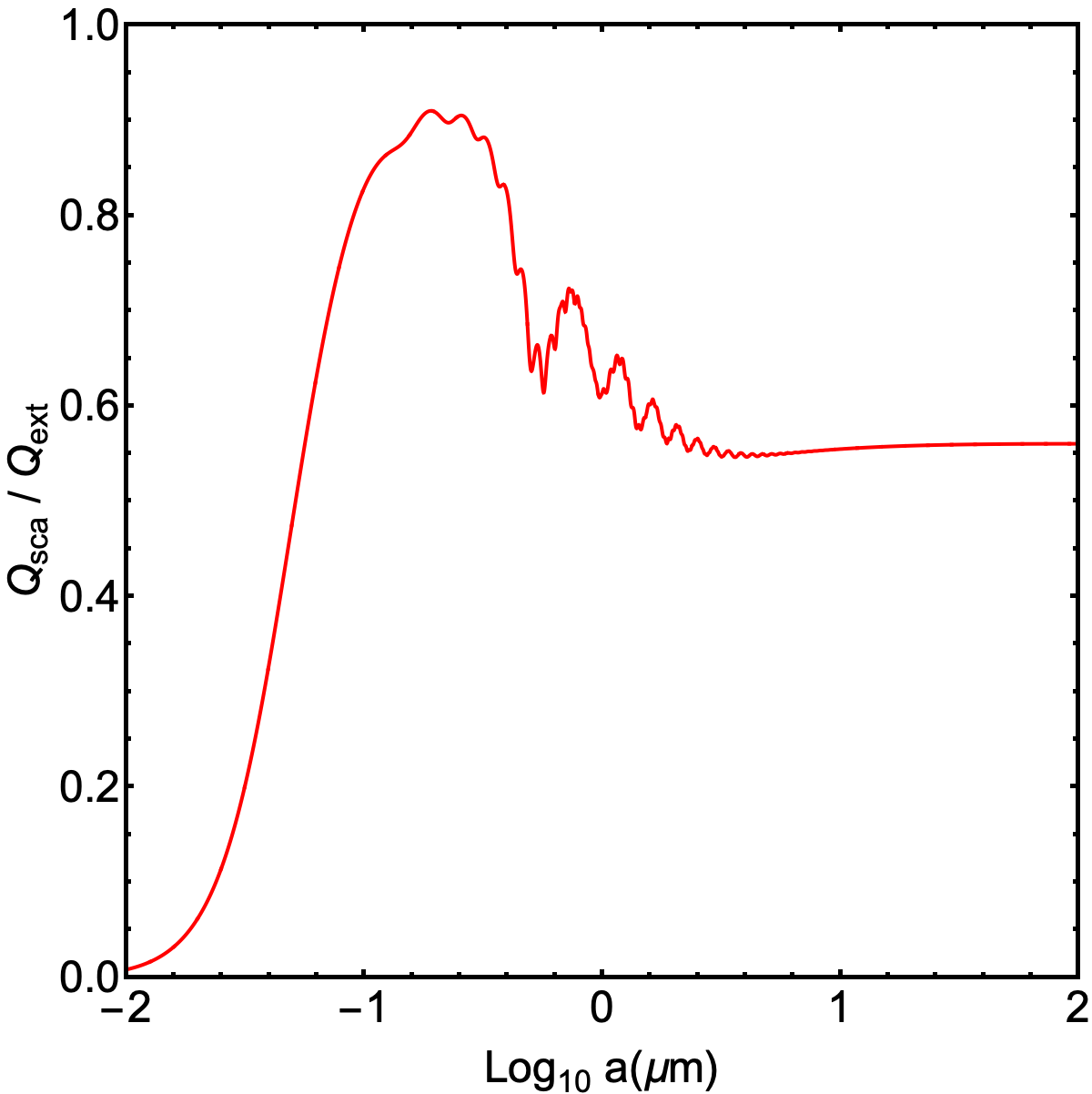}}%
    \caption{Optical properties of spherical dust grains, at $\lambda=600\,{\rm nm}$, as a function of grain radius. The materials are: top row, solid \htwo; middle row, graphite; and, bottom row, silicate. The properties are: left column, opacity to momentum transfer; middle column, the ratio of extinction efficiency to momentum-transfer efficiency; and, right column, scattering albedo.}
    \label{fig:opticalproperties}%
\end{figure*}

We model all dust particles as tiny spheres, because it is straightforward to calculate the optical properties of spheres. We denote the particle radius by $a$, and we quantify the optical properties as a function of particle size for a single wavelength only: $\lambda=600\,{\rm nm}$. That wavelength was chosen as representative of the wavelengths at which most RCBs (with photospheric temperatures in the range $5{,}000\;{\rm K}$ to $8{,}000\;{\rm K}$) put out most of their power, but is otherwise an arbitrary choice. At $\lambda=600\,{\rm nm}$ the refractive indices of our two conventional materials are: \citep{1984ApJ...285...89D} $2.33- 0.21i$ ($2.78 - 1.60i$) for graphite parallel (perpendicular) to the crystal axis, and $1.69-0.03i$ for ``astro-silicate''. For graphite spheres we evaluated the cross-sections for parallel and perpendicular cases separately, and then averaged them with respective weights of $1/3$, $2/3$.

In the optical band the real part of the refractive index of pure solid \htwo\ is close to unity and varying only slowly with wavelength \citep{2015MNRAS.450.1032K}. The value is approximately 1.129 at $\lambda=600\,{\rm nm}$, and larger (smaller) by only $0.004$ ($0.001$) at a wavelength that is $200\,{\rm nm}$ shorter (longer). By contrast, the imaginary part of the refractive index of solid \htwo\ is very small at all frequencies well below the far-UV resonances, because the underlying rovibrational lines of the \htwo\ molecule are quadrupole transitions and thus extremely weak. What absorption there is arises mainly in distinct, highly structured bands. The laboratory data used by \citet{2015MNRAS.450.1032K} were confined to the infrared, longward of $\lambda=1\,{\rm\mu m}$, and thus in the optical band their model refractive index only exhibits absorption due to the wings of the far-UV resonances. However, the strength of the rovibrational absorptions in the optical can be constrained from the measured infrared absorptions, as follows.
\twocolumn

For our purposes the relevant quantity is the average absorption across a broad chunk of the spectrum centred on $\lambda=600\,{\rm nm}$. Now the \citet{2015MNRAS.450.1032K} model refractive index for solid \htwo\ has an average imaginary component of $\sim 8\times 10^{-6}$ in the wavenumber range $4{,}000\le \lambda^{-1}\;{\rm(cm^{-1})}\le 6{,}000$, around the fundamental vibrational transition. In the range $8{,}000\le \lambda^{-1}\;{\rm(cm^{-1})}\le 10{,}000$, around the first vibrational overtone, the average is only $\sim 10^{-7}$. Those averages are both from spectral regions where the absorption is measured, so they are reliable. And the rovibrational contribution is expected to decrease further as one considers bands around progressively higher vibrational overtones --- because of the rapidly decreasing strength of those molecular transitions. We therefore expect that the average value of the imaginary part of the refractive index of solid \htwo\ in the vicinity of $\lambda=600\,{\rm nm}$ is $< 10^{-7}$, and for our purposes that is negligible. Hydrogen dust is thus expected to scatter optical light, but not absorb it, so for typical RCBs there is no strong heating of hydrogen dust or the \htwo\ snow cloud, despite the high luminosity of the star.

The foregoing remarks relate to pure \htwo\ grains. If one imagines solid \htwo\ coating a more conventional dust material -- e.g. a silicate grain -- then the resulting composite grain is expected to absorb strongly. But such grains do not appear to be relevant to the modelling we are undertaking because, over time, refractory grains would sediment out of the atmosphere, resulting in a single, solid lump in the core of the cloud. Rather, the dust near the surface should be predominantly \htwo, which can grow in situ. However, it does seem likely that there would be impurities trapped within the bulk of each hydrogen grain --- as a consequence of the various atoms and small molecules that are presumably present in gas phase in the atmosphere of the cloud. At the temperatures of interest though their saturated vapour pressures are very tiny in comparison with that of \htwo, so the appropriate picture is one of very low concentrations of isolated impurities within the hydrogen matrix. Such impurities may be of interest in themselves \citep[e.g.][]{2013ApJ...768...84B}. But absorption of radiation in their spectral lines is not expected to cause strong heating of the grains. That is because the excited states are typically long-lived, with line-widths even less than those seen in the gas phase --- implying radiative decay of the excited state, rather than a coupling of the energy into phonons. And it is for that reason that solid para-\htwo\ is used for matrix isolation spectroscopy \citep[e.g.][]{2009JChPh.130x4508F}. When a photon is absorbed and then re-emitted the result is in effect equivalent to scattering, so impurities do not have a major impact on the optical properties of the solid \htwo\ grains.

To calculate the optical properties of the grains we employ Mie Theory \citep[e.g.][]{1981lssp.book.....V,1983asls.book.....B}, because it is applicable to all particle sizes and materials of interest here. The characterisations we require are the cross-sections for extinction, for scattering, and for momentum transfer (radiation pressure). Expressed in units of the geometric cross-section these quantities are $Q_{ext}$, $Q_{sca}$ and $Q_{pr}$, respectively. For all three materials, figure \ref{fig:opticalproperties} shows the three optical properties of dust that are actually used in our calculations: (i) the opacity for momentum transfer, $\kappa_{pr}=3 Q_{pr}/(4\,a\,\rho_s)$, where $\rho_s$ is the density of the solid \citep[$0.087\;{\rm g\,cm^{-3}}$ for solid \htwo,][]{1980RvMP...52..393S}; (ii) the ratio $Q_{ext}/Q_{pr}$; and, (iii) the scattering albedo, $Q_{sca}/Q_{ext}$.

For our reference calculation, in which the dust is composed of solid \htwo, we assume that the atmosphere of the cloud contains a broad spectrum of grain sizes. Figure \ref{fig:opticalproperties} then leads us to expect that the wind will be mainly composed of dust in the size range $a\sim 0.1 - 1\;{\rm \mu m}$, where $\kappa_{pr}$ peaks, as these particles experience the highest accelerations. We therefore fix the opacity of our model \htwo\ dust grains at $\kappa_{pr}= 10{,}000\;{\rm cm^2\,g^{-1}}$. However, in the region of the opacity peak the efficiency ratio $Q_{ext}/Q_{pr}$ increases from $\simeq 1$ to $\sim 30$, so depending on the details of the grain spectrum a broad range of extinction properties could apply. We have therefore calculated two separate sets of models with distinct values of the efficiency ratio: one set with $Q_{ext}/Q_{pr} = 2$, appropriate to grains on the small side of the opacity peak, and another with $Q_{ext}/Q_{pr} = 10$ appropriate to large grains. When compared with data on the fading events of RCBs, our models with small grains appear to be a better match than those with large grains; in particular our models with large grains exhibit high levels of scattered light, and therefore yield eclipses that are quite shallow. We therefore present only the small grain models in the main body of this paper; light curves for large \htwo\ grains are given in Appendix B.

\begin{figure}[h]%
    \centering
    {\includegraphics[width=8cm]{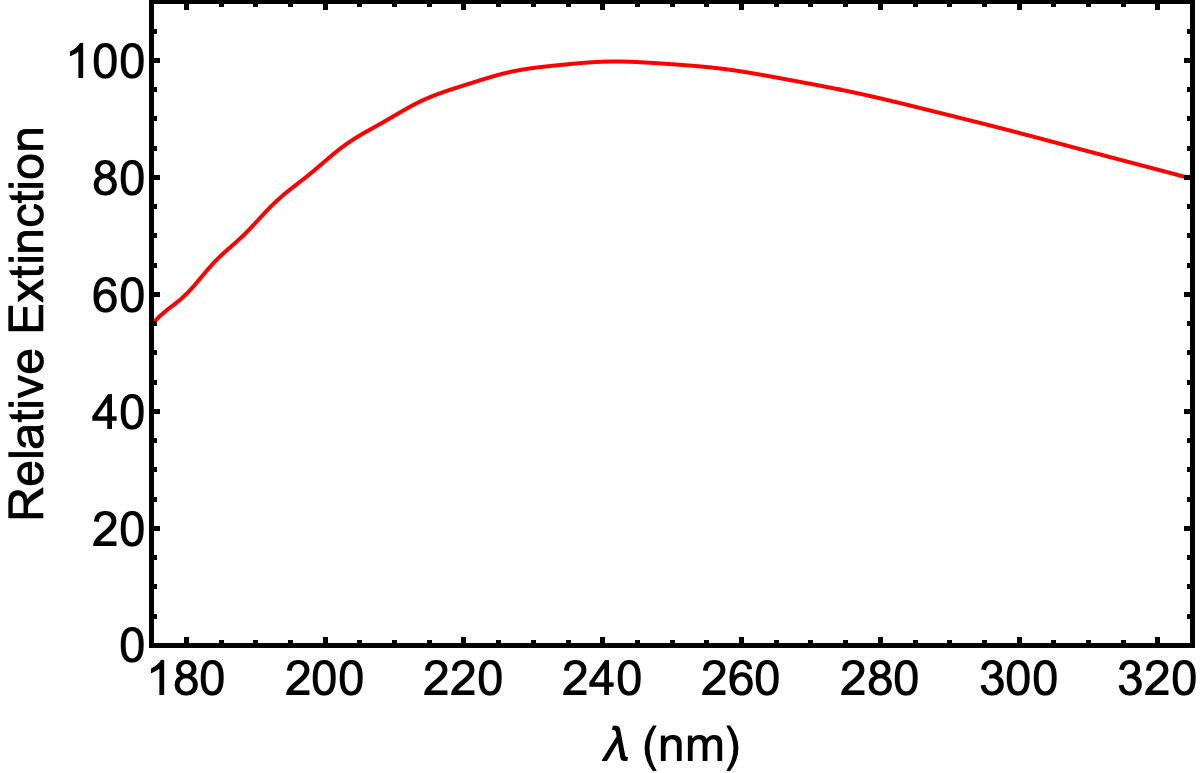}}%
    \caption{The UV extinction curve, i.e. $Q_{ext}$, for solid \htwo\ spheres of radius $a=0.55\;{\rm \mu m}$.}%
    \label{fig:uvextinction}%
\end{figure}

In this paper we will not attempt any detailed comparison between our model and the observed variation of extinction with wavelength for RCBs. The extinction properties of a population of grains depend not only on the material composition of the dust, but also on the distribution of sizes and shapes therein, and without a better understanding of those aspects we cannot hope to arrive at an accurate prediction. It is, however, straightforward to compute extinction curves for grains of the type that we are using --- i.e. spheres of solid \htwo\ within the size range $a\sim 0.1 - 1\;{\rm \mu m}$. At both extremes of that range the calculated extinction curves are unlike the UV extinction curve reported by \citet{1984ApJ...280..228H}. However, less extreme values produce a better match. Figure \ref{fig:uvextinction} shows the UV extinction curve for $a=0.55\;{\rm \mu m}$, which does resemble the data \citep[see figure 1 of][]{1984ApJ...280..228H}. We emphasise that this is simply a calculation of $Q_{ext}$, whereas -- as discussed in \S\ref{sec:scatteredlight} -- our model implies that scattered light is important during declines.

\section{Photo-dissociation of solid \htwo}\label{sec:photodiss}
In this appendix we consider the effect of UV photo-dissociation of solid \htwo, as a result of exposure to radiation from the RCB star at a distance of only a few ${\rm AU}$. Photo-dissociation rates specific to solid \htwo\ have not been measured or calculated, as far as we are aware; here, therefore, we make use of the description given by \citet{1996ApJ...468..269D} for \htwo\ in gas phase. The calculation requires a Doppler parameter, $b$, as input, and we adopt $b=0.3\;{\rm km\,s^{-1}}$ --- corresponding to temperatures of just a few Kelvin.

At low column-densities the dissociation rate per molecule, $\zeta_0$, is given in terms of $\nu u_\nu(\lambda=1{,}000\,{\rm \AA})$ --- the energy-density in the radiation field at a wavelength $\lambda=1{,}000\,{\rm \AA}$, expressed in units of ${\rm erg\,cm^{-3}}$:
\begin{equation}
\zeta_0 \simeq 10^3\, \nu u_\nu(\lambda=1{,}000\,{\rm \AA}) \quad {\rm s^{-1}}.
\end{equation}
However, at depths $\gta 2{\rm \AA}$ in the solid the \htwo\ column is sufficiently large that self-shielding is important. By integrating the shielding function \citep[equation 37 of][]{1996ApJ...468..269D} over depth we find a total rate per unit area of ${\cal F}\simeq 4.1\times 10^{16}\,\zeta_0\;{\rm cm^{-2}\,s^{-1}}$. This dissociation rate must then be integrated up over time as the solid particle moves away from the star.

For a macroscopic lump of solid \htwo\ in the tidal debris stream we proceed by neglecting radiation pressure and approximating the orbit as parabolic. Because the disruption of the cloud is post-periastron, an upper limit on the dissociated column is obtained by integrating from periastron to radial infinity. That limit, ${\cal N}(macro)$, can be simply expressed in terms of the dissociation rate per unit area at periastron ${\cal F}_p$, and the particle speed, ${\cal V}_p$, of a parabolic orbit at periastron:
\begin{equation}
{\cal N}(macro) = \pi\;{\cal F}_p\,\frac{D_p}{{\cal V}_p}.\label{eq:macroscopic}
\end{equation}
The scalings of ${\cal F}_p$ and ${\cal V}_p$ with $D_p$ show that ${\cal N}(macro)\propto 1/\sqrt{D_p}$, and there is thus only a modest variation across the full range of periastron distances considered in \S\ref{sec:modellightcurves}; we adopt a fiducial periastron distance of $D_p=10\,{\rm AU}$.

In the case of dust particles we can undertake an analogous calculation by neglecting gravity, in comparison with radiation pressure, and setting the initial particle speed to zero. That permits a straightforward radial integration, using eq. \ref{eq:velocities}, with the result
\begin{equation}
{\cal N}(micro) =2\;{\cal F}_p\,\frac{D_p}{U}.\label{eq:microscopic}
\end{equation}
In the intense radiation field of an RCB star the characteristic speed of dust particles, $U$, is much higher than Keplerian speeds at the same distance from the star, and thus the estimated fluence is much smaller. As gravitational and radiation forces both vary as $1 / r^2$, the variation of ${\cal N}$ with $D_p$ is the same in both cases.

If we characterise the stellar photosphere as a black-body emitter at $T_*=6{,}750\,{\rm K}$ we arrive at ${\cal N}(macro)\simeq 6.6\times 10^{20}\,{\rm cm^{-2}}$, corresponding to a depth of roughly $250\;{\rm \mu m}$ in the solid. That estimate is valid for lumps that are ${\rm cm}$-sized or larger, and is therefore always a tiny fraction of the total amount of solid material. For dust, on the other hand, the estimate is ${\cal N}(micro)\simeq 4\times 10^{18}\,{\rm cm^{-2}}$ (a depth of approximately $1.5\;{\rm \mu m}$) at $\kappa_{pr}=10^4\;{\rm cm^2\,g^{-1}}$, and scaling as $1/\sqrt{\kappa_{pr}}$. With reference to figure \ref{fig:opticalproperties} we see that the implication of this is that the photo-dissociated column is comparable to the total column of an individual grain for particles that are close to the opacity peak. Smaller grains are completely destroyed, whereas larger grains experience progressively smaller fractional loss (scaling roughly as $1/\sqrt{a}$).

The photo-dissociation rates for all particle sizes increase greatly in the case where the star is one of the few RCBs with a hot photosphere --- in direct proportion to the far-UV intensity. If we again assume a black-body spectrum, with photospheric temperatures now in the range $T_*=15{,}000-20{,}000\;{\rm K}$, the expected far-UV intensity is some five-to-six orders of magnitude larger than the more typical $T_*=6{,}750\,{\rm K}$ case described above. That is such a large factor that, in the context of the model we have presented, it makes hot RCBs far more interesting than warm ones for atomic hydrogen searches. 

A black-body is a simple and convenient model to use. However, we expect the true far-UV photospheric intensities to be far below that of the corresponding black-bodies, because the continuous opacity in RCB atmospheres is dominated by neutral carbon, and that atom has strong absorption edges just longward of the far-UV transitions of \htwo.\footnote{The relevant edges are at $\lambda\simeq 1{,}101\,{\rm \AA}$ (from the carbon ground state), and at $\lambda\simeq 1{,}240\,{\rm \AA}$ (from a low-lying metastable state) \citep{2011piim.book.....D} --- the latter being on the long wavelength side of {\it all\/} the far-UV absorptions of ground state \htwo.} A detailed stellar atmosphere model is required to obtain an accurate value of the far-UV intensity; estimates based on black-body spectra serve as an upper limit.

\section{light curves for large \htwo\ grains}
If the dust particles are large spheres of solid \htwo\ they are strongly forward scattering, and for this circumstance we adopt $Q_{ext}/Q_{pr}= 10$. The Monte Carlo simulations of photon transport -- used to determine the wind launch conditions (\S\ref{sec:flowfromsurface}) and the intensity of light scattered by the wind (\S\ref{sec:scatteredlight}) -- need the single-scattering angular distribution to be specified. We have employed the distribution given by \cite{1981lssp.book.....V} (his section 12.43), which for solid \htwo\ at $\lambda=600\;{\rm nm}$ (refractive index $1.129$) has the analytic form
\begin{equation}
I(\chi) \propto \left(4\times0.129^2 + \chi^2\right)^{-2}\; ,
\end{equation}
for scattering angle $\chi$.

\begin{figure}[h!]%
    \centering
    {\includegraphics[width=8cm]{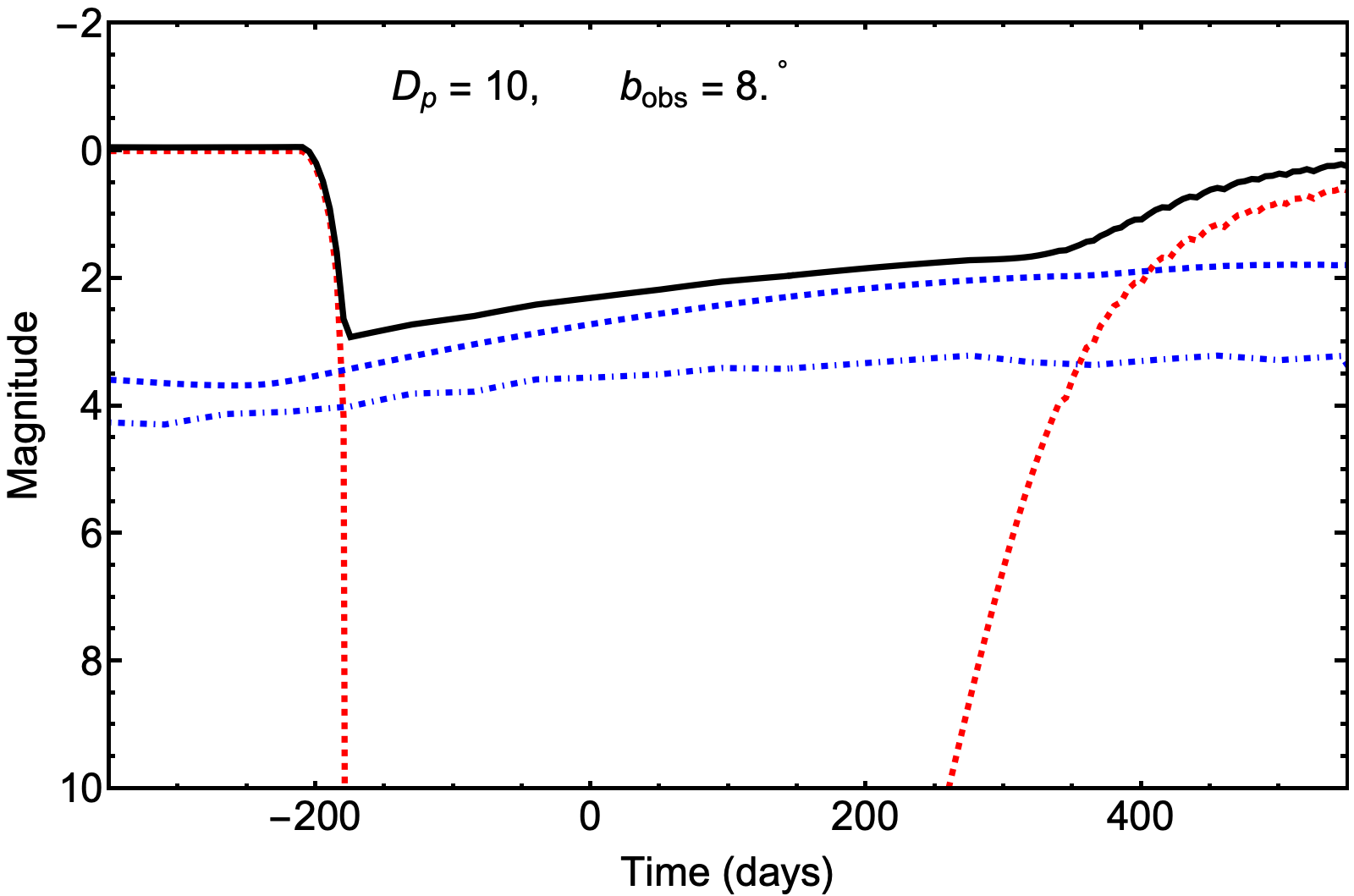}}%
    \caption{Light curve for an eclipse by a cloud on the same orbit as in figure \ref{fig:medianeclipse}, but calculated for large \htwo\ dust grains.}%
    \label{fig:largegrainmedianeclipse}%
\end{figure}

\begin{figure*}[t!]%
    \centering
    \hbox{\qquad{\includegraphics[width=8cm]{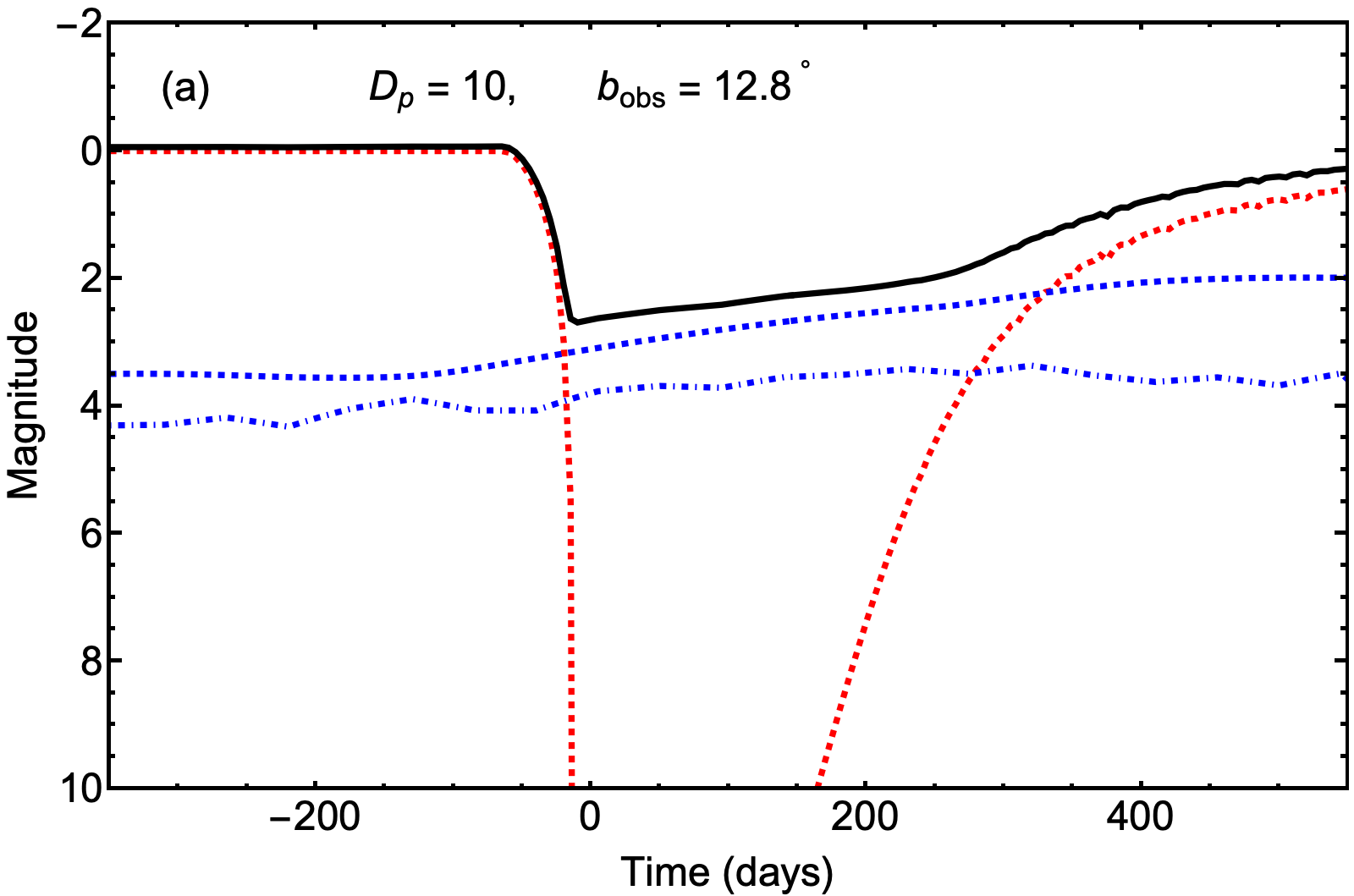}}\qquad{\includegraphics[width=8cm]{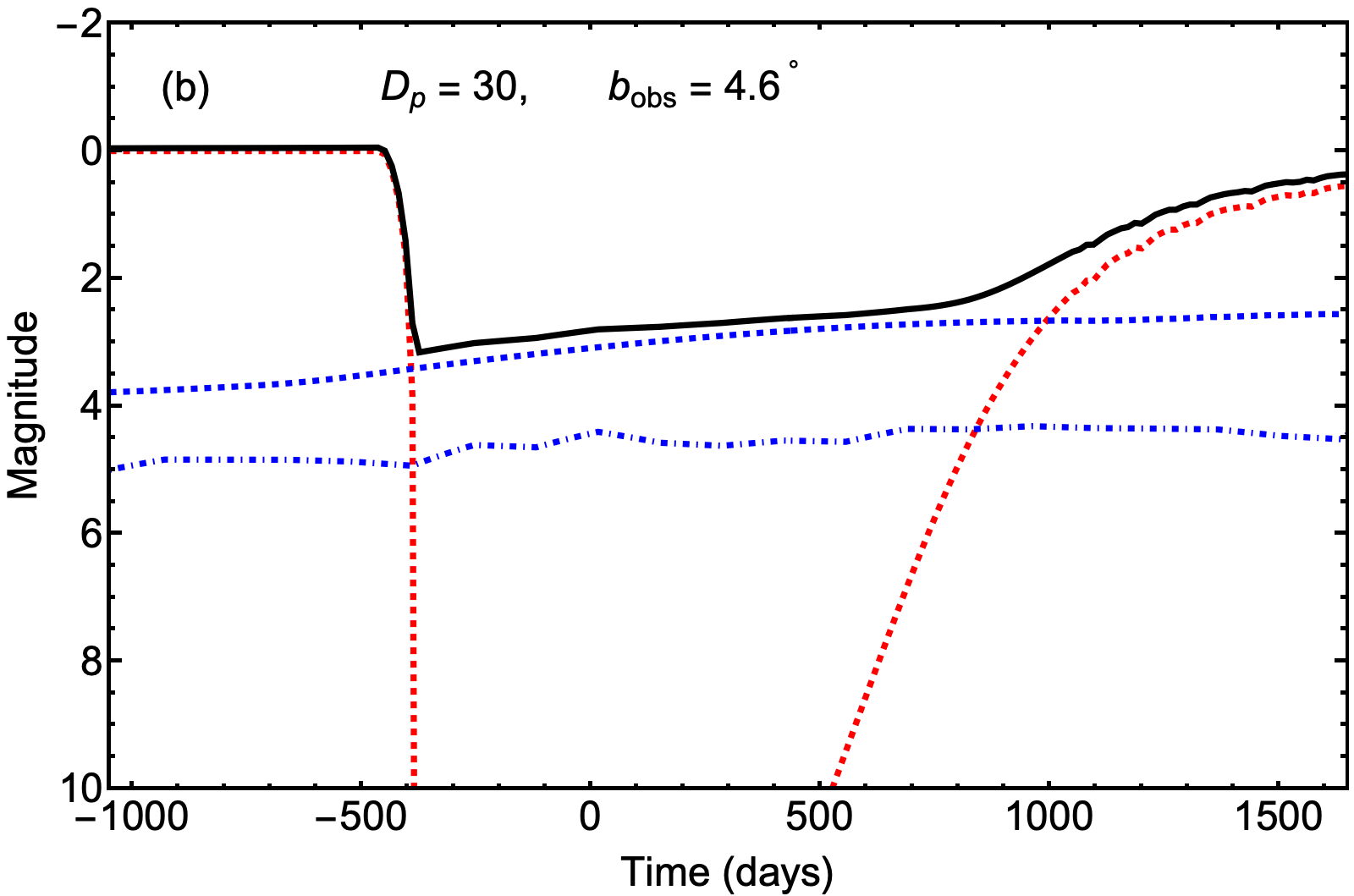}}}%
    \medskip
    \hbox{\qquad{\includegraphics[width=8cm]{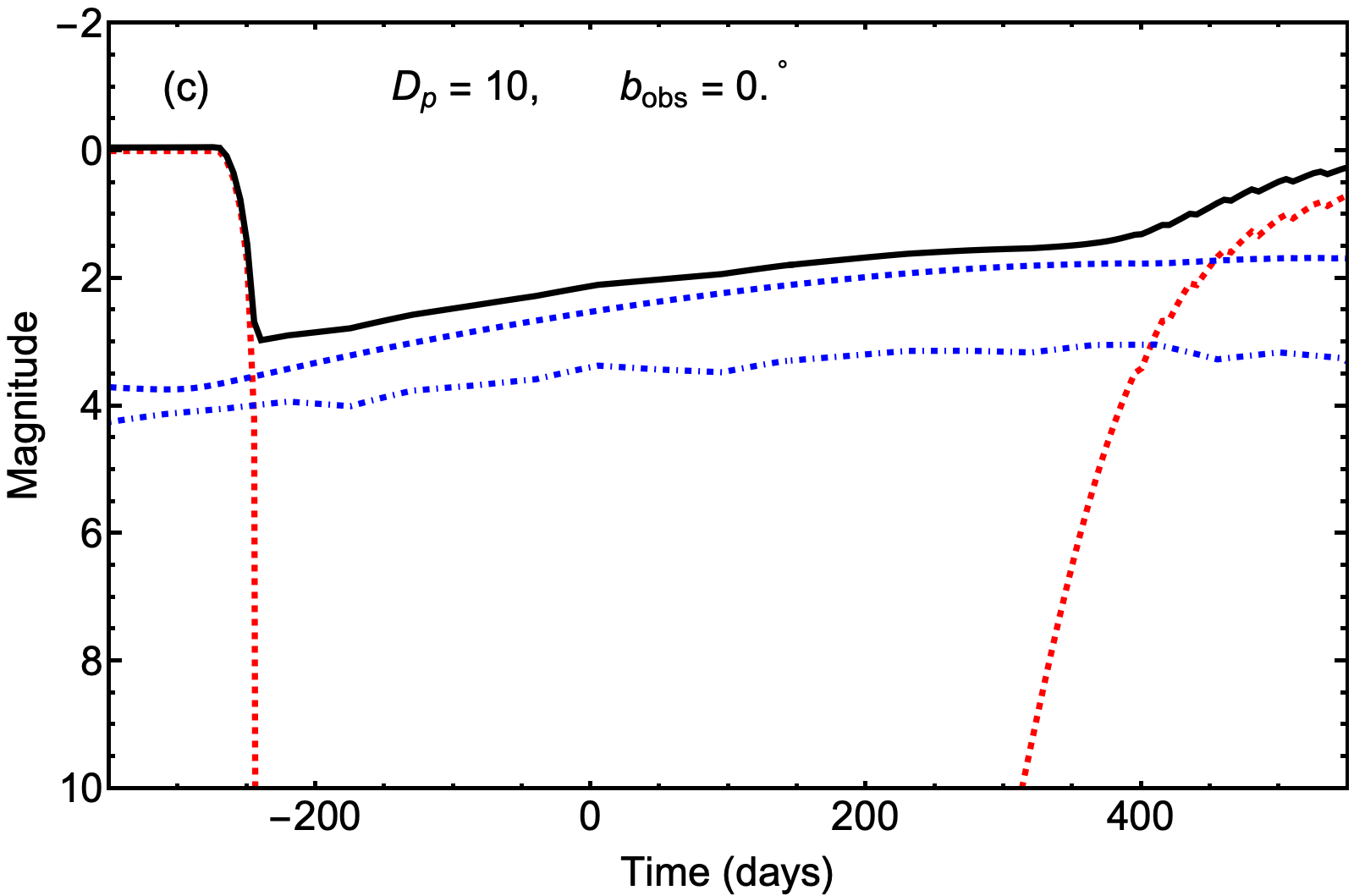}}\qquad{\includegraphics[width=8cm]{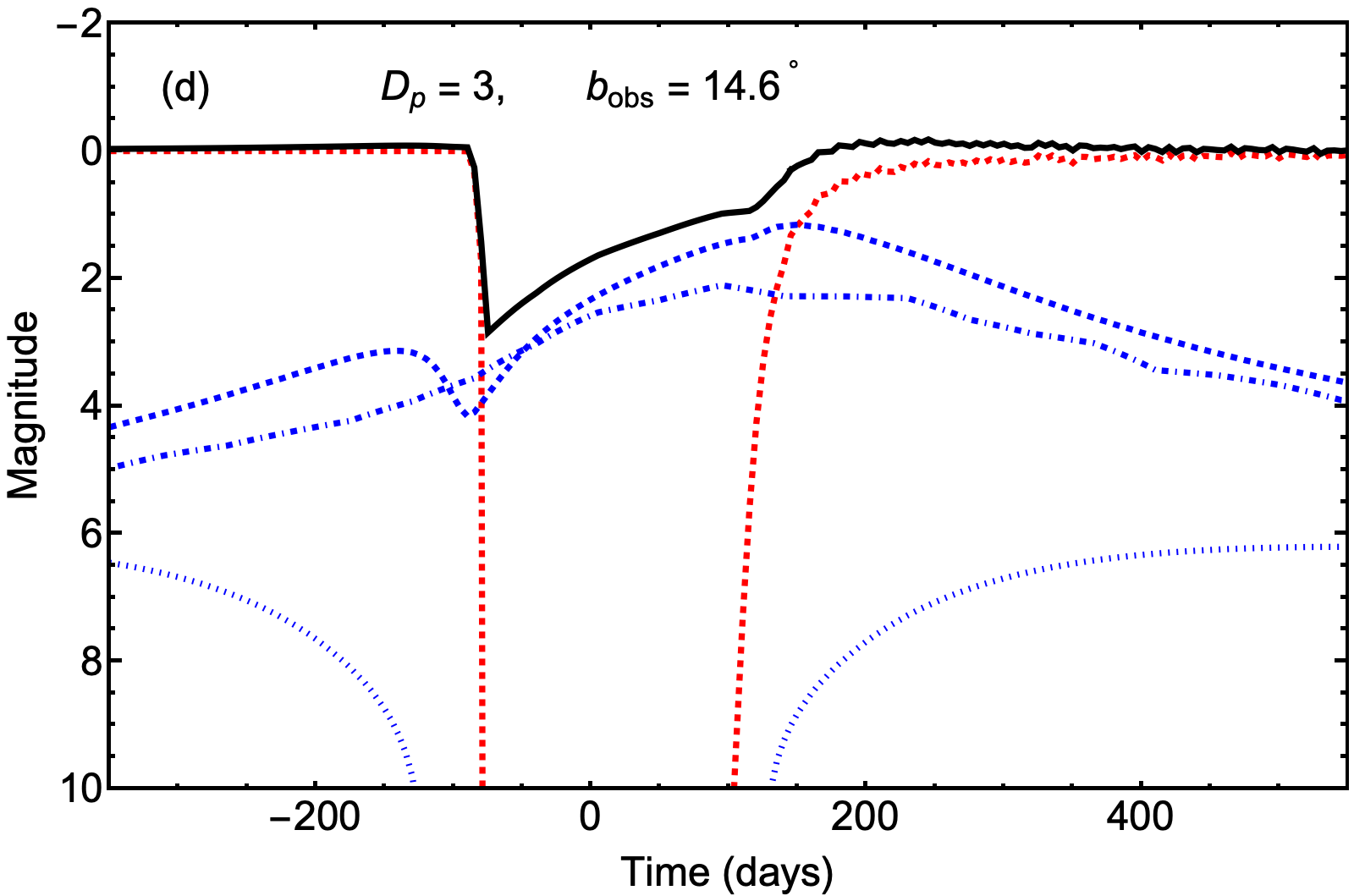}}}%
    \caption{Light curves for eclipses by clouds on the same orbits as in figure \ref{fig:distancelatitudevariation}, but here calculated for large \htwo\ dust grains. Note the increased temporal coverage in panel (b).}
    \label{fig:largegraindistancelatitudevariation}%
\end{figure*}

As noted earlier (\S\ref{sec:flowfromsurface}), and evident in figure \ref{fig:surfaceforces}, the launch conditions for large \htwo\ grains prove to be similar to those for small \htwo\ grains, so in the present calculations we have employed the same $\psi_\circ$ (and $\kappa_{pr}$), and the wind structure is unchanged relative to that described in \S\S\ref{sec:flowfromsurface},\ref{sec:orbit}. Our various large-grain models are computed for the same parameter combinations as shown in figures \ref{fig:medianeclipse} and \ref{fig:distancelatitudevariation}; and, in common with those light curves, these calculations assume spherical clouds (i.e. tidal deformations are neglected). Figures \ref{fig:largegrainmedianeclipse} and \ref{fig:largegraindistancelatitudevariation} display much higher levels of extinction than their small-grain counterparts, shown earlier, but they also have higher levels of scattered light and the resulting eclipses are shallower (see the summaries in table \ref{tab:largegrainmetrics}).

\begin{table}[h!]
\centering
\begin{tabular}{|c|c||c|c|c|c|c|}
\hline
$D_p$ & $b_{obs}$ & Depth & Onset & Duration & Centroid \\
 & & & Time & & Shift \\
$({\rm AU})$ & $({}^\circ)$ & (mag)&(days) & (years) & (AU)\\
\hline
\hline
3 & 14.6 & 2.9 & 13 & 0.5 & 26\\
\hline
10 & 0 & 3.0 & 25 & 1.9 & 53\\ 
\hline
10 & 8.0 & 2.8 & 27 & 1.6 & 53\\ 
\hline
10 & 12.8 & 2.7 & 41 & 1.1 & 51\\
\hline
30 & 4.6 & 3.1 & 67 & 4.6 & 86\\ 
\hline
\end{tabular}
\caption{Properties of the light curves in figures \ref{fig:largegrainmedianeclipse} and \ref{fig:largegraindistancelatitudevariation}. This  parallels table \ref{tab:lightcurvemetrics}, but is for models with large \htwo\ grains.}
\label{tab:largegrainmetrics}
\end{table}

\FloatBarrier

\section{The influence of tides}\label{sec:tidal}
\begin{figure}[!h]%
    \centering
    {\includegraphics[width=8cm]{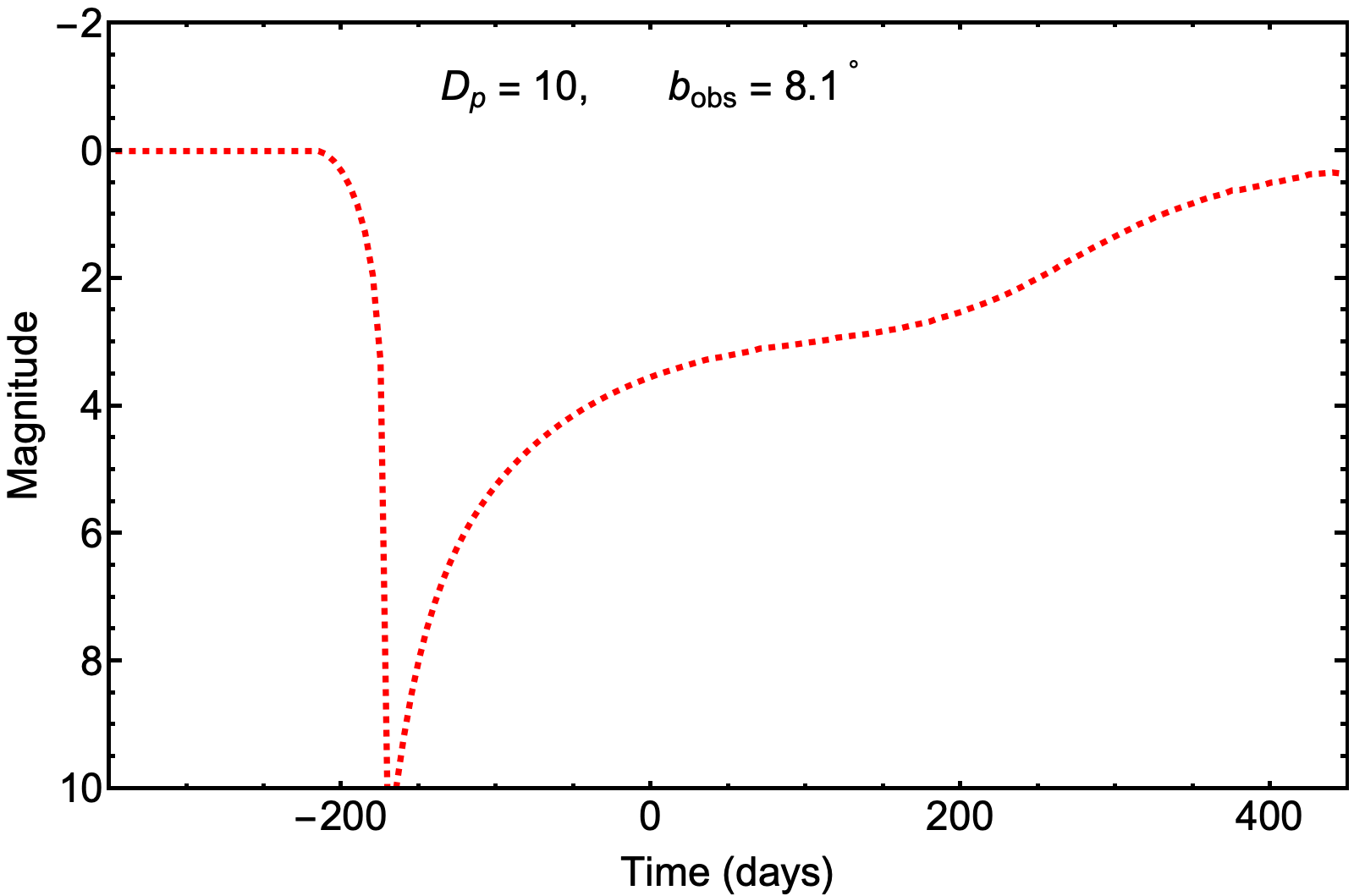}}%
    \caption{Extinction light curve for an eclipse by a cloud on the same orbit as in figure \ref{fig:medianeclipse}, but here including the effects of the tidal deformation of the cloud by the star. This calculation does not include scattered light and should therefore be compared to the red, dashed curve in figure \ref{fig:medianeclipse}.}%
    \label{fig:deformedmedianeclipse}%
\end{figure}

\begin{figure*}[!ht]%
    \centering
    \hbox{\qquad{\includegraphics[width=8cm]{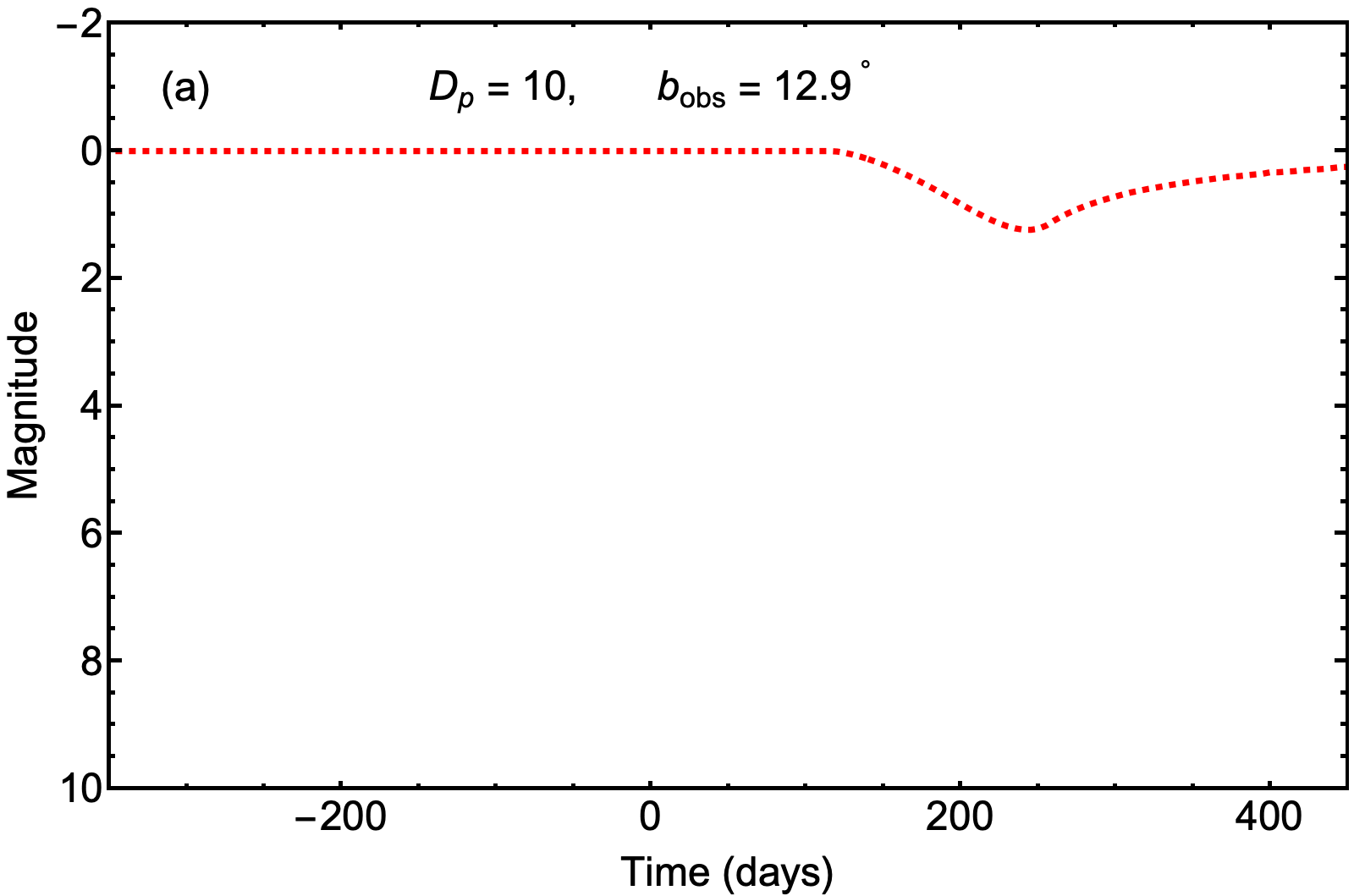}}\qquad{\includegraphics[width=8cm]{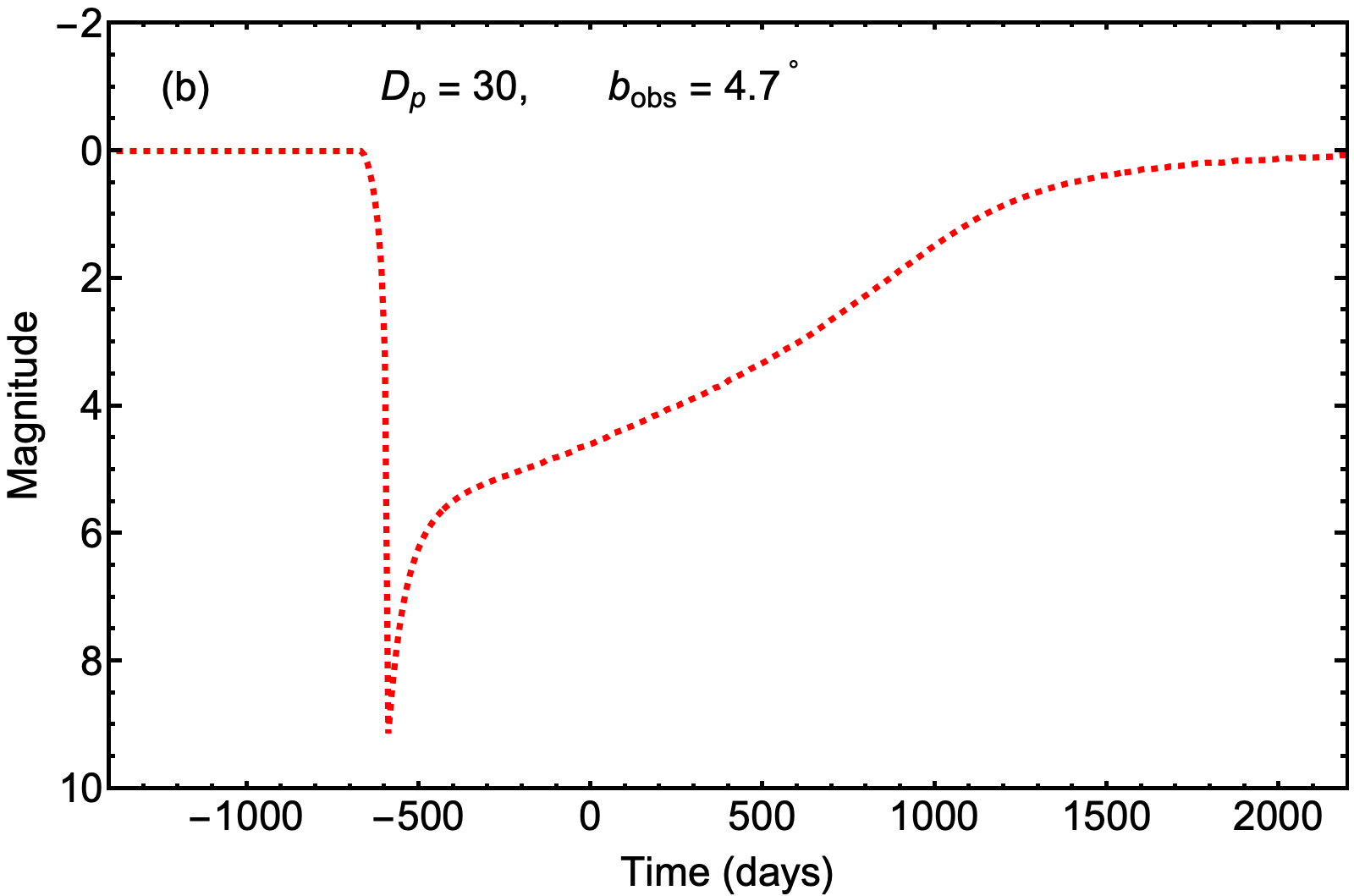}}}%
    \medskip
    \hbox{\qquad{\includegraphics[width=8cm]{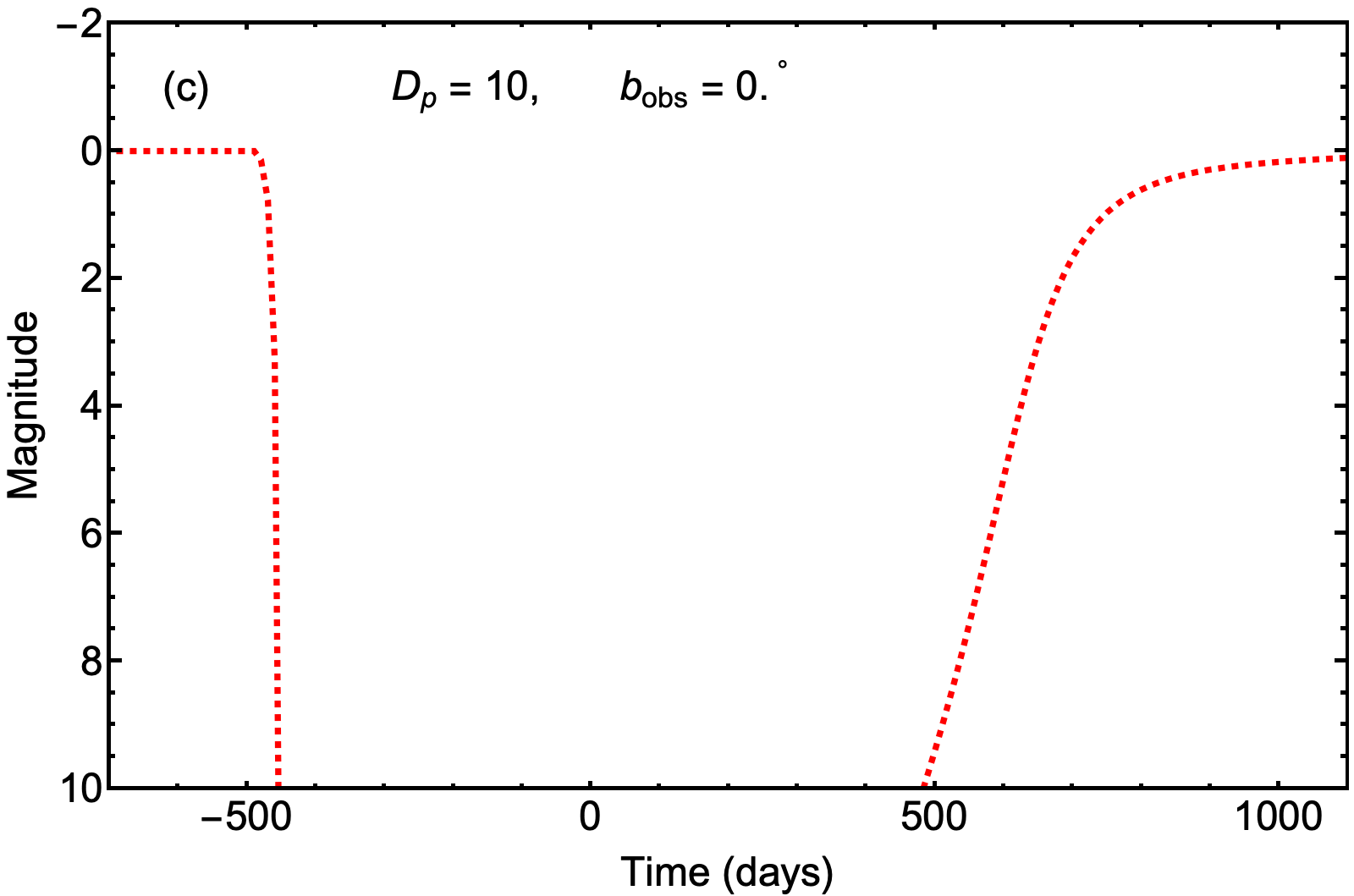}}\qquad{\includegraphics[width=8cm]{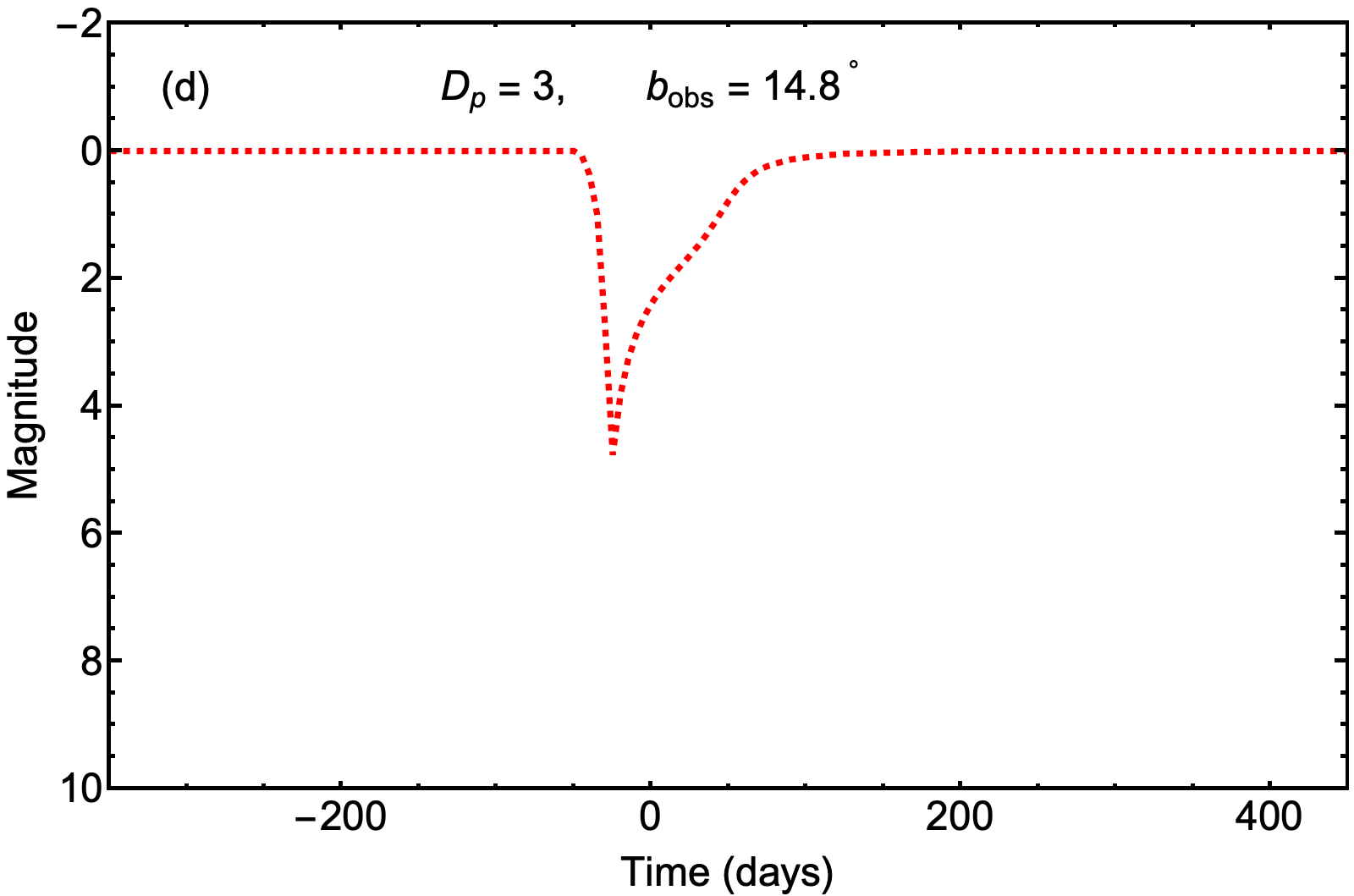}}}%
    \caption{Extinction light curves for eclipses by clouds on the same orbits as in figure \ref{fig:distancelatitudevariation}, but here including the effects of the tidal deformation of the cloud by the star. These calculations do not include scattered light and should therefore be compared to the red, dashed curves in figure \ref{fig:distancelatitudevariation}; but note the differences in temporal coverage.}
    \label{fig:deformeddistancelatitudevariation}%
\end{figure*}

Up to this point all light curves have been computed under the assumption of a spherical cloud. But in fact each cloud is distorted by the tidal gravitational field of the star, and that distortion modifies the initial (launch) conditions of the wind. Tides add significant additional complexity to the modelling, because they may cause the shape and size of the cloud to evolve strongly through the course of the orbit, and the tidal response depends on the internal structure of the cloud as well as its mass and radius, and of course the orbit itself. But tides are largely familiar, and they are not our principal concern here, so rather than exploring those various effects in detail we confine ourselves to illustrations of how tidal deformation modifies the light curves presented in figures \ref{fig:medianeclipse} and \ref{fig:distancelatitudevariation}. Only extinction light curves have been computed for the deformed clouds; they are presented as figures \ref{fig:deformedmedianeclipse} and \ref{fig:deformeddistancelatitudevariation}, and those curves can be compared to their counterparts, the red, dashed curves in figures \ref{fig:medianeclipse} and \ref{fig:distancelatitudevariation}.

The calculations presented here follow our previous work \citep{suvorovwalker2025} in utilising the restricted hydrodynamic scheme of \citet{cart85}, implemented by us in {\it Python\/}. We also use the same ($n=3/2$ polytrope) structural model as before. In common with our light curves computed for spherical clouds, the wind is launched from the limb of the deformed cloud and thus the cross-section of the dust sheet is initially just a reflection of the shape of the limb. That initial condition is evident in the light curves in figures \ref{fig:deformedmedianeclipse} and \ref{fig:deformeddistancelatitudevariation} in two respects. First, in the plane of the orbit the cloud is stretched in one direction, so observers at low latitudes see an earlier eclipse onset and the extinction remains high for a longer time --- compare the lower-left panels in figures \ref{fig:distancelatitudevariation} and \ref{fig:deformeddistancelatitudevariation}, and note that the latter covers a $2\times$ longer interval. Secondly, perpendicular to the orbital plane the cloud is initially compressed by the tidal forces, and thus the line-of-sight for observers at ``typical'' latitudes never gets as close to the cloud surface as it does in the case of a spherical cloud. That difference is evident when comparing the lower-right panels in figures \ref{fig:distancelatitudevariation} and \ref{fig:deformeddistancelatitudevariation}, where the former displays much higher levels of extinction around the time of conjunction ($t=0$).

Of all the light curves shown in figures \ref{fig:deformedmedianeclipse} and \ref{fig:deformeddistancelatitudevariation}, it is the top-right panel (b) in figure \ref{fig:deformeddistancelatitudevariation} that most closely resembles its counterpart for the case of a spherical cloud. That is as expected, because tidal forces decline rapidly with distance and that example is the case with the largest periastron distance.

All light curves in this manuscript are for observers located at the same longitude as periastron. But the nature of the influence of the tides can be anticipated for other longitudes, as follows. Observers located at longitudes that are pre-periastron would see the same pattern of tidal effects as described above; but the deformation, and thus the magnitude of those effects would be smaller. However, for observers who are located post-periastron the situation is different: out-of-plane tidal compression may create a ``pancake'', leading to a rebound in which the out-of-plane dimension can eventually become much larger than the initial cloud radius. Thus, in cases where the conjunction occurs long after periastron, the character of the eclipse may be very different from that of our simplified model, with the tidal debris itself -- not just the dust wind -- being a major influence on the structure of the observed light curve.

A caveat must accompany our wind models for the tidally deformed clouds: the radiation forces at the surface of the cloud change in response to the deformation, so in principle one should simulate the Monte Carlo transport (figure \ref{fig:forcedensity}) for each cloud shape as the deformation proceeds. That is computationally prohibitive, so we have simply assumed that the forces, and thus the launch velocities, are similar to those for a spherical cloud.

\section{A disruption shower scenario}\label{sec:eventshower}
The model eclipse light-curves shown in \S 5 of this paper can, on an individual basis, plausibly account for the fading events that are observed in RCB stars. However, the rate at which such fading events occur is observed to be $\sim 0.3\;{\rm yr^{-1}}$ in active RCB stars \citep{2025MNRAS.537.2635C}, and that raises the question of how a swarm of hydrogen snow clouds could populate disrupting orbits around a WD? That is a substantial topic in its own right, and it is not possible to give a comprehensive discussion here; in this appendix we present just one idea.

\begin{figure*}[ht]%
    \centering
    {\includegraphics[width=8cm]{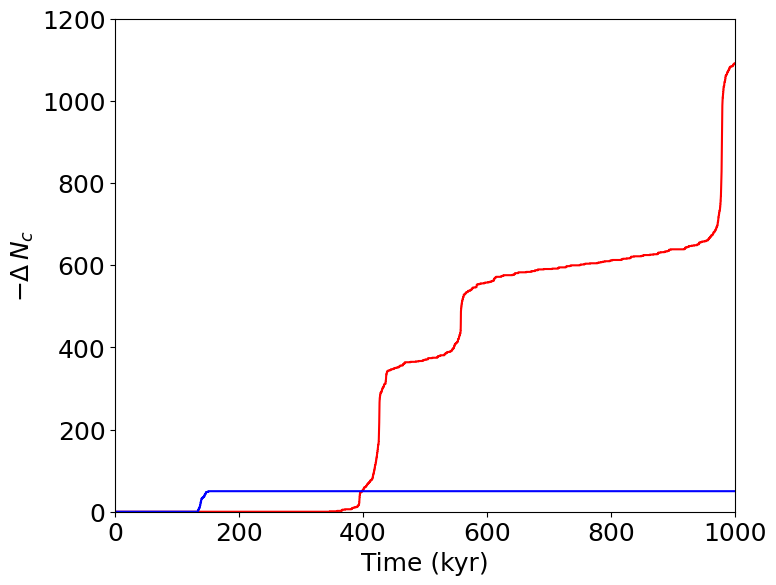}\hskip 1cm\includegraphics[width=9cm]{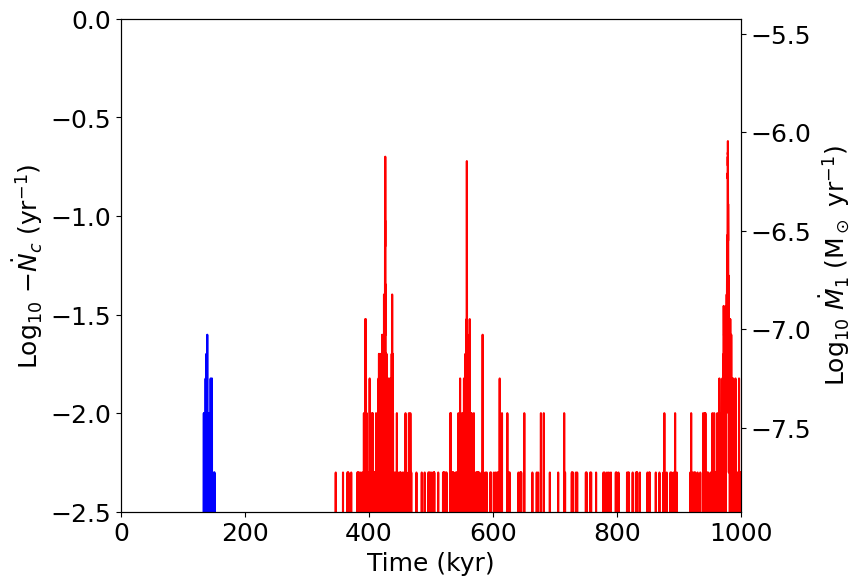}}%
    \caption{Snow cloud disruptions caused by the WD (red) and MS star (blue) in the $N$-body simulation described in the text. The left-hand panel shows the cumulative number of clouds disrupted as the simulation progresses. The right-hand panel shows the rate of disruptions, averaged over a $200\,{\rm yr}$ window, and the corresponding accretion rate in the case of the WD.}%
    \label{fig:disruptions}%
\end{figure*}

In \S\ref{sec:discussion} we recalled the large number of low-mass clumps of molecular \htwo\ seen in the Helix Nebula (and other planetaries), and noted the possibility that those structures could be a primordial population of \htwo\ snow clouds. Building on that observation, our suggested scenario starts from a wide binary of intermediate-mass main-sequence stars inside such a halo. The primary evolves more rapidly than the secondary, becoming an asymptotic giant branch (AGB) star and finally shedding its envelope in a ``superwind'' over the course of a few thousand years \citep{1983ARA&A..21..271I}. Providing the envelope is ejected in a time that is less than the orbital period, and most of the mass of the binary is shed, the binary dissolves and a CO-WD is born on an orbit that takes it through the snow-cloud halo --- thus giving rise to disruptions, accretion and eclipses. In this picture all the clouds are initially at large distances from the binary star, so after the binary has dissolved any clouds that have velocities close to that of the WD will plunge towards it on near-parabolic orbits.

As a specific example we take a main-sequence binary with a $5.56\;{\rm M_\odot}$ primary -- so that normal stellar evolution yields a $1\;{\rm M_\odot}$ CO-WD \citep{2008ApJ...676..594K} -- and a $1\;{\rm M_\odot}$ secondary in a circular orbit at a separation of $650\;{\rm AU}$. We have used the $N$-body code {\tt pytreegrav\/} \citep{2021JOSS....6.3675G} to compute cloud and star orbits and the resulting disruption rates for a model halo of clouds around the binary.  Initial cloud positions are selected randomly, under the assumption of a uniform average density for the halo, and their initial velocities are specified as for circular, Keplerian orbits in the same plane as the binary orbit and with the same sense of circulation. A planar model halo might seem surprising; our motivation for the choice is as follows. Individual clouds have a low mean density and tides are correspondingly important: the relative motions of the clouds will be damped, and over time that drives the halo towards a minimum energy state. In practice gravitational interactions with passing stars would perturb the orbits, and there would be a competition between gravitational kicks and tidal dissipation. We also note that clustering would likely be present in the distribution of clouds in the halo, though none is included in our model. Our model is a limiting case, chosen for its simplicity.

Our model halo notionally contains $10^5$ clouds, each of mass $3\times 10^{-5}\;{\rm M_\odot}$, distributed between $30{,}000\;{\rm AU}$ and $40{,}000\;{\rm AU}$ from the binary. But to save on computational resources we took the following steps. We first simulated a halo containing $10^4$ clouds each of mass $3\times 10^{-4}\;{\rm M_\odot}$, which allowed us to identify the angular sectors where disruptions occur. Then we placed ($35{,}701$) clouds of mass $3\times 10^{-5}\;{\rm M_\odot}$ within those sectors, and ($6{,}430$) clouds of mass $3\times 10^{-4}\;{\rm M_\odot}$ elsewhere.

Shortly after the start of the $N$-body calculation, mass-loss is initiated from the  primary, simulating the ejection of the star's envelope at an assumed rate of $10^{-3}\;{\rm M_\odot\;yr^{-1}}$. The wind speed is assumed to be $\gta 10\;{\rm km\;s^{-1}}$, so that the envelope mass moves beyond the limits of the halo on a timescale that is short compared to any of the clouds' orbital periods. Consequently we remove mass from the simulation as soon as it is shed by the primary.

The {\tt pytreegrav\/} code tracks pointwise gravitational accelerations, but not tidal deformations of the particles themselves. We therefore use a prescription for determining which clouds are disrupted: any cloud that approaches to within the tidal disruption radius of either star is assumed to be disrupted by that star. The disrupted cloud is then removed from the simulation, and one eighth of its mass (i.e. half of the helium content) is added to the disrupting star. We adopted a tidal disruption radius of $30\;{\rm AU}$ --- appropriate to an $n=3/2$ polytropic cloud of mass $3\times 10^{-5}\;{\rm M_\odot}$ and radius $0.78\;{\rm AU}$, and a $1\;{\rm M_\odot}$ star \citep{suvorovwalker2025}.

Our results are displayed in figure \ref{fig:disruptions}. There we see a single, prompt shower of disruptions caused by the MS star, followed by a series of three more intense showers on the WD. At peak, the latter showers show disruption rates comparable to the observed rate of fading events in active RCB stars, with 37, 27 and 45 disruptions, respectively, recorded in bins of width two hundred years. Examining the trajectories of these disrupting clouds reveals that each of the three showers is caused by a cluster that has developed in the cloud distribution, under their mutual gravitational attraction, even though no clustering was present initially.

Unlike the WD, the MS star moves at a speed that is much higher than any of the clouds in the halo, and all cloud orbits around the MS star are much further from the marginally-bound (parabolic) condition. It follows that the MS star accretes a smaller fraction of the tidal debris it creates. Moreover, a given accretion rate yields a luminosity that is $\sim 10^3\times$ smaller for the MS star than for the WD, so the MS star in this simulation would never look like an RCB. Dusty winds, and eclipses by those winds, may arise from any clouds that come close to the MS star. However, because of the lower luminosity the wind speeds are much smaller and the eclipse shapes are therefore different to those of RCBs --- ingress would be much slower, for example. Wind speeds are in fact not much different from the orbital speed of the cloud, and as a result the dust wind would look quite different from the structure shown in figure \ref{fig:windstructure} --- e.g. it may contain self-intersections. Finally, with much less radiation pressure acting on the lumps of condensed \htwo\ in the tidal debris, any material that does accrete onto the MS star will not be greatly depleted in hydrogen.

\end{appendix}
\end{document}